\documentclass[a4paper,11pt]{article}
\usepackage{jheppub} % for details on the use of the package, please
\usepackage{etoolbox}
\usepackage[T1]{fontenc} % if needed
\usepackage{latexsym,amsmath,amsfonts,amssymb,amsthm}
\usepackage{amsmath}
\usepackage{mathtools}
\usepackage{xcolor}
\usepackage{tikz}
\usepackage{setspace}

\title{Modularity of Schur index, modular differential equations, and high-temperature asymptotics}

\author{Yiwen Pan$^1$,}

\affiliation{$^1$ School of Physics, Sun Yat-sen University,\\No. 135 Xingangxi Road, Guangzhou, Guangdong, China}

\author{Peihe Yang$^2$}

\affiliation{$^2$ Center for Joint Quantum Studies and Department of Physics, School of Science,\\Tianjin University, 135 Yaguan Road, Tianjin 300350, P. R. China}

\emailAdd{panyw5@mail.sysu.edu.cn}
\emailAdd{peiehe\_yang@tju.edu.cn}
\abstract{
In this paper we analytically explore the modularity of the flavored Schur index of 4d $\mathcal{N} = 2$ SCFTs. We focus on the $A_1$ theories of class-$\mathcal{S}$ and $\mathcal{N} = 4$ theories with $SU(N)$ gauge group. We work out the modular orbit of the flavored index and defect index, compute the dimension of the space spanned by the orbit, and provide complete basis for computing modular transformation matrices. The dimension obtained from the flavored analysis predicts the minimal order of the unflavored modular differential equation satisfied by the unflavored Schur index. With the help of modularity, we also study analytically the high-temperature asymptotics of the Schur index. In the high-temperature limit $\tau \to +i0$, we identified the (defect) Schur index of the genus-zero $A_1$ theories of class-$\mathcal{S}$ with the $S^3$-partition function of the $SU(2) \times U(1)^n$ star-shape quiver (with Wilson line insertion). In the identification, we observe an interesting relation between the linear-independence of defect indices and the convergence of the Wilson line partition functions.
}

\makeatletter
\patchcmd{\maketitle}{\@fpheader}{}{}{}
\makeatother

\begin{document} 
\maketitle

\flushbottom

%!TEX root = ../Schur index and modularity.tex

\section{Introduction}

Modularity is an important piece of data in two dimensional (2d) conformal field theory (CFT). For rational theories, modularity of the chiral algebra characters combined with the Verlinde formula yields the fusion rules \cite{Verlinde:1988sn}. Modularity connects the high and low temperature behavior of partition function, which leads to the Cardy formula for the density of states and the entropy of black holes through the holography principle \cite{CARDY1986186}. Modularity is also heavily exploited in the ``holomorphic modular bootstrap'' approach to classifying rational CFTs and analyze correlation functions \cite{Mathur:1988rx,Eguchi:1987qd,Naculich:1989NuPhB.323..423N,Chandra:2018pjq,Bae:2020xzl,Das:2020wsi,Mukhi:2020gnj,Kaidi:2021ent,Bae:2021mej,Das:2021uvd,Mahanta:2022fvl}.

In this paper, we study analytically the modularity and related aspects of Schur index of four dimensional (4d) $\mathcal{N} = 2$ superconformal field theories (SCFTs) \cite{Eleftheriou:2022kkv,ArabiArdehali:2023bpq,Jiang:2024baj}. The Schur index $\mathcal{I}$ is defined as the Schur limit of the $\mathcal{N} = 2$ superconformal index $\mathcal{I}(p, q, t)$,
\begin{equation}
  \mathcal{I}(\mathbf{b},q) = q^{\frac{c_\text{4d}}{2}}\operatorname{tr}_{\mathcal{H}} (-1)^F q^{E - R} \mathbf{b}^\mathbf{f} = \mathcal{I} (\mathbf{b};p, q, t)\Big|_{t \to q} \ .
\end{equation}
Here $E$ is the scaling dimension and $R$ is the $SU(2)_\mathcal{R}$ charge of a state (or a local operator) in the full Hilbert space $\mathcal{H}$, and $ \mathbf{f}, \mathbf{b}$ collectively denote the flavor Cartans and the flavor fugacities. The superconformal index and in particular the Schur index are important invariants of 4d $\mathcal{N} = 2$ SCFTs, which is robust against exactly marginal deformations. They are invaluable tools in checking dualities. 

Any 4d $\mathcal{N} = 2$ SCFT $\mathcal{T}$ contains a non-trivial $\frac{1}{4}$-BPS sector of local operators. Operators in the sector are referred to the Schur operators, which form a 2d chiral algebra $\mathbb{V}[\mathcal{T}]$ when restricted and twisted-translated on a plane \cite{Beem:2013sza,Beem:2014rza,Lemos:2014lua,Xie:2016evu,Kiyoshige:2020uqz}. The Schur index is simply counting the Schur operators with sign, graded by $E - R$ and flavor charges. Hence, the Schur index is identified with the vacuum character of $\mathbb{V}[\mathcal{T}]$, and therefore is of particular interest. Furthermore, when computing the Schur index one may additionally insert BPS non-local operators that preserve the supercharges used to define the Schur index. It is believed that these operators correspond to the non-vacuum modules of the chiral algebra $\mathbb{V}[\mathcal{T}]$, and a defect index is the character of the corresponding module \cite{Cordova:2015nma,Cordova:2016uwk,Cordova:2017mhb,Nishinaka:2018zwq,Pan:2021mrw,Guo:2023mkn}.

Similar to the characters of RCFTs, or those of the admissible Kac-Moody algebras \cite{Kac:1988qc}, the Schur index is expected to enjoy certain modularity. In \cite{Beem:2017ooy}, by identifying the Higgs branch of $\mathcal{T}$ with the associated variety of $\mathbb{V}[\mathcal{T}]$, a special null $\mathcal{N}_T$ implementing the nilpotency of the stress tensor $T$ is shown to exist in $\mathbb{V}[\mathcal{T}]$. The null leads to a unflavored modular differential equation, constraining the unflavored Schur index $\mathcal{I}(\mathbf{b} \to 1, q)$, and is useful in classifying 4D $\mathcal{N} = 2$ SCFT \cite{Kaidi:2022sng}. Additional solutions to the equation are likely to be unflavored non-vacuum module characters, and potentially correspond to insertion of non-local BPS operators. All the unflavored solutions form a vector valued modular form (vvmf), transforming in a representation of a suitable modular group. Additional null states may also lead to more unflavored modular differential equations that further constrain the unflavored characters. Among all the independent equations, the equation from $\mathcal{N}_T$ may not assume the minimal order: the equation with the minimal order could have non-zero Wronskian index, meaning that the modular form coefficients may contain poles in $\tau$ \cite{Beemetal}.

The associated chiral algebra of a 4d $\mathcal{N} = 2$ SCFT often harbors more complicated spectra compared with an RCFT \cite{2016arXiv161207423K,Arakawa:2016hkg,Beem:2017ooy,2023arXiv230409681L,Hatsuda:2023iwi}. To deal with the complexity, flavor refinement becomes necessary and flavored modular differential equations are expected to play an important role. Although this type of equations lacks systematic study, examples have shown that they enjoy some generalization of modularity, which we simply call quasi-modularity \cite{Zheng:2022zkm,Pan:2023jjw}. These observations encourage further investigation of the modularity of the flavored Schur index, in the absence or presence of non-local operators.

To study modularity, it is most convenient to work with closed-form expressions of the relevant index. Thanks to a series of recent development, the Schur index in closed-form of several classes of theories have become available \cite{Bourdier:2015wda,Pan:2021mrw,Huang:2022bry,Hatsuda:2022xdv,Du:2023kfu,Guo:2023mkn}.\footnote{See also some work on rewriting the Schur index with giant graviton partition function \cite{Beccaria:2024szi,Gaiotto:2021xce}} In particular, the flavored Schur index $\mathcal{I}_{g, n}$ and vortex defect \cite{Gaiotto:2012xa} indices $\mathcal{I}_{g, n}^\text{defect}(\kappa)$ of the $A_1$ class-$\mathcal{S}$ theories \cite{Gaiotto:2009we}, as well as $\mathcal{I}_{\mathcal{N} = 4}$ of several low rank $\mathcal{N} = 4$ theories can be concisely written as a finite combinations of the Jacobi $\vartheta$-functions and Eisenstein series. A simple formula for all the unflavored $\mathcal{N} = 4$ $SU(N)$ index is also available. With the well-known modularity of these special functions, we are able to work out explicitly the modular transformations of all the relevant indices. From these results, we construct the linear space $\mathcal{V}(\operatorname{ch})$ spanned by the modular orbit of an index $\operatorname{ch}$. In particular, for the $A_1$ class-$\mathcal{S}$ theories, we find that all the vortex defect index with even vorticity (including the index without defect) belong to one vector space $\mathcal{V}(\mathcal{I}_{g, n})$, and those with odd vorticity belong to another. The dimension of $\mathcal{V}(\mathcal{I}_{g, n})$ can be worked out,
\begin{align}
	\dim \mathcal{V}(\mathcal{I}_{g, n}) = & \ \sum_{\substack{k = 1\\ k = n \mod 2}}^{2g - 2 + n} (k + g) \ , & \ \text{when} ~ n > & \ 0\\
	\dim \mathcal{V}(\mathcal{I}_{g, n}) = & \ \sum_{\substack{k = 0\\ k = n \mod 2}}^{2g - 2 + n} (k + g) & \text{when} ~ n = & \ 0\ .
\end{align}
This dimension coincides the predicted minimal order of the unflavored modular differential equation of the $A_1$ theories in \cite{Beemetal}, thus providing a flavored proof of their conjectural formula. Some simple basis for $\mathcal{V}(\mathcal{I}_{g, n})$ is also constructed. We also investigate the modular orbit of the flavored and unflavored $\mathcal{N} = 4$ $SU(N)$ index, and propose a conjecture for the dimension and complete basis of $\mathcal{V}(\mathcal{I}_{\mathcal{N} = 4})$, which also predicts the minimal order of the equation.

The Schur index $\mathcal{I}_\mathcal{T}$ with $\tau = i \beta$ can be viewed as an $S^3 \times S_\beta^1$ partition function \cite{Pan:2019bor,Dedushenko:2019yiw}, where $\beta$ is the circumference of $S^1$, or the inverse temperature $\beta = 1/kT$. Using modularity, we compute analytically the high temperature asymptotics of the flavored (defect) Schur index of $A_1$ class-$\mathcal{S}$ theories as $\beta \to 0$, or $\tau \to +i0$, generalizing the discussions in \cite{Ardehali:2015bla,ArabiArdehali:2023bpq}. In the limit $\beta \to 0$, the 4D $\mathcal{N} = 2$ theory $\mathcal{T}$ effectively dimensionally reduces to a 3d $\mathcal{N} = 4$ theory $\mathcal{T}^\text{3d}$, which has a Lagrangian mirror dual $\check{\mathcal{T}}^\text{3d}$ \cite{Gadde:2011ia,Nishioka:2011dq,Buican:2015hsa}. The Higgs branch $\mathcal{H}$ remains unchanged under the reduction. The associated chiral algebra $\mathbb{V}[\mathcal{T}]$ is mapped to a deformation quantization of the coordinate ring of $\mathcal{H}$ \cite{Dedushenko:2016jxl,Dedushenko:2019mzv,Pan:2020cgc,Dedushenko:2019mnd}, and the Schur index in four dimension is expected to reduce to the $S^3$ partition function of $\mathcal{T}^\text{3d}$ or $\check{\mathcal{T}}^\text{3d}$. For $g = 0, n \ge 3$, we explicitly perform the reduction starting from the closed-form expression of $\mathcal{I}_{g, n}$, and match the result with the $S^3$-partition function $Z^{S^3}$ of the $SU(2) \times U(1)^n$ star-shape quiver \cite{Benini:2010uu}, providing a closed-form expression for $Z^{S^3}$. Furthermore, we consider the defect index $\mathcal{I}^\text{defect}_{g = 0, n}(\kappa)$ and its reduction. Exploiting the two-dimensional $q$-Yang-Mills description of the $\mathcal{I}^\text{defect}_{g, n}(\kappa)$ \cite{Gadde:2011ik,Gaiotto:2012xa,Alday:2013kda}, we establish an equality between the reduced defect index and $S^3$-partition function of a Wilson line $W_{j = \kappa/2}$ in the 3d mirror dual $\check{\mathcal{T}}^\text{3d}$,
\begin{equation}
  \mathcal{I}_{g = 0,n}^\text{defect}(\kappa) \xrightarrow{\tau \to +i0}
  Z^{S^3}_{\check{\mathcal{T}}^\text{3d}}(W_{j = \kappa/2}) \ , \qquad \kappa = 0, 1,2, ..., n - 3\ .
\end{equation}
In particular, we observe a peculiar relation between the linear-independence of $\mathcal{I}_{0, n}^\text{defect}(\kappa)$ and the convergence of the $S^3$-partition function of the Wilson loop $W_{j = \kappa/2}$.

This paper is organized as follows. In section \ref{section:introduction}, we review the Schur index and its modularity. In section \ref{section:modularity} we analyze the modular property of the flavored (defect) Schur index $\mathcal{I}_{g, n}$, $\mathcal{I}_{g, n}^\text{defect}(\kappa)$ for $A_1$ class-$\mathcal{S}$ theories and $\mathcal{N} = 4$ theories, establishing the vector space spanned by the modular orbit and its basis. In section \ref{section:high-temperature-behavior}, we analytically compute the high temperature asymptotics of the flavored (defect) Schur index of $A_1$ class-$\mathcal{S}$. We study the high temperature limit of $\mathcal{I}_{g = 0, n}$, $\mathcal{I}_{g = 0, n}^\text{defect}(\kappa)$ and identify the result from the viewpoint of $S^3$-partition function of the 3d mirror dual.

%!TEX root = ../Schur index and modularity.tex

\section{Chiral algebra\label{section:introduction}}

\subsection{Chiral algebra and Schur index\label{section:chiral-algebra}}

Any 4d $\mathcal{N} = 2$ local unitary SCFT $\mathcal{T}$ has been shown to contain a nontrivial set of local BPS operators that are annihilated by the four supercharges $Q_-^1, S_1^-, \tilde Q_{2\dot -}, \tilde S^{2 \dot -}$, referred to as the Schur operators \cite{Beem:2013sza}. These Schur operators, viewed as suitable cohomology classes restricted on a two dimensional plane in the spacetime, form a nontrivial two dimensional chiral algebra (or, vertex operator algebra, and VOA for short) $\mathbb{V}[\mathcal{T}]$. We will refer to this fact simply as the SCFT/VOA correspondence.

The chiral algebra $\mathbb{V}[\mathcal{T}]$ is always nontrivial. It must contain a Virasoro subalgebra with the central charge $c_\text{2d} = -12 c_\text{4d}$, where $c_\text{4d}$ is the four dimensional c-central charge. If the 4d SCFT enjoys a flavor symmetry $G$ with flavor anomaly $k_\text{4d}$, $\mathbb{V}[\mathcal{T}]$ must contain a corresponding affine Kac-Moody algebra $\widehat{\mathfrak{g}}_{k_\text{2d}}$, where the level $k_\text{2d}$ is determined by $k_\text{2d} = - \frac{1}{2}k_\text{4d}$. The chiral algebra serves as an important invariant of 4d $\mathcal{N} = 2$ SCFT, and is identified as a 2d topological field theory (TQFT) for theories of class-$\mathcal{S}$ \cite{Beem:2014rza}. 

4d unitarity enforces 2d non-unitarity condition $c_\text{2d} < 0$ \footnote{Some non-unitary chiral algebras in the SCFT/VOA correspondence may have close relation to unitary ones, for example, through unitarisation \cite{MUKHI1990263,Chandra:2018pjq,Buican:2017rya,MdAbhishek:2023vgp}.}. In general, the chiral algebra $\mathbb{V}[\mathcal{T}]$ is believed to be quasi-lisse \cite{Arakawa:2016hkg,Beem:2017ooy,2023arXiv230409681L}\footnote{A chiral algebra is quasi-lisse if its associated variety has finitely many symplectic leaves. lisse chiral algebra has a zero-dimensional associated variety. Rational chiral algebras are special cases of lisse algebras.}, which is a significant generalization of the well-known rational chiral algebra underlying rational conformal field theories (RCFTs). This implies that theories with a nontrivial Higgs branch is automatically non-rational, since it is identified with the associated variety of the associated chiral algebra. These general chiral algebras have much more complicated representation theory compared to the rational ones, as one may need to account for the logarithmic or non-ordinary modules.

The Schur index $\mathcal{I}_\mathcal{T}(q)$ is $t \to q$ limit \cite{Gadde:2011uv} of the full superconformal index $\mathcal{I}(p,q,t)$ \cite{Kinney:2005ej,Rastelli:2014jja}
\begin{align}
  \mathcal{I}(p,q,t)
  \coloneqq \operatorname{tr} (-1)^F
    p^{\frac{1}{2}(\Delta - 2j_1 - 2 \mathcal{R} - r)}
    q^{\frac{1}{2}(\Delta + 2j_1 - 2 \mathcal{R} - r)}
    t^{\mathcal{R} + r}
    e^{- \beta \{\tilde Q_{2 \dot -}, \tilde S^{2 \dot -}\}} b^f\ .
\end{align}
Here $f$ collectively denotes flavor Cartan generators, and $b$ are the corresponding flavor fugacities. The Schur index $\mathcal{I}_\mathcal{T}(q)$ counts the Schur operators with sign. As an index, $\mathcal{I}_\mathcal{T}(q)$ is independent of marginal deformations of the theory, providing an invariants of superconformal field theories. For the theory of class-$\mathcal{S}$, this invariant property can be reinterpreted as a 2d Topological Field Theory (TQFT) given by the zero-area $q$-deformed Yang-Mills theory on punctured Riemann surface $\Sigma_{g,n}$. From the point of view of the associated chiral algebra $\mathbb{V}[\mathcal{T}]$, the Schur index is simply the character of the vacuum module. The Schur index for a wide range of 4d $\mathcal{N} = 2$ SCFTs have been computed \cite{Beem:2013sza}. In particular, exploiting the invariance under marginal deformation or the technique of supersymmetric localization, the Schur index for Lagrangian theories can be organized into a multivariate contour integral of elliptic functions. Recently, there have been efforts to evaluate these contour integral analytically in closed form \cite{Bourdier:2015wda,Bourdier:2015sga,Pan:2021mrw,Hatsuda:2022xdv,Huang:2022bry}. In the following discussions we will mainly exploit the analytic results in \cite{Pan:2021mrw} where the Schur index of Lagrangian theories is reorganized in to finite sums of products of Eisenstein series, where the modular property of the latter has been studied extensively.

The theories we focus on will be the $A_1$ theories of class-$\mathcal{S}$, and $\mathcal{N} = 4$ theories with $SU(N)$ gauge groups. The $A_1$ theories $\mathcal{T}_{g, n}$ of class-$\mathcal{S}$ are 4d $\mathcal{N} = 2$ SCFTs from compactifying the 6d $(0,2)$ SCFT of type $A_1$ on a genus-$g$ Riemann surface $\Sigma_{g,n}$ with $n$ punctures. In this simple setup, there is only one type of regular puncture, and therefore the SCFT $\mathcal{T}_{g, n}$ is completely determined by $(g, n)$. Each puncture corresponds to an $SU(2)$ flavor subgroup, and also an $\widehat{\mathfrak{su}}(2)_{-2}$ subalgebra in the associated chiral algebra. The theory $\mathcal{T}_{g, n}$ usually admits different mutually $S$-dual gauge theory descriptions corresponding to different pants decompositions of the Riemann surface, where each long tube or handle corresponds to an $SU(2)$ gauge group. The flavored Schur index $\mathcal{I}_{g, n}$ of $\mathcal{T}_{g, n}$ computed in \cite{Pan:2021mrw} reads,
\begin{align}\label{Ign}
  \mathcal{I}_{g, n \ge 1} = & \ \frac{i^n}{2} \frac{\eta(\tau)^{n + 2g - 2}}{\prod_{j = 1}^{n} \vartheta_1(2 \mathfrak{b}_j)}
  \sum_{k = 1}^{n + 2g - 2}\lambda_k^{(n + 2g - 2)} 
  \sum_{\alpha_j = \pm}\bigg(\prod_{j = 1}^{n}\alpha_j\bigg)
  E_k\begin{bmatrix}
    (-1)^n \\ \prod_{j = 1}^{n}b_j^{\alpha_j}
  \end{bmatrix} \ , \\
  \mathcal{I}_{g \ge 1, 0} = & \ \frac{1}{2}\eta(\tau)^{2g - 2}\sum_{k = 1}^{g - 1}
  \lambda_{2k}^{2g - 2}\left(E_{2k} + \frac{B_{2k}}{(2k)!}\right) \ ,
\end{align}
where $b_{1, 2, ..., n}$ are the flavor fugacities associated to each puncture. See appendix \ref{app:special-functions} for a quick review of Jacobi $\vartheta_i$ functions and the Eisenstein series. The $\lambda_k^{(2g - 2 + n)}$ denote rational numbers determined by a set of recursive algebraic equations,
\begin{align}
	\lambda^{(\text{even})}_0 = & \ \lambda^{(\text{odd})}_\text{even} = \lambda^{(\text{even})}_\text{odd} = 0\;, \qquad \lambda^{(1)}_1 = \lambda^{(2)}_2 = 1 \; , \\
	\lambda^{(2k + 1)}_1 = & \ \sum_{\ell = 1}^{k} \lambda_{2 \ell}^{2k} \left( \mathcal{S}_{2\ell} - \frac{B_{2\ell}}{(2\ell)!} \right)  \;,  \\
	\lambda^{(2k + 1)}_{2m + 1} = & \ \sum_{\ell = m}^{k}\lambda_{2\ell}^{(2k)} \mathcal{S}_{2(\ell - m)} \;, \quad
	\lambda^{(2k + 2)}_{2m + 2} = \sum_{\ell = m}^{k}\lambda^{(2k + 1)}_{2\ell + 1}\mathcal{S}_{2(\ell - m)}\; ,
\end{align}

The Schur index of the $\mathcal{N} = 4$ theories are also studied extensively, and closed forms are available using different methods \cite{Bourdier:2015wda,Pan:2021mrw,Hatsuda:2022xdv,Huang:2022bry,Du:2023kfu}. In terms of Eisenstein series, the flavored Schur index of the $SU(3)$ theory is given by
\begin{align}
  \mathcal{I}_{\mathcal{N} = 4 \ SU(3)} = - \frac{1}{8} \frac{\vartheta_4(\mathfrak{b})}{\vartheta_4(3 \mathfrak{b})}
  \biggl(- \frac{1}{3}
  + 4 E_1 \begin{bmatrix}
    -1 \\ b
  \end{bmatrix}^2
  - 4 E_2 \begin{bmatrix}
    + 1 \\ b^2
  \end{bmatrix}
  \biggr) \ .
\end{align}
The index for the $SU(4)$ case is
\begin{align}
    \mathcal{I}_{\mathcal{N}=4 SU(4)}
    = & \ \frac{\vartheta_4(\mathfrak{b})}{\vartheta_1(4 \mathfrak{b})}\Bigg(-\frac{i}{3} E_3\left[\begin{array}{c}
    -1 \\
    b^3
    \end{array}\right]+\frac{i}{2} E_1\left[\begin{array}{c}
    -1 \\
    b
    \end{array}\right] E_2\left[\begin{array}{c}
    1 \\
    b^2
    \end{array}\right]-\frac{i}{6} E_1\left[\begin{array}{c}
    -1 \\
    b
    \end{array}\right]^3 \nonumber\\
    & \qquad \qquad\qquad \qquad \qquad\qquad\qquad \ +\frac{i}{24} E_1\left[\begin{array}{c}
    -1 \\
    b
    \end{array}\right]+\frac{i}{24} E_1\left[\begin{array}{c}
    -1 \\
    b^3
    \end{array}\right]\Bigg) \ .
\end{align}
General closed-form formula for all $SU(N)$ gauge group is available in \cite{Hatsuda:2022xdv}, though the modular property of the involved special functions remain to be written down explicitly. In \cite{Pan:2021mrw}, the unflavored Schur index for all $\mathcal{N} = 4$ $SU(N)$ theories is also conjectured,

\subsection{Modular differential equations \label{section:modular-differential-equations}}

The goal of this paper is to study the representation theory of $\mathbb{V}[\mathcal{T}]$ from the modularity perspective. A rational chiral algebra has finitely many irreducible modules corresponding to the chiral primary operators. The unflavored characters of these modules form a weight-zero vector-valued modular form, for example in the bosonic case,
\begin{equation}
  \operatorname{ch}_i \bigl(- \frac{1}{\tau}\bigr) = \sum_{j} S_{ij} \operatorname{ch}_j(\tau) \ , \qquad
  \operatorname{ch}_i (\tau + 1) = \sum_{j}T_{ij} \operatorname{ch}_j(\tau) \ .
\end{equation}
Here $S_{ij}$, $T_{ij}$ belongs to certain representation of the modular group $SL(2, \mathbb{Z})$. Using $S_{ij}$ one may extract the fusion algebra using the Verlinde formula \cite{Verlinde:1988sn}.

For a rational theory with $n$ characters, modularity of characters is closely tied to the notion of modular differential equations: modularity, combined with the Wronskian method, leads to a modular differential equation constraining all the unflavored characters \cite{Eguchi:1987qd,Mathur:1988na},
\begin{equation}\label{eq:unflavored-MDE}
  \Bigg[D_q^{(n)} + \sum_{r = 0}^{n - 1}\phi_r(\tau) D^{(n - r)}_q\Bigg] \operatorname{ch} = 0 \ .
\end{equation}
Here $\phi_r$ are weight-$(2r)$ modular forms with respect to a suitable modular group, and $D_q^{(n)}$ is the $n$-th order Serre derivative,
\begin{equation}
  D_q^{(n)} \coloneqq \partial_{(2n - 2)} \circ \partial_{(2n - 4)} \circ \cdots \circ \partial_{(2)} \circ \partial_{(0)}\ , \qquad
  \partial_{(k)} \coloneqq q \partial_q + k E_2(\tau) \ .
\end{equation}
The operator $D_q^{(n)}$ maps a weight-$2k$ modular form to a weight-$(2k + 2n)$ modular form. Obviously, (\ref{eq:unflavored-MDE}) transforms covariantly under the relevant modular group. This modular differential equation has been utilized in classifying RCFTs \cite{Chandra:2018pjq,Mukhi:2020gnj,Das:2020wsi,Kaidi:2021ent,Das:2021uvd,Bae:2020xzl,Bae:2021mej,Duan:2022kxr}.

In the context of SCFT/VOA correspondence, the relevant chiral algebras are in general non-rational \cite{Arakawa:2016hkg,Arakawa:2018egx,Arakawa:2023cki,Arakawa:2024ejd}. These algebras contain modules whose weight spaces are infinite dimensional, and flavor refinement becomes necessary since the corresponding characters has no unflavoring limit. Hence, flavored modular differential equations are expected to play a crucial role in this correspondence for better spectroscopy \cite{Peelaers,Pan:2021ulr,Zheng:2022zkm,Pan:2023jjw}. To derive these equations, one applies the Zhu's recursion relation to the (super-)trace of a level-$n$ null state $|\mathcal{N}\rangle$ with respect to a module $M$ of $\mathbb{V}[\mathcal{T}]$ \cite{zhu1996modular,Gaberdiel:2008pr,Tuite:2014fha,Beem:2017ooy},
\begin{equation}
  0 = \operatorname{tr}_{M} \mathcal{N}_0 q^{L_0 - \frac{c_\text{24}}{24}} \mathbf{b}^\mathbf{f} \ ,
\end{equation}
turning it into a weight-$(2n)$ partial differential equation of the form
\begin{equation}
  \sum_{\substack{k, \ell_1, ..., \ell_r \\ 2k + \ell_1 + ... + \ell_r \le 2n}} \phi_{k, \ell_1, ..., \ell_r}(\tau,b_1, ..., b_r)D_q^{(k)} D_{b_i}^{\ell_1} ... D_{b_r}^{\ell_r} \operatorname{ch}_M = 0 \ ,
\end{equation}
where $\phi_{k, \ell_1, ..., \ell_r}$ are quasi-Jacobi forms with weight $2n - ( 2k + \ell_1 + ... + \ell_r)$, such as products of the Eisenstein series, and $D_{b_i} \coloneqq b_i \partial_{b_i}$ are derivatives with respect to the flavor fugacities $b_i$. Generically, there may be several null states that lead to different and independent nontrivial  equations, and they together constrain all the characters of the theory.

Some of these flavored equations can be unflavored. For example, the special level-$(2n_T)$ null $\mathcal{N}_T$ that implements the nilpotency of the stress tensor $T \in \mathbb{V}[\mathcal{T}]$,
\begin{equation}
  (L_{-2})^{n_T} |0\rangle = \mathcal{N}_T + \varphi, \quad
  \varphi \in C_2(\mathbb{V}[\mathcal{T}]), \quad n_T \in \mathbb{N}_{\ge 1}  \ ,
\end{equation}
is expected to give rise to a unflavored modular differential equation \cite{Beem:2017ooy,Kaidi:2022sng}. Here $C_2(\mathbb{V}[\mathcal{T}])$ is the $C_2$ subspace spanned by all the states of the form $a_{-h_a - 1}|b\rangle$. A unflavored equation takes the same general form as in (\ref{eq:unflavored-MDE}), enjoying the usual modularity under $SL(2, \mathbb{Z})$ or a suitable congruence subgroup. In general $\phi_r$ can have pole in $\tau$, contributing a nonzero Wronskian index, however, the weight-$(2n_T)$ equation from $\mathcal{N}_T$ has a zero Wronskian index, following from the Zhu's recursion. When more than one modular differential equations exist at a given weight, say, $2n$, they may combine into giving a lower weight equation with non-zero Wronskian index. Schematically,
\begin{align}
  \Bigg[D_q^{(n)} + \sum_{r = 0}^{n - 1}\phi_r(\tau) D^{(n - r)}_q\Bigg] \operatorname{ch} = & \ 0, \qquad
  \Bigg[D_q^{(n)} + \sum_{r = 0}^{n - 1}\varphi_r(\tau) D^{(n - r)}_q\Bigg] \operatorname{ch} =  0 \nonumber \\
  \Rightarrow \Bigg[\sum_{r = 0}^{r_\text{max}}(\varphi_r - \phi_r)(\tau) D^{(n - r)}_q\Bigg] \operatorname{ch} = & \ 0 \ ,
\end{align}
and one may divide the last equation by the non-zero $\varphi_{r_\text{max}} - \phi_{r_\text{max}}$ to obtain a weight $2(n - r_\text{max})$ equation with a non-zero Wronskian index. Importantly, it may happen that $n - r_\text{max} < n_T$ \cite{Beemetal}.

There are also equations that do not have well-defined unflavoring limit, and they will be important to constrain those characters that cannot be unflavored. Flavored modular differential equations have quasi-Jacobi forms as coefficients, and therefore do not possess the usual modularity. Luckily, although their general property remains under-explored, known examples of flavored equations have been shown to enjoy quasi-modularity (or, almost covariance): under the action of the modular group a higher weight equation transforms into a linear combination of lower weight equations \cite{Zheng:2022zkm,Pan:2023jjw}. Such property also suggests hidden modular relations among the null states in a chiral algebra. As a result, flavored solutions to the equations also form a representations of the modular group. We will be interested in the dimension and basis of this representation, and the counterpart in the unflavoring limit.

\subsection{Defect index \label{section:defect-index}}

In a 4d $\mathcal{N} = 2$ superconformal field theory, one can consider non-local operators that preserve certain amount of supersymmetry. 

The first type of non-local operators are surface defects that are perpendicular to the plane where the associated chiral algebra resides, preserving an a $\mathcal{N} = (2,2)$ subalgebra \cite{Cordova:2017mhb,Bianchi:2019sxz,Dedushenko:2019yiw}. This subalgebra contains the four supercharges used to define the chiral algebra, and therefore a defect index $\mathcal{I}^\text{defect}$ counting the defect operators in the relevant cohomology can be defined. These defect operators are acted on by the bulk Schur operators through bulk-defect operator product expansion (OPE), forming a (twisted-)module of the associated chiral algebra. Therefore it is believed that the defect index should be interpreted as the character of the corresponding module; if this is the case, the defect index satisfy all the flavored modular differential equations. 

Consider an $A_1$ theory $\mathcal{T}_{g, n}$ of class-$\mathcal{S}$ associated to an $n$-punctured genus-$g$ Riemann surface. Following \cite{Gaiotto:2012xa}, we consider inserting a vortex surface defect with vorticity $\kappa \in \mathbb{N}$ by first gauging in a free trinion theory $\mathcal{T}_{0,3}$ to obtain an UV theory $\mathcal{T}_{g, n + 1}$, and subsequently giving a position-dependent background $\sim z^\kappa$ to a baryonic operator. The background triggers an RG-flow that leads to an IR superconformal theory coupled to a vortex surface defect. The Schur index can be easily computed by residue,
\begin{equation}
	\mathcal{I}^\text{defect}_{g, n}(\kappa) \coloneqq 2(-1)^\kappa q^{- \frac{(\kappa + 1)^2}{2}}\mathop{\operatorname{Res}}_{b_{n + 1} \to q^{\frac{1}{2} + \frac{\kappa}{2}}}  \frac{\eta(\tau)^2}{b_{n + 1}} \mathcal{I}_{g, n + 1}(b_1, \ldots, b_{n + 1}) \ .
\end{equation}
Substituting the closed form expression for $\mathcal{I}_{g, n + 1}$, the surface defect index can be written down in closed form,
\begin{align}\label{eq:defect-index}
  \mathcal{I}^{\text{defect}}_{g, n}(\kappa) = (-1)^\kappa & \ \frac{\eta(\tau)^{n + 2g - 2}}{\prod_{i = 1}^n\vartheta_1(2 \mathfrak{b}_i)}\\
  & \ \times \sum_{\alpha_i = \pm}\left(\prod_{i = 1}^{n}\alpha_i\right)
  \sum_{\ell = 1}^{n + 1 + 2g - 2}\tilde\lambda^{n + 1 + 2g - 2}_\ell(\kappa + 1) E_\ell \left[\begin{matrix}
      (-1)^{n + \kappa}\\ \prod_{i = 1}^{n}b_i^{\alpha_i}
  \end{matrix}\right]\ , \nonumber
\end{align}
where $\tilde \lambda$ are rational numbers defined by
\begin{align}
  \tilde \lambda_\ell^{(n + 1 + 2g -2)}(\kappa) \coloneqq \sum_{\ell' = \ell}^{n + 1 + 2g -2} \left(\frac{\kappa}{2}\right)^{\ell' - \ell} \frac{1}{(\ell' - \ell)!} \lambda_{\ell'}^{(n + 1 + 2g - 2)}\ .
\end{align}
When $\kappa = 0$, the defect is trivial, and $\mathcal{I}^\text{defect}_{g, n}(\kappa) = \mathcal{I}_{g, n}$. In many known cases, the vortex defect index with $\kappa = \text{even}$ has been shown to satisfy the same flavored or unflavored modular differential equations as the Schur index \cite{Zheng:2022zkm}, therefore supporting the above mentioned interpretation. When $\kappa = \text{odd}$, the defect index satisfies the twisted version of the equations, which indicates that they such defects give rise to twisted modules of the chiral algebra.

The superconformal index in the presence of vortex surface defects can be computed by acting on the index with a difference operator $\mathfrak{G}_{r, s}$ \cite{Gaiotto:2012xa,Alday:2013kda}. For vortex surface defects that survive the Schur limit, the defect index can be computed by \footnote{Here we have renormalized the difference operator to match with our convention.}
\begin{align}
  \mathfrak{G}_{0,\kappa} \mathcal{I}_{g, n} \coloneqq & \ q^{ - \frac{\kappa(\kappa+2)}{2}}
  \sum_{m = 0}^{\kappa} \Bigg[
  \frac{
    \prod_{n = 0}^{m - 1} (1 - q^{n + 1}) \prod_{n = 0}^{\kappa - m - 1}(1 - q^{n + 1})
  }{
    \prod_{n = 1}^{m}(1 - q^{-n}) \prod_{n = 1}^{\kappa - m}(1 - q^{-n})
  } \nonumber\\
  & \ \times \prod_{n = 0}^{m - 1} \frac{1 - q^{\kappa - 2m + n + 1}b^2}{1 - q^{-\kappa + m + n}b^{-2}}
  \prod_{n = 0}^{\kappa - m - 1} \frac{1 - q^{- \kappa + 2m + n + 1}b^{-2}}{1 - q^{-m + n}b^2} \mathcal{I}_{g, n}(q^{\frac{1}{2}\kappa - m}b)\Bigg]\\
  = & \ \mathcal{I}^\text{defect}_{g,n}(\kappa) \ . \nonumber
\end{align}
Here $b$ can be any one of the $n$ fugacities $b_i$, and $\mathcal{I}(q^{\frac{1}{2}\kappa - m}b)$ means shifting the one chosen fugacity $b$ to $q^{\frac{1}{2}\kappa - m}b$ while the other $b_i$'s remains unchanged. Note that to establish the equality concretely as $q$-series, the shift $b \to q^{\frac{1}{2}\kappa - m}b$ should be performed at the level of analytic function using properties of $E_k\big[\substack{\pm 1\\b}\big]$ before series expansion.

The Schur index of class-$\mathcal{S}$ theory can be identified as 2d $q$-Yang-Mills partition function on the punctured Riemann Riemann surface \cite{Gadde:2011ik}. For $A_1$ type theories with only regular punctures, the identification reads
\begin{equation}
  \mathcal{I}_{g, n} = q^{- \frac{c_\text{2d}}{24}} \sum_{j \in \frac{1}{2}\mathbb{N}} C_j(q)^{2g - 2 + n} \prod_{i = 1}^{n}\psi_j(b_i,q) \ .
\end{equation}
Here $\psi_j(b_i)$ denotes the wave function specifying the boundary condition of the 2d Yang-Mills gauge field on the $i$-th puncture,
\begin{equation}
  \psi_j(b_i) = \frac{\chi_j(b_i)}{(q;q)(q b_i^2;q)(q b_i^{-2};q)} \ ,
\end{equation}
and $C_j(q)$ denotes the three-point function of the 2d theory,
\begin{equation}
  C_j(q) = \frac{(q^2 ;q)}{\dim_q \mathcal{R}_j} \ , \qquad
  \dim_q \mathcal{R}_j \coloneqq \frac{q^{\frac{2j+1}{2}} - q^{- \frac{2j+1}{2}}}{q^{\frac{1}{2}} - q^{- \frac{1}{2}}} \ .
\end{equation}

As shown in \cite{Gaiotto:2012xa,Alday:2013kda}, the wave function $\psi_j(b_i)$ is actually an eigenfunction of the difference operator $\mathfrak{G}_{0,\kappa}$ (when acting on $b_i$),
\begin{equation}
  \mathfrak{G}_{0, \kappa} \psi_j(b_i) = (-1)^\kappa \frac{q^{\frac{(\kappa + 1)(2j + 1)}{2}} - q^{ - \frac{(\kappa + 1)(2j + 1)}{2}}}{
    q^{\frac{2j + 1}{2}} - q^{ - \frac{2j + 1}{2}}
  } \psi_j(b_i) \coloneqq S_{\kappa j} \psi_j(b_i) \ .
\end{equation}
Therefore, one may rewrite the defect index in $q$-Yang-Mills theory as
\begin{equation}
  \mathcal{I}_{g,n}^\text{defect}(\kappa)
  = q^{- \frac{c_\text{2d}}{24}}
  \sum_{j \in \frac{1}{2}\mathbb{N}}
  S_{\kappa j}
  C_j(q)^{2g - 2 + n} \prod_{i = 1}^{n} \psi_j(b_i) \ .
\end{equation}
Series expanding both sides easily verifies the statement.

Besides BPS surface operators, one may also insert BPS line operators into the SCFT $\mathcal{T}$ that preserve the supercharges used to define the Schur index. As discussed in \cite{Gang:2012yr,Cordova:2016uwk}, such line operators extend along rays or a full lines through the origin that are perpendicular to the chiral algebra plane. The simplest such operators are Wilson line operators charged under the gauge groups of the theory. For a theory of class-$\mathcal{S}$, there may be multiple gauge groups in each gauge theory description, and the Wilson line operator may be charged under any of them.

\begin{figure}[t]
  \centering
  \includegraphics[width=0.8\textwidth]{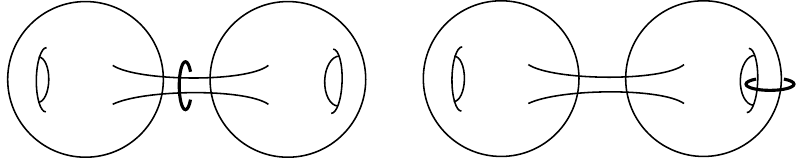}
  \caption{The figure on the left depicts a type-2 Wilson line. The one on the right is type-1.\label{fig:Wilson-line}}
\end{figure}

For $A_1$ theories of class-$\mathcal{S}$, we may compute analytically the Schur correlators of any BPS Wilson lines. These theories are labeled by a punctured Riemann surface $\Sigma_{g, n}$, whose gauge theory descriptions are correspond to pants decompositions. A Long tube or a handle in a decomposition corresponds an $SU(2)$ gauge group under which a Wilson line may be charged. As discussed in \cite{Guo:2023mkn}, for the purpose of calculating Schur correlators, we may consider Wilson lines charged under only one such $SU(2)$ as the basic objects. There are two types of such Wilson lines. A type-1 Wilson line corresponds to a tube or handle that when cut, the Riemann surface remains connected, while for a type-2 Wilson line the Riemann surface will become disconnected. A type-1 Wilson line index will be denoted by $\langle W_j\rangle_{g, n}^{(1)}$, and $\langle W_j\rangle_{g_1, n_1; g_2, n_2}^{(2)}$ for type-2; here $g_i, n_i$ are the genus and puncture numbers of the two connected components of the Riemann surface after cutting the tube corresponding to the Wilson line. See Figure \ref{fig:Wilson-line} for an illustration.

Consider a Wilson line transforming in the spin-$j$ representation of $SU(2)$. If the line is type-1, the line index reads \cite{Guo:2023mkn}
\begin{align}\label{eq:type-1-Wilson-line-index}
  \langle W_j \rangle_{g,n}^{(1)} = & \ \delta_{j \in \mathbb{Z}}\delta_{j,0} \mathcal{I}_{g, n} \\
  & \ - \frac{\delta_{j \in \mathbb{Z}}}{2}
  \biggl(\prod_{i = 1}^{n}\frac{\eta(\tau)}{\vartheta_1(2 \mathfrak{b}_i)}\biggr)
  \sum_{\substack{m = -j \\  m \ne 0}}^{+ j}
  \bigg[\frac{\mathfrak{\eta(\tau)}}{q^{m /2} - q^{- m /2}}\bigg]^{2g - 2} \prod_{i = 1}^{n} \frac{b_i^m - b_i^{-m}}{q^{m/2} - q^{- m/2}}\ .\nonumber
\end{align}
Note that even though for a given $(g, n)$ there may be different choices of gauge theory description, and for each description there may be different choices of an $SU(2)$ gauge group, the Schur index of a type-1 Wilson line is independent of any of these choices. Note also that whenever $j \in \mathbb{Z} + \frac{1}{2}$, the index vanishes. The Type-2 Wilson line index generally takes a more complicated form and involves the following type of integrals:
\begin{equation}\label{type2integral}
	\oint\frac{dz}{2\pi iz} z^n E_k\begin{bmatrix}(-1)^{n_1}\\ z\end{bmatrix}E_\ell\begin{bmatrix}(-1)^{n_2}\\az\end{bmatrix}.
\end{equation}
For example, when $n_1, n_2$ are both even, the integral can be performed as 
\begin{align}\label{eq:type-2-Wilson-line-index}
	& \ \langle W_j\rangle^{(2)}_{g_1, n_1; g_2, n_2}
	= \delta_{j \in \mathbb{Z}}\mathcal{I}_{g, n} \\
	& \ + \frac{\eta(\tau)^{2g - 2 + n}}{2 \prod_{i = 1}^{n}\vartheta_1(2 \mathfrak{b}_i)} \sum_{\substack{m = -j \\ m\ne 0}}^{j}\sum_{\substack{\ell = 0 \\ \ell = n_i \operatorname{mod} 2}}^{\max(2g_i - 2 + n_i)} \Lambda_\ell^{(g_1, n_1; g_2, n_2)}
	\sum_{\alpha_j=\pm}\biggl(\prod_{i = 1}^{n}\alpha_i\biggr) E_\ell \begin{bmatrix}
		1 \\ \prod_{i = 1}^{n}b_i^{\alpha_i}
	\end{bmatrix} \ , \nonumber
\end{align}
where $g = g_1 + g_2$, $n = n_1 + n_2 - 2$, $\Lambda_\ell$ are some rational functions of $q, b_i$ whose exact form is not important for our discussion. Note that the index does not vanish even when $j \in \mathbb{Z}+ \frac{1}{2}$ since only the $\mathcal{I}_{g, n}$ term is proportional to $\delta_{j \in \mathbb{Z} }$. For other cases, the results from \eqref{onem} and \eqref{twom} imply that the type-2 Wilson index contains a linear combination of $\sum_{\alpha} (\prod_j \alpha_j)E_\ell \Big[\substack{1\\\prod_{i=1}^nb_i^{\alpha_i}}\Big]$ when both $n_i$ are odd, and a combination of $\sum_{\alpha} (\prod_j \alpha_j)E_\ell \Big[\substack{-1\\\prod_{i=1}^nb_i^{\alpha_i}}\Big]$ when one $n_i$ is odd and the other is even, where the coefficients are again rational functions of $q, b_i$.
%!TEX root = ../Schur index and modularity.tex

\section{Modularity\label{section:modularity}}
As reviewed in section \ref{section:modular-differential-equations}, the unflavored Schur index $\mathcal{I}^\text{unflavored}$ of a 4d $\mathcal{N} = 2$ theory must satisfy a certain unflavored modular differential equation of the form
\begin{equation}
	(D_q^{(n)} + \sum_{r = 0}^{n - 1}\phi_{2r} D_q^{(n - r)}) \mathcal{I}^\text{unflavored} = 0 \ ,
\end{equation}
where $\phi_{2r}$ denote $SL(2, \mathbb{Z})$ or $\Gamma^0(2)$ modular forms of weight-$2r$. Under appropriate modular transformations, the equation is covariant, and therefore a solution must transform into another. Solutions to the equation form a vector-valued modular form (vvmf).

For a 4d $\mathcal{N}=2$ SCFT with flavor symmetry, one may construct flavored modular differential equations that constrain all the flavored characters of the associated chiral algebra. The modular property of these equations remains largely unexplored, however, it is natural to expect that they satisfy more general form of modularity, as have been examined in several examples \cite{Zheng:2022zkm,Pan:2023jjw}. Such conjectural quasi-modularity implies that the solutions continue to transform in certain representation of the modular group. Therefore, in the subsequent discussions we will be interested in the modular orbit of the solutions to both the flavored and unflavored modular differential equations. Concretely, starting with any solution $\operatorname{ch}$, the orbit $\{\gamma \cdot \operatorname{ch} \ | \ \gamma \in SL(2, \mathbb{Z}) \text{ or } \Gamma^0(2)\}$ spans a linear space $\mathcal{V}(\operatorname{ch})$, and we will explore the dimension and content of $\mathcal{V}(\operatorname{ch})$. For this, it is helpful to know the closed-form expression of at least a few solutions. Hence we will mainly focus on the $A_1$-theories of class-$\mathcal{S}$ where the closed-form expressions are known explicitly in terms of Eisenstein series. In the last subsection, we will comment on the case of $\mathcal{N} = 4$ theories with $SU(N)$ gauge group.

%!TEX root = ../Schur index and modularity.tex
\subsection{Modular transformation}

We begin our discussions with the 4d $A_1$ theories of class-$\mathcal{S}$. The detail of their Schur index is reviewed in section \ref{section:chiral-algebra}. It is easy to write down the modular transformation of $\mathcal{I}_{g, n}$. Recall that under the $S$-transformation,
\begin{align}
	\eta(\tau) \xrightarrow{S} & \ \sqrt{- i \tau}\eta(\tau), \qquad
	\vartheta_1(2 \mathfrak{b}_j) \xrightarrow{S} - i \sqrt{- i \tau}e^{\frac{4\pi i}{\tau}\mathfrak{b}_j^2}\vartheta_1(2 \mathfrak{b}_j) \ ,
\end{align}
and (\ref{Eisenstein-S-transformation}), (\ref{Eisenstein-T-transformation}) for the Eisenstein series. Under the $T$ transformation,
\begin{align}
	& \ \eta(\tau) \xrightarrow{T} e^{\frac{\pi i}{12}} \eta(\tau), \quad
	\vartheta_1(2 \mathfrak{b}_j) \xrightarrow{T} e^{\frac{\pi i}{4}} \vartheta_1(2 \mathfrak{b}_j) \ , \\
	& \ E_k \begin{bmatrix}
  	1 \\ \prod_{j}b_j^{\alpha_j} 
	\end{bmatrix}
	\xrightarrow{T} E_k \begin{bmatrix}
  	1 \\ \prod_{j}b_j^{\alpha_j}  
	\end{bmatrix} \ .
\end{align}

The $T$ transformation of $\mathcal{I}_{g, n}$ is fairly simple,
\begin{equation}\label{eq:T-transformation-of-Ign}
	\mathcal{I}_{g, n} \xrightarrow{T} \frac{ie^{\frac{\pi i(g - n - 1)}{6}}}{2}
	\frac{\eta(\tau)^{2g - 2 + n}}{\prod_{j = 1}^{n}\vartheta_1(2 \mathfrak{b}_j)}
	\sum_{k = 1}^{2g - 2 + n}\lambda_k^{2g - 2 + n} \sum_{\alpha_j = \pm}
	\biggl(\prod_{j = 1}^{n}\alpha_j\biggr)
	E_k \begin{bmatrix}
		(-1)^n \\ (-1)^nb^\alpha
	\end{bmatrix} \ , \nonumber
\end{equation}
where we abbreviate $b^\alpha \coloneqq \prod_{j = 1}^{n} b_j ^{\alpha_j}$. We can also write down the $S$-transformation of $\mathcal{I}_{g, n}$,
\begin{align}\label{eq:S-transformation-of-Ign}
	\mathcal{I}_{g, n} & \ \xrightarrow{S} \frac{(-i\tau)^{g - 1}\eta(\tau)^{2g - 2 + n}}{(-i)^n e^{\frac{4\pi i}{\tau}\sum_{j = 1}^{n}\mathfrak{b}_j^2} \prod_{j = 1}^{n}\vartheta_1(2 \mathfrak{b}_j)}\\
	& \ \sum_{k = 1}^{2g - 2 + n}
	\lambda^{(2g - 2 + n)}_k \sum_{\alpha_j = \pm}
	\biggl(\prod_{j = 1}^{n}\alpha_j\biggr)
	\sum_{\ell = 0}^{k}\frac{(-1)^{k - \ell}}{(k - \ell)!} \biggl(\sum_{j = 1}^{n}\alpha_j\mathfrak{b}_j\biggr)^{k - \ell} \tau^\ell
	E_\ell
	\begin{bmatrix}1 \\ (-1)^n b^\alpha\end{bmatrix} \ . \nonumber
\end{align}
The factor $e^{\frac{4\pi i}{\tau} \sum_{j = 1}^{n}\mathfrak{b}_j^2}$ can be absorbed by introducing additional fugacities $y_j = e^{2\pi i \mathfrak{y}_j}$ that transform under $S$ as $\mathfrak{y}_j \to \mathfrak{y}_j - \frac{1}{\tau}\mathfrak{b}_j^2$, and redefining $\mathcal{I}_{g, n} \to (\prod_{j = 1}^{n}y_j^{-2}) \mathcal{I}_{g, n}$. We will often leave the fugacities $y_j$ implicit, and omit the factor $e^{\frac{4\pi i}{\tau}\sum_{j}\mathfrak{b}_j^2}$ in the following discussion.

In general, the highest power of $\tau$ in the $S$-transformation is $3g - 3 + n$, which coincides with the rank of the theory, namely, the dimension of the complex moduli space of the punctured Riemann surface $\Sigma_{g, n}$. The term with the lowest power of $\tau$ reads
\begin{equation}\label{eq:lowest-power-of-tau}
	\frac{(-1)(-i\tau)^{g - 1}\eta(\tau)^{2g - 2 + n}}{(-1)^n \prod_{j = 1}^{n}\vartheta_1(2 \mathfrak{b}_j)}
	\sum_{k = 1}^{2g - 2 + n}\lambda_k^{2g - 2 + n}
	\sum_{\alpha_j = \pm} \biggl(\prod_{j = 1}^{n}\alpha_j\biggr)
	\frac{(-1)^k}{k!}\biggl(\sum_{j = 1}^{n}\alpha_j \mathfrak{b}_j\biggr)^k \ .\nonumber
\end{equation}
However, for $g = 0$, this term actually vanishes identically due to the identity
\begin{equation}
	\sum_{\alpha_j = \pm}\biggl(\prod_{j = 1}^{n}\alpha_j\biggr) \biggl(\sum_{j = 1}^{n}\alpha_j \mathfrak{b}_j\biggr)^k = 0 \ , \qquad k = 0, 1, 2, ..., n - 1 \ .
\end{equation}
For $g > 0$, this $\tau^{g - 1}$ term is nontrivial, and is proportional to $\frac{\eta(\tau)^{2g - 2 + n}}{\prod_{j = 1}^{n}\vartheta_1(2 \mathfrak{b}_j)}$. For example,
\begin{equation}
	\sum_{\alpha_j = \pm} \Big(  \prod_{j = 1}^n\alpha_j  \Big) (\sum_{j = 1}^n\alpha_j \mathfrak{b}_j)^n
	= \frac{1}{2} \frac{1}{4^n n (2n - 1)!} \big(\prod_{j = 1}^n \mathfrak{b}_j\big)\ .
\end{equation}
This fact will be important later.
%!TEX root = ../Schur index and modularity.tex
\subsection{Modular orbit}

It is straightforward to compute general modular transformations of the Schur index. To connect with the structure of the modular differential equations, we may focus on the full orbit $SL(2, \mathbb{Z}) \cdot \mathcal{I}_{g, n}$ of the index $\mathcal{I}_{g, n}$ when $n$ is even, or the orbit under $\Gamma^0(2) \subset SL(2, \mathbb{Z})$\footnote{Note that the subgroup $\Gamma^0(2)$ is generated by $STS, T^2$.} when $n$ is odd. Like in the case of rational CFT, only a finite number of these infinitely many expressions are linear independent. The linear space that the orbit span will simply be denoted as $\mathcal{V}_{g, n} \coloneqq \mathcal{V}_{g, n}(\mathcal{I}_{g,n})$. Sometimes for brevity we will simply call $\mathcal{V}_{g, n}$ as the modular orbit.

Let us start with a simple example with $g = 0, n = 4$. The index $\mathcal{I}_{0,4}$ reads
\begin{equation}
	\mathcal{I}_{0, 4} = \Big(  \prod_{j = 1}^{4}y_i^{-2}  \Big) \frac{\eta(\tau)^2}{\prod_{j = 1}^4 \vartheta_1(2 \mathfrak{b}_j)} \sum_{\alpha_j = \pm} E_2 \begin{bmatrix}
  	1 \\ \prod_{j = 1}^4 b_j^{\alpha_j}  
	\end{bmatrix} \ .
\end{equation}
Under the $S$-transformation,
\begin{align}
	\eta(\tau) \xrightarrow{S} & \ \sqrt{- i \tau}\eta(\tau),\qquad
	\vartheta_1(2 \mathfrak{b}_j) \xrightarrow{S} - i \sqrt{- i \tau}e^{4\pi i \frac{\mathfrak{b}_j^2}{\tau}}\vartheta_1(2 \mathfrak{b}_j) \ ,
	\\
	E_2\begin{bmatrix}
  	1 \\ \prod_{j = 1}^{4}b_j^{\alpha_j}  
	\end{bmatrix}\xrightarrow{S} & \
	- \frac{1}{2}(\sum_{j}\alpha_j \mathfrak{b}_j)^2
	- \tau \sum_{j}\alpha_j \mathfrak{b}_j E_1 \begin{bmatrix}
  	1 \\ \prod_{j = 1}^{4}b_j^{\alpha_j}  
	\end{bmatrix}
	+ \tau^2 E_2\begin{bmatrix}
  	1 \\ \prod_{j = 1}^{4}b_j^{\alpha_j}  
	\end{bmatrix}\ , \nonumber
\end{align}

Naively, after the $S$-transformation the $E_2$ in $\mathcal{I}_{0,4}$ generates a term independent on $\tau$, while the factor $\eta(\tau)^2/\prod_{j = 1}^{4} \frac{y_j^{-2}}{\vartheta_1(2 \mathfrak{b}_j)} $ generates a $\tau^{-1}$ factor. A negative power of $\tau$ in a modular transformation would suggest the orbit spans an infinite dimensional space, since subsequent $T$ transformations will create infinitely many rational terms in $\tau$. Luckily this is not the case here, as we have observed earlier that for $g = 0$ the $\tau^{g - 1}$ term actually vanishes,
\begin{equation}
	\sum_{\alpha_j = \pm} (\prod_{j = 1}^4\alpha_j) (\sum_{j = 1}^{4}\alpha_j \mathfrak{b}_j)^2
	= 0 \ .
\end{equation}
Overall, under $S, T$, 
\begin{align}
	\mathcal{I}_{0,4} \xrightarrow{S} & \
	- \sum_{\alpha_j = \pm} \Big(  \prod_{j = 1}^{4} \alpha_j  \Big) (\sum_{j = 1}^{4}\alpha_j \mathfrak{b}_j)E_1 \begin{bmatrix}
  	1 \\ \prod_{j}b_j^{\alpha_j}  
	\end{bmatrix}
	+ \tau \sum_{\alpha_j = \pm} \Big(  \prod_{j = 1}^{4} \alpha_j  \Big) E_2 \begin{bmatrix}
  	1 \\ \prod_{j}b_j^{\alpha_j}  
	\end{bmatrix} \ , \nonumber\\
	\mathcal{I}_{0,4} \xrightarrow{T} & \
	e^{\frac{\pi i}{6}(g - n - 1)}\mathcal{I}_{0,4} \ .
\end{align}
The $S \mathcal{I}_{0,4}$ further transforms,
\begin{align}
	S^2 \mathcal{I}_{0,4} = & \ \mathcal{I}_{0,4}, \\
	T S\mathcal{I}_{0,4} = & \ i e^{\frac{\pi i}{6}(g - n - 1)}\mathcal{I}_{0,4}
	+ e^{\frac{\pi i}{6}(g - n - 1)}S\mathcal{I}_{0,4} \ .
\end{align}
To summarize, $\mathcal{I}_{0,4}, S\mathcal{I}_{0,4}$ generate the $SL(2, \mathbb{Z})$-orbit of $\mathcal{I}_{0,4}$ through linear combinations, giving rise to a two-dimensional representation of the modular group, hence $\dim \mathcal{V}_{0,4} = 2$.

The index with odd $n$ undergoes a similar analysis, with the main difference being that we consider the $\mathcal{I}_{g=0, n=\text{odd}}$ orbit under the modular subgroup $\Gamma^0(2)$ generated by $STS$ and $T^2$.\footnote{The $T,S$ transformation properties of twisted Eisenstein series imply a mixture of twisted Eisenstein series in the case of odd $n$, whereas this mixture does not occur when considering $\Gamma^0(2)$. Therefore, we investigate the orbit of $\Gamma^0(2)$ in the case of odd $n$.}

Again, the factor $\frac{\eta(\tau)^{2g-2+n}}{\prod_{j=1}^{n}\vartheta_{1}(2{\mathfrak{b}}{j})}$ generates a $(\tau-1)^{g-1}$ under the $STS$ transformation. This factor becomes $1/(\tau-1)$ when the genus equals zero. Under the action of $\Gamma^0(2)$, the index $\mathcal{I}_{g=0, n=\text{odd}}$ behaves as
\begin{equation}
	\begin{aligned}
		\mathcal{I}_{g=0, n=\text{odd}}\xrightarrow{T^2} &e^{\frac{1}{3}\pi i(g-n-1)}	\mathcal{I}_{g=0, n=\text{odd}}\\
		\mathcal{I}_{g=0, n=\text{odd}}\xrightarrow{STS}& (-e^{\pi i/6})^{-1-n}(\tau-1)^{-1}\frac{\eta(\tau)^{2g-2+n}}{\prod_{j=1}^{n}\vartheta_{1}(2{\mathfrak{b}}_{j})} \sum_{k=1}^{2g-2+n}\lambda_{k}^{2g-2+n} \\
		&\sum_{\alpha_{j}=\pm}\left(\prod_{j=1}^{n}\alpha_{j}\right) \sum_{\ell\geq 0}\frac{1}{\ell !} \left(-\sum_j\alpha_j\mathfrak{b}_j\right)^\ell (\tau-1)^{k-\ell}E_{k-\ell}\begin{bmatrix}-1\\+z\end{bmatrix}.
	\end{aligned}
\end{equation}
It is then observed that the lowest degree of $\tau$ in $STS (\mathcal{I}_{g = 0, n=\text{odd}})$ involves
\begin{equation}
	\sum_{\alpha_{j}=\pm}\left(\prod_{j=1}^{n}\alpha_{j}\right)\left(\sum_{j}\alpha_{j}\mathfrak{b}_{j}\right)^{k}=0 \quad\text{for odd } k,n,
\end{equation}
where we utilize the fact that $\lambda_{\mathrm{even}}^{\mathrm{(odd)}}=\lambda_{\mathrm{odd}}^{\mathrm{(even)}}=0$, hence the summation only contains the odd $k$ terms. Therefore, we can conclude that $\dim \mathcal{V}_{g = 0, n=\text{odd}}$ is finite dimensional.

\subsection{Dimension of the orbit}
We would like to compute the dimension of $\mathcal{V}_{g, n}$. To do so we briefly analyze the modular orbit of the simple ingredients that appear in $\mathcal{I}_{g, n}$. See also \cite{Beemetal} for a similar analysis in the unflavoring limit. The first ingredient is the Eisenstein series $E_k \big[\substack{\pm 1\\b}\big]$ which are quasi-Jacobi forms, transforming inhomogeneously under the $S$-transformation,
\begin{align}
	S E_k \begin{bmatrix}
		1 \\ b
	\end{bmatrix}
	= & \ \tau^k E_k \begin{bmatrix}
		1 \\ b
	\end{bmatrix}
	- \tau^{k - 1} \mathfrak{b} E_{k - 1} \begin{bmatrix}
		1 \\ b
	\end{bmatrix}
	+ \dots - \frac{(-1)^k}{k!}\mathfrak{b}^k \\
	STS E_k \begin{bmatrix}
		-1 \\ b
	\end{bmatrix} = & \
	\tau^k E_k \begin{bmatrix}
		-1 \\ b
	\end{bmatrix}
	- \tau^{k - 1} \biggl(\mathfrak{b} E_{k - 1} \begin{bmatrix}
		-1 \\ b
	\end{bmatrix}
	+ k E_k \begin{bmatrix}
		1 \\ b
	\end{bmatrix}
	\biggr) + \dots \nonumber \\
	& + (-1)^k \sum_{\ell = 0}^{k}\frac{\mathfrak{b}^\ell}{\ell!}E_{k - \ell}\begin{bmatrix}
		-1 \\ b
	\end{bmatrix} \ .
\end{align}

Let us define $\mathbf{T}f \coloneqq Tf - f$. Further $\mathbf{T}$ actions on $SE_k \big[\substack{1\\b}\big]$ give rise to $k - 1$ additional expressions with lower maximal $\tau$ power, since $\tau^\ell E_k\big[\substack{1\\b}\big] \xrightarrow{\mathbf{T}} \ell \tau^{\ell - 1}E_k\big[\substack{1\\b}\big]$. All together, the linear space spanned by the $SL(2, \mathbb{Z})$ orbit of $E_k \big[\substack{1 \\ b}\big]$ is given by
\begin{equation}
	\operatorname{span}\biggl\{E_k \begin{bmatrix}
		1 \\ b
	\end{bmatrix},\quad 
	S E_k \begin{bmatrix}
		1 \\ b
	\end{bmatrix},
	\quad
	\mathbf{T}^\ell S E_k \begin{bmatrix}
		1 \\ b
	\end{bmatrix}, \quad \ell = 1, 2, \dots, k - 1\biggr\} \ ,
\end{equation}
which is $(k + 1)$-dimensional. In particular, $\mathbf{T}^k S E_k \big[\substack{1 \\ b}\big] = k! E_k \big[\substack{1 \\ b}\big]$. Similarly, the linear space associated to $E_k \big[\substack{-1 \\ b}\big]$ is given by
\begin{equation}
	\operatorname{span}\biggl\{E_k \begin{bmatrix}
		-1 \\ b
	\end{bmatrix}, \quad
	S_{(2)} E_k \begin{bmatrix}
		-1 \\ b
	\end{bmatrix}, \quad
	\mathbf{T}_{(2)}^\ell S_{(2)} E_k \begin{bmatrix}
		-1 \\ b
	\end{bmatrix}, \quad
	\ell = 1, 2, \dots, k - 1
	\biggr\} \ ,
\end{equation}
where
\begin{equation}
	S_{(2)} \coloneqq STS, \qquad
	\mathbf{T}_{(2)}f \coloneqq T^2 f - f
\end{equation}

More generally, we can consider a linear combination of Eisenstein series $E_{k_i} \big[\substack{+1\\b}\big]$ of different weights $k_1 < k_2 < ... < k_n$. It is easy to check that
\begin{equation}
	\mathcal{E}_n \coloneqq \sum_{k = 0}^{n} a_k E_k \begin{bmatrix}
		+1 \\ b
	\end{bmatrix} \ , \qquad
	\mathbf{T}^{k_n} S \mathcal{E}_n = a_n k_n! E_{k_n} \begin{bmatrix}
		+1 \\ b
	\end{bmatrix} \ .
\end{equation}
As a result, 
\begin{equation}
	\mathcal{E}_n - \frac{1}{k_n!} \mathbf{T}^{k_n} S \mathcal{E}_n
	= \sum_{i = 1}^{n - 1} a_i E_i \begin{bmatrix}
		1 \\ b
	\end{bmatrix} \coloneqq \mathcal{E}_{n - 1} \ .
\end{equation}
This logic implies that each individual $E_{k_i} \big[\substack{+1\\b}\big]$ in $\mathcal{E}_n$ can be expressed as some linear combination of $S, T$ acting on $\mathcal{E}_n$, and therefore is in the same modular orbit as $\mathcal{E}_n$. Hence each $(k_i + 1)$-dimensional linear space spanned by the $SL(2, \mathbb{Z})$ orbit of $E_{k_i} \big[\substack{+1\\b}\big]$ is contained as a subspace in the space spanned by the orbit of $\mathcal{E}_n$. These subspaces do not intersect except at the origin, since the weights $k_i$ are mutually different. To summarize, the linear space spanned by the $SL(2, \mathbb{Z})$ orbit of $\mathcal{E}_n$ is simply the direct sum of these subspaces.

On the other hand, individual terms in a linear combination of the same weight do not appear in the modular orbit separately. For example, consider
\begin{equation}
	\mathcal{E}_2(b_1, b_2) = \lambda_1 E_2 \begin{bmatrix}
		1 \\ b_1
	\end{bmatrix} + \lambda_2 E_2 \begin{bmatrix}
		1 \\ b_2
	\end{bmatrix} \ .
\end{equation}
Under the $S$ transformation,
\begin{equation}
	S \mathcal{E}_2 = - \frac{1}{2} \sum_{\ell = 0}^{2} \frac{1}{\ell!} \tau^\ell
	\sum_{i = 1}^{2}\lambda_i \mathfrak{b}_i^{2 - \ell} E_\ell \begin{bmatrix}
		1 \\ b_i
	\end{bmatrix}\ ,
\end{equation}
or, under the $\mathbf{T}$ action,
\begin{equation}
	\mathbf{T}S \mathcal{E}_2
	= - \sum_{i = 1}^2 \mathfrak{b}_i\lambda_i E_1 \begin{bmatrix}
		1 \\ b_i
	\end{bmatrix}
	+ (1 + 2 \tau) \sum_i \lambda_i E_2 \begin{bmatrix}
		1 \\ b_i
	\end{bmatrix}
\end{equation}
Obviously, objects that appear in the modular orbit will always be the linear combinations
\begin{equation}
	\tau^\ell\sum_{i = 1}^{2} \lambda_i \mathfrak{b}_i^{2 - \ell} E_\ell \begin{bmatrix}
		1 \\ b_i
	\end{bmatrix} \ , \qquad \ell = 0, 1, 2 \ ,
\end{equation}
instead of the individual $E_\ell \big[\substack{1 \\ b_i}\big]$. This observation suggests when analyzing the modular orbit of $\mathcal{I}_{g, n}$ one should treat each linear combination
\begin{equation}
	\mathbf{E}_k(b_1, ..., b_n) \coloneqq \sum_{\alpha_j = \pm}(\prod_{j = 1}^n \alpha_j) E_k \begin{bmatrix}
		1 \\ \prod_{j = 1}^n b_j^{\alpha_j}
	\end{bmatrix} \nonumber
\end{equation}
as a whole. Hence, the $SL(2, \mathbb{Z})$-orbit of $\mathbf{E}_k(b_1, ..., b_n)$ spans a $(k + 1)$-dimensional space.

A final ingredient in $\mathcal{I}_{g, n}$ is the Dedekind eta function $\eta$ and the Jacobi theta functions,
\begin{equation}
	\frac{\eta(\tau)^{2g - 2 + n}}{\prod_{j = 1}^{n}\vartheta_1(2 \mathfrak{b}_j)} \xrightarrow{S} \frac{(-i\tau)^{g - 1}\eta(\tau)^{2g - 2 + n}}{(-i)^n \prod_{j = 1}^{n}\vartheta_1(2 \mathfrak{b}_j)} \ .
\end{equation}
When multiplying to the Eisenstein $E_k \big[\substack{+1\\b}\big]$, it increases the maximal power of $\tau$ in the $S$-transform from $k$ to $g - 1 + k$, enlarging the spanned linear space from dimension $k + 1$ to $k + g$ since the space is obtained by action of $\mathbf{T}$ that gradually decreases the powers of $\tau$.

Putting all together, we argue that the dimension of $\mathcal{V}_{g, n}$ is simply given by
\begin{align}\label{eq:dimension-formula}
	\dim \mathcal{V}_{g, n} = & \ \sum_{\substack{k = 1\\ k = n \mod 2}}^{2g - 2 + n} (k + g) \ , & \ \text{when} ~ n > & \ 0\\
	\dim \mathcal{V}_{g, n} = & \ \sum_{\substack{k = 0\\ k = n \mod 2}}^{2g - 2 + n} (k + g) & \text{when} ~ n = & \ 0\ .
\end{align}
The two formula differ only in the lower bound of the sum, due to the lack of a sum $\sum_{\alpha_j = \pm}$ when $n = 0$. Said differently, when $n = 0$ there is a non-trivial $\eta(\tau)^{2g - 2 + n}$ term, while when $n > 0$, the $\frac{\eta(\tau)^{2g - 2 + n}}{\prod_{j = 1}^{n}\vartheta_1(2 \mathfrak{b}_j)}$ term is absent. This formula is actually identical to the equation (37) in \cite{Beemetal}, which is written in a different way and we reproduce here,
\begin{equation}
	(1 - \delta_{g, 0})g ( 2g + n - 1) + (1 - \delta_{n, 0}) \Big\lfloor \frac{n - 1}{2} \Big\rfloor \Bigl(g + n - 1 - \Big\lfloor \frac{n - 1}{2} \Big\rfloor\Bigr) \ .
\end{equation}
This formula in \cite{Beemetal} is based on a conjectural unflavored Schur index of the $A_1$ class-$\mathcal{S}$ theories. The dimension predicts the minimal order of the unflavored modular differential equation satisfied by the unflavored Schur index: such an equation may have a nonzero Wronskian index. For example, when $(g = 1, n = 4)$, the first equation with zero Wronskian index is at order-$10$, but there is another equation at order-$8$ with a nonzero Wronskian index,
\begin{align}
	\Bigg[ & \ D^{(8)}_q E_4
	- 14 E_6 D^{(7)}_q - 720 E_4^2 D^{(6)}_q
	- 21000 E_4 E_6 D^{(5)}_q 
	+ 1200 (-293 E_4^3 + 245 E_6^2)D^{(4)}_q 
	\nonumber \\
	& \ - 2788800 E_4^2 E_6 D^{(3)}_q
	+ 2000  (10120 E_4^4 + 3087 E_4 E_6^2) D^{(2)}_q
	\\
	& \ + 28000 (43814 E_4^3 E_6 - 3087 E_6^3)D^{(1)}_q
	+ 2048000  E_4^2 (1860 E_4^3 + 3283 E_6^2)
  \Bigg]\mathcal{I}_{1,4}^\text{unflavored} = 0 \ , \nonumber
\end{align}
where one may divide the equation by $E_4$ to get singular coefficients. The dimension also predicts the the number of rational indicial roots of the lowest order zero-Wronskian equation \cite{Beemetal}.

In principal one would expect $\dim \mathcal{V}_{g, n} \ge $ minimal order. The agreement between our flavored results and that in \cite{Beemetal} means that linear independent elements in $\mathcal{V}_{g, n}$ remain independent after unflavoring. This not at all obvious, as we will later encounter disagreement in the $\mathcal{N} = 4$ examples. This is also unlike the frequent observation in rational CFTs that different flavored characters may be unrefined to identical unflavored characters.

\bigskip

Now that we have a linear space $\mathcal{V}_{g, n}$, we may construct some basis. One possible basis can be inspired by the very formula (\ref{eq:dimension-formula}). Take the cases with $g > 0$ and even $n > 0$ as an example. One can check that the following set of expressions form a complete basis of $\mathcal{V}_{g, n}$,
\begin{align}
	\Bigl\{\mathbf{T}_{g,n}^{\ell} S\mathcal{I}_{g, n}\Bigr\}_{\ell = 0}^{g + 1} \cup
	\Bigl\{\mathbf{T}_{g,n}^{\ell} &S\mathbf{T}_{g,n}^{g + 1} S\mathcal{I}_{g, n}\Bigr\}_{\ell = 0}^{g + 3}
	\cup
	\Bigl\{\mathbf{T}_{g,n}^{\ell} S\mathbf{T}_{g,n}^{g + 3}S\mathbf{T}_{g,n}^{g + 1} S\mathcal{I}_{g, n}\Bigr\}_{\ell = 0}^{g + 5} \cup \cdots \nonumber\\
	& \ \cdots \cup \Bigl\{\mathbf{T}_{g,n}^{\ell} S \mathbf{T}_{g,n}^{3g - 3 + n - 2} \cdots S\mathbf{T}_{g,n}^{g + 1} S\mathcal{I}_{g, n}\Bigr\}_{\ell = 0}^{3g - 3 + n}  \ .
\end{align}
Here we define $\mathbf{T}_{g, n} f \coloneqq e^{ - \frac{\pi i}{6}(g - n - 1)}Tf - f$\footnote{One may alternatively use the usual $T$ instead of $\mathbf{T}$, which results in slightly more complicated expression for each individual basis element.}. Note that $\mathcal{I}_{g, n}$ itself is not included in the basis. Instead, $\mathcal{I}_{g, n}$ is a linear combination of the basis elements. Alternatively, one may choose to dropped the last expression with the largest number of $\mathbf{T}_{g,n}$ from the basis and trade it with $\mathcal{I}_{g, n}$: the resulting set still furnishes a complete basis. As an example, in the case with $g = 1, n = 4$, one may compute the $S$, $T$ matrices explicitly using the following basis,
\begin{align}
	\mathcal{I}_{1,4}, & \quad S\mathcal{I}_{1,4}, \quad \mathbf{T}_{g,n}S\mathcal{I}_{1,4}, \quad \mathbf{T}_{g,n}^2S\mathcal{I}_{1,4} ,\nonumber\\
	& \quad S \mathbf{T}_{g,n}^2 S\mathcal{I}_{g, n},
	\quad\mathbf{T}_{g,n}S \mathbf{T}_{g,n}^2 S\mathcal{I}_{g, n},
	\quad\mathbf{T}_{g,n}^2S \mathbf{T}_{g,n}^2 S\mathcal{I}_{g, n},
	\quad\mathbf{T}_{g,n}^3S \mathbf{T}_{g,n}^2 S\mathcal{I}_{g, n} \ .
\end{align}
With respect to this basis, the $S$-matrix reads
\begin{equation}
	S = \left(
		\begin{array}{cccccccc}
		 0 & 1 & 0 & 0 & 0 & 0 & 0 & 0 \\
		 1 & 0 & 0 & 0 & 0 & 0 & 0 & 0 \\
		 \frac{79}{17} & 1 & -1 & -\frac{39}{17} & 0 & 0 & \frac{4}{17} & \frac{1}{34} \\
		 0 & 0 & 0 & 0 & 1 & 0 & 0 & 0 \\
		 0 & 0 & 0 & 1 & 0 & 0 & 0 & 0 \\
		 \frac{518}{17} & 2 & -2 & -\frac{288}{17} & 0 & 0 & \frac{23}{17} & \frac{5}{17} \\
		 \frac{496}{17} & -8 & 2 & -\frac{312}{17} & 6 & -1 & \frac{32}{17} & \frac{4}{17} \\
		 -\frac{1860}{17} & -60 & -16 & \frac{1170}{17} & 30 & 8 & -\frac{120}{17} & -\frac{15}{17} \\
		\end{array}
		\right)
\end{equation}
and
{\small\begin{equation}
	T = \left(
		\begin{array}{cccccccc}
		 e^{\frac{4\pi i}{3}} & 0 & 0 & 0 & 0 & 0 & 0 & 0 \\
		 0 & e^{\frac{4\pi i}{3}} & e^{\frac{4\pi i}{3}} & 0 & 0 & 0 & 0 & 0 \\
		 0 & 0 & e^{\frac{4\pi i}{3}} & e^{\frac{4\pi i}{3}} & 0 & 0 & 0 & 0 \\
		 \frac{72}{17} e^{\frac{4\pi i}{3}} & 0 & 0 & \frac{25}{17}e^{\frac{\pi i}{3}} & 0 & 0 & -\frac{3}{17} e^{\frac{\pi i}{3}} & -\frac{5}{34} e^{\frac{\pi i}{3}} \\
		 0 & 0 & 0 & 0 & e^{\frac{4\pi i}{3}} & e^{\frac{4\pi i}{3}} & 0 & 0 \\
		 0 & 0 & 0 & 0 & 0 & e^{\frac{4\pi i}{3}} & e^{\frac{4\pi i}{3}} & 0 \\
		 0 & 0 & 0 & 0 & 0 & 0 & e^{\frac{4\pi i}{3}} & e^{\frac{4\pi i}{3}} \\
		 -\frac{2352}{17} e^{\frac{\pi i}{3}} & 0 & 0 & \frac{1372}{17}e^{\frac{\pi i}{3}} & 0 & 0 & -\frac{98}{17} e^{\frac{\pi i}{3}} & -\frac{59}{17} e^{\frac{\pi i}{3}} \\
		\end{array}
		\right) \ . \nonumber
\end{equation}}

One may also find a complete but slightly more complicated basis for all $\mathcal{V}_{g, n}$ with general $g, n$, by replacing $\mathbf{T}, S$ with $\mathbf{T}_{(2)},S_{(2)}$, where $\mathbf{T}_{(2)}\equiv e^{-\frac{\pi i}{3}(g-n-1)} T^2 f-f$, $S_{(2)} \coloneqq STS$. Then the complete basis for $\mathcal{V}_{g,n=\text{odd}}$ is given by
\begin{equation}
	\begin{aligned} & \left\{\mathbf{T}_{(2)}^{\ell}S_{(2)}\mathcal{I}_{g,n}\right\}_{\ell=0}^{g}\cup\left\{\mathbf{T}_{(2)}^{\ell}S_{(2)}\mathbf{T}_{(2)}^{g}S_{(2)}\mathcal{I}_{g,n}\right\}_{\ell=0}^{g+2}\cup\left\{\mathbf{T}_{(2)}^{\ell}S_{(2)}\mathbf{T}_{(2)}^{g+2}S_{(2)}\mathbf{T}_{(2)}^{g+1}S_{(2)}\mathcal{I}_{g,n}\right\}_{\ell=0}^{g+4}\\ &\cup\cdots  \cdots\cup\left\{\mathbf{T}_{(2)}^{\ell}S_{(2)}\mathbf{T}_{(2)}^{3g-3+n-2}\cdots S_{(2)}\mathbf{T}_{(2)}^{g+1}S_{(2)}\mathcal{I}_{g,n}\right\}_{\ell=0}^{3g-3+n}.\end{aligned}
\end{equation}
As an example, in the case $g=1, n=3$ the complete basis is given by
\begin{equation}
	\begin{aligned}
		\mathcal{I}_{1,3},\quad & S_{(2)}\mathcal{I}_{1,3}, \quad \mathbf{T}_{(2)}S_{(2)}\mathcal{I}_{1,3}, \quad S_{(2)}\mathbf{T}_{(2)}S_{(2)}\mathcal{I}_{1,3}, \\
		&\mathbf{T}_{(2)}S_{(2)}\mathbf{T}_{(2)}S_{(2)}\mathcal{I}_{1,3}, \quad \mathbf{T}^2_{(2)}S_{(2)}\mathbf{T}_{(2)}S_{(2)}\mathcal{I}_{1,3}
	\end{aligned}
\end{equation}
With this basis, we can express the matrices for $STS, T^2$ transformations as follows:
\begin{equation}
	\mathbf{T}_{(2)}=\left(
	\begin{array}{cccccc}
		-1 & 0 & 0 & 0 & 0 & 0 \\
		0 & 1 & 1 & 0 & 0 & 0 \\
		3 i & 0 & -1 & \frac{i}{4} & 0 & -\frac{i}{16} \\
		0 & 0 & 0 & 1 & 1 & 0 \\
		0 & 0 & 0 & 0 & 1 & 1 \\
		-96 & 128 i & 64 i & -32 & -28 & -7 \\
	\end{array}
	\right)
\end{equation}
and
\begin{equation}
	S_{(2)}=\left(
	\begin{array}{cccccc}
		0 & 1 & 0 & 0 & 0 & 0 \\
		-1 & 0 & 0 & -1 & 0 & 0 \\
		0 & 0 & 0 & 1 & 0 & 0 \\
		-8 i & -8 & -1 & -4 i & 0 & 0 \\
		23 i & 32 & 12 & \frac{29 i}{4} & 2 i & \frac{3 i}{16} \\
		-480 i & -624 & -324 & -64 i & -48 i & -4 i \\
	\end{array}
	\right).
\end{equation}

A remark follows. The above result does not account for all possible solutions to the flavored modular differential equations that govern the associated chiral algebra. There may be additional solutions that do not have unflavoring limit. For example, in the case of $\mathcal{I}_{0,4}$, the $\mathcal{V}_{0, 4}$ is two-dimensional, however, there are four additional solutions to the flavored modular differential equations $R_j$ given by
\begin{equation}
	R_j \coloneqq \frac{i}{2} \frac{\vartheta_1(2 \mathfrak{m}_j)}{\eta(\tau)}
	\prod_{\substack{\ell = 1 \\ \ell \ne j}}^{4} \frac{\eta(\tau)}{\vartheta_1(\mathfrak{m}_j + \mathfrak{m}_\ell)}
	\frac{\eta(\tau)}{\vartheta_1(\mathfrak{m}_j - \mathfrak{m}_\ell)}, \qquad
	j = 1,2,3,4 \ .
\end{equation}
Here $\mathfrak{m}_1 = \mathfrak{b}_1 + \mathfrak{b}_2$, $\mathfrak{m}_2 = \mathfrak{b}_1 - \mathfrak{b}_2$, $\mathfrak{m}_3 = \mathfrak{b}_3 + \mathfrak{b}_4$, $\mathfrak{m}_4 = \mathfrak{b}_3 - \mathfrak{b}_4$.
Obviously these four solutions do not have unflavoring limit, and they are not included in $\mathcal{V}_{0, 4}$. Each of the four solutions forms a one-dimensional representation of the modular group $SL(2, \mathbb{Z})$,
\begin{equation}
	S R_j = i R_j, \qquad
	T R_j = e^{-\frac{5 \pi i}{6}}R_j \ .
\end{equation}
Another $A_1$ class-$\mathcal{S}$ example is the case with $g = 1, n = 2$, where it has been shown that the ratio $\frac{\eta(\tau)^{2}}{\prod_{j = 1}^{2}\vartheta_1(2 \mathfrak{b}_j)}$ is a solution to the flavored modular differential equations, which does not have smooth unflavoring limit and form a one-dimensional representation of the modular group. Another infinite family of examples come from the 4d $\mathcal{N} = 4$ theories with a simple gauge group $G$ of rank $r$. It has been shown that the ratio
\begin{equation}
	q^{\frac{\dim G}{8}}
	\prod_{i = 1}^{r} \frac{(b^{d_i - 1};q)(b^{-d_i + 1}q^{\frac{1 - d_i}{2}};q)}{(b^{d_i}q^{\frac{d_i}{2}};q)(b^{-d_i}q^{\frac{1 - d_i}{2}};q)}
\end{equation}
is the vacuum character of a free $bc \beta \gamma$ system that provides a special free field realization of the associated chiral algebra \cite{Bonetti:2018fqz}, and therefore solves all relevant flavored modular differential equations. Again, this solution does not have a smooth unflavoring limit, and by itself forms a one-dimensional representation of the suitable modular group.

It would be very interesting to understand the structure of the full space of solutions to the flavored modular differential equations, and its relation to the subspace $\mathcal{V}_{g, n}$. We leave this for future work.

%!TEX root = ../Schur index and modularity.tex
\subsection{Defect index}

Besides the Schur index $\mathcal{I}_{g,n}$, there are other solutions to the flavored modular differential equations, which are potential linear combinations of the module characters of $\mathbb{V}[\mathcal{T}_{g,n}]$. For example, the vortex defect index $\mathcal{I}_{g, n}^\text{vortex}(\kappa = \text{even})$ is believed to be a solution to all the equations and the character of the corresponding $\mathbb{V}[\mathcal{T}_{g,n}]$-module, where $\kappa$ denotes the vorticity of the defect. We will now show that all the vortex defect index are elements of $\mathcal{V}_{g, n}$.

For simplicity, we start with $n = \text{even}$. The vortex defect index $\mathcal{I}_{g, n}^\text{vortex}(\kappa)$ and $\mathcal{I}_{g, n}$ are all just different linear combinations of $E_2, E_4, \dots, E_{2g - 2 + n}$ (with the characteristics suppressed) times an overall factor $\frac{\eta(\tau)^{2g - 2 + n}}{\prod_{j = 1}^n \vartheta_1(2 \mathfrak{b}_j)}$. Using $\mathbf{T}_{g, n} f \coloneqq e^{ - \frac{\pi i}{6}(g - n - 1)}Tf - f$, we have a simple identity
\begin{align}
	&  \mathbf{T}_{g,n}^{3g - 3 + n} S\mathcal{I}_{g,n} \\
	& \ = - (3g - 3 + n)! i^{3g - 3 + 2n} \frac{\eta(\tau)^{2g - 2 + n}}{2\prod_{j = 1}^{n}\vartheta_1(2 \mathfrak{b}_j)} \sum_{\alpha_j = \pm}\bigl(\prod_{j = 1}^n\alpha_j\bigr) E_{2g - 2 + n}\begin{bmatrix}
		1 \\ \prod_{j = 1}^n b_j^{\alpha_j}
	\end{bmatrix} \ . \nonumber
\end{align}
In other words, the term with the highest weight inside $\mathcal{I}_{g,n}^\text{vortex}(\kappa = \text{even})$ is some linear combination of $S, T$ acting on $\mathcal{I}_{g,n}$. By simple iteration, the logic then further implies that all the terms of different weight
\begin{equation}
	\frac{\eta(\tau)^{2g - 2 + n}}{\prod_{j = 1}^{n}\vartheta_1(2 \mathfrak{b}_j)} \sum_{\alpha_j = \pm} \bigl(\prod_{j = 1}^n\alpha_j\bigr) E_{k}\begin{bmatrix}
		1 \\ \prod_{j = 1}^n b_j^{\alpha_j}
	\end{bmatrix} \ , \qquad k = 2, 4, \dots, 2g - 2 + n \ ,
\end{equation}
can be separately expressed as some linear combinations of $S, T$ acting on $\mathcal{I}_{g,n}$. Thus, all the vortex defect index $\mathcal{I}_{g, n}^\text{vortex}(k = \text{even}) \in \mathcal{V}_{g, n}$ as they are simply linear combinations of these terms. Here we list a few examples with vorticity $\kappa = 2$, 
\begin{equation}
	\mathcal{I}^\text{defect}_{1,4}(\kappa = 2) = - 3 \mathcal{I}_{1,4} + \frac{1}{4}(\mathbf{T}_{1,4})^4 S\mathcal{I}_{1,4}  \ ,
\end{equation}
\begin{equation}
	\mathcal{I}^\text{defect}_{2,2}(\kappa = 2) = 3 \mathcal{I}_{2,2} + \frac{i}{20}(\mathbf{T}_{2,2})^5 S \mathcal{I}_{2,2}  \ ,
\end{equation}
and for $g = 3, n = 2$,
\begin{align}
	\mathcal{I}^\text{defect}_{3,2}(\kappa = 2) = & \ 2 \mathcal{I}_{3,2}
	- \frac{1}{8064}(\mathbf{T}_{3,2})^{3g - 3 + n}\mathcal{I}_{3,2} \\
	& \ - \frac{1}{720} (\mathbf{T}_{3,2})^{3g - 3 + n - 2} S \Big(\mathcal{I}_{3,2} - \frac{1}{8!} (\mathbf{T}_{3,2})^{3g - 3 + n} \mathcal{I}_{3,2} \Big) \ . \nonumber
\end{align}
The analysis for odd $n$ is similar. Here, we have the formula
\begin{align}
	\mathbf{T}_{(2)}^{3g-3+n}S_{(2)}\mathcal{I}_{g,n} = & \ (-1)^{\frac{1}{6}(5+8n+7g)}2^{3g-4+n}(3g-3+n)! \nonumber\\
	& \times \frac{\eta(\tau)^{2g-2+n}}{\prod_{j=1}^n\vartheta_1(2\mathfrak{b}_j)}\sum_{\alpha_j=\pm}\left(\prod_{j=1}^n\alpha_j\right)E_{2g-2+n}
	\begin{bmatrix}
		 -1\\\prod_{j=1}^nb_j^{\alpha_j}
	\end{bmatrix} \ .
\end{align}
where $\mathbf{T}_{(2)}\coloneqq e^{-\frac{\pi i}{3}(g-n-1)} T^2 f-f$. Using the same logic as in the case of even $n$, the term with different weight can be expressed as a combination of  $S_{(2)},\mathbf{T}_{(2)}$ acting on $\mathcal{I}_{g,n}$.

For example, let's consider the case $\kappa=2, g=1, n=3$. Now, the defect index relates to $\mathcal{I}_{1,3}$ as follows:
\begin{equation}
	\mathcal{I}_{1,3}^{\mathrm{defect}}(\kappa=2)=-5\mathcal{I}_{1,3}+\frac{i}{6}(\mathbf{T}_{(2)})^3S_{(2)}\mathcal{I}_{1,3}.
\end{equation}
Similarly, for the case $\kappa=2, g=0, n=5$:
\begin{equation}
	\mathcal{I}_{0,5}^{\mathrm{defect}}(\kappa=2)=-5\mathcal{I}_{0,5}-(\mathbf{T}_{(2)})^2S_{(2)}\mathcal{I}_{0,5}.
\end{equation}

Now we turn to the Wilson line index and its relation with the modular property of the Schur index. First we note that the line index itself is outside of the modular orbit of the Schur index, since it contains coefficients that are rational function of $b, q$. Another way of interpreting this is that the Wilson line index is not solution to the modular differential equations, which can be checked easily in all known examples. However, once the functional coefficients are removed, we can extract analytic expressions that are closely related to the modular orbit of $\mathcal{I}_{g, n}$.

For example, when $g \ge 1$, a non-vanishing type-1 Wilson line index (\ref{eq:type-1-Wilson-line-index}) contains two terms (where we have suppressed the coefficients rational in $b_i$ and $q$),
\begin{equation}
	\mathcal{I}_{g, n}, \qquad
	\frac{\eta(\tau)^{2g - 2 + n}}{\prod_{j = 1}^n \vartheta_1(2 \mathfrak{b}_j)} \ .
\end{equation}
The first term is simply the Schur index itself. The second term itself is not in the orbit of $\mathcal{I}_{g, n}$, however, it does always appear in the $S$-transformation of $\mathcal{I}_{g, n}$. Recall that $S\mathcal{I}_{g,n}$ is given by (\ref{eq:S-transformation-of-Ign}). Precisely when $g \ge 1$, there is always a non-zero $\tau^{g - 1}$ term (\ref{eq:lowest-power-of-tau}) proportional to $\frac{\eta(\tau)^{2g - 2 + n}}{\prod_{j = 1}^{n} \vartheta_1(2 \mathfrak{b}_j)}$ with a coefficient
\begin{equation}
	\tau^{g - 1} \sum_{k = 1}^{2g - 2 + n} \frac{(-1)^k}{k!} \lambda_k^{2g - 2 + n} \sum_{\alpha_j = \pm} \biggl(\prod_{j = 1}^{n} \alpha_j\biggr) \biggl(\sum_{j = 1}^{n}\alpha_j \mathfrak{b}_j\biggr)^k \ .
\end{equation}

A similar statement can be made for the type-2 Wilson line index $\langle W_j\rangle_{g_1, n_1; g_2, n_2}^{(2)}$. Recall that the type-2 index takes a general form (\ref{eq:type-2-Wilson-line-index}). After dropping the functional coefficients $\Lambda_\ell$, we may immediately extract the following expressions
\begin{equation}
	\mathcal{I}_{g, n}, \qquad
	\frac{\eta(\tau)^{2g - 2 + n}}{\prod_{j = 1}^n \vartheta_1(2 \mathfrak{b}_j)}\sum_{\alpha_j = \pm} \biggl(\prod_{j = 1}^{n} \alpha_j\biggr) E_k \begin{bmatrix}
		1 \\ \prod_{j = 1}^{n} b_j^{\alpha_j}
	\end{bmatrix} \ ,
\end{equation}
for $k = 2, 4, \dots, \max(2g_i - 2 + n_i)$. As we have analyzed for the vortex defect index, the second series of terms are all elements in $\mathcal{V}_{g, n}$.

%!TEX root = ../Schur index and modularity.tex

\subsection{\texorpdfstring{$\mathcal{N} = 4$ theories}{}}

Let us also comment on a few $\mathcal{N} = 4$ examples where the Schur index is known in closed form in terms of the Eisenstein series \cite{Pan:2021mrw}.\footnote{In \cite{Hatsuda:2022xdv} the Schur index of general $\mathcal{N} = 4$ theories with $SU(N)$ gauge group has been computed analytically, it would be interesting to investigate the modular property with the results therein.} Consider the $\mathcal{N} = 4$ theory with $SU(3)$ gauge group. The unflavored Schur index reads
\begin{equation}
  \mathcal{I}_{\mathcal{N} = 4 \ SU(3)}^\text{unflavored}(q)
  = \frac{1}{24} + \frac{1}{2}E_2(\tau) \ .
\end{equation}
The index satisfies a modular differential equation of the form \cite{Beem:2017ooy}
\begin{equation}\label{eq:unflavored-MDE-N4-SU3}
  [D_q^{(4)} - 220E_4 D_q^{(2)} + 700 E_6D_q^{(1)}]\operatorname{ch} = 0 \ .
\end{equation}
The equation is covariant under the modular group $SL(2, \mathbb{Z})$.
The $SL(2, \mathbb{Z})$-orbit of the unflavored index spanned a linear space with the basis
\begin{equation}\label{eq:unflavored-basis-1-N4-SU3}
  \mathcal{I}_{\mathcal{N} = 4 \ SU(3)}^\text{unflavored}(q), \qquad
  S\mathcal{I}_{\mathcal{N} = 4 \ SU(3)}^\text{unflavored}, \qquad
  \mathbf{T}S\mathcal{I}_{\mathcal{N} = 4 \ SU(3)}^\text{unflavored}, \qquad
  \mathbf{T}^2S\mathcal{I}_{\mathcal{N} = 4 \ SU(3)}^\text{unflavored} \ .
\end{equation}
In particular, $\mathbf{T}^2 S\mathcal{I}^\text{unflavored}_{\mathcal{N} = 4 \ SU(3)} = E_2(\tau)$, and a constant solution to (\ref{eq:unflavored-MDE-N4-SU3}) can be obtained
\begin{equation}
  2 \mathcal{I}^\text{unflavored}_{\mathcal{N} = 4 \ SU(3)}
  - \mathbf{T}^2S\mathcal{I}_{\mathcal{N} = 4 \ SU(3)}^\text{unflavored} = \frac{1}{12} \ .
\end{equation}
The $S, T$ matrices are given by
\begin{equation}
  S = \left(
    \begin{array}{cccc}
     0 & 1 & 0 & 0 \\
     1 & 0 & 0 & 0 \\
     -1 & 1 & -1 & 1 \\
     -2 & 2 & 0 & 1 \\
    \end{array}
    \right), \qquad
  T = \begin{pmatrix}
    1 & 0 & 0 & 0 \\
    0 & 1 & 1 & 0 \\
    0 & 0 & 1 & 1 \\
    0 & 0 & 0 & 1 
  \end{pmatrix} \ .
\end{equation}
One may also consider another basis,
\begin{equation}\label{eq:unflavored-basis-2-N4-SU3}
  \mathcal{I}_{\mathcal{N} = 4 \ SU(3)}^\text{unflavored}(q), \quad
  S_{(2)}\mathcal{I}_{\mathcal{N} = 4 \ SU(3)}^\text{unflavored}, \quad
  \mathbf{T}_{(2)}S_{(2)}\mathcal{I}_{\mathcal{N} = 4 \ SU(3)}^\text{unflavored}, \quad
  \mathbf{T}^2_{(2)}S_{(2)}\mathcal{I}_{\mathcal{N} = 4 \ SU(3)}^\text{unflavored} \ , 
\end{equation}
where $\mathbf{T}_{(2)}f \coloneqq T^2 f - f$, $S_{(2)} \coloneqq STS$. With this basis, $S, T$ matrices can also be computed,
\begin{equation}
  S = \left(
    \begin{array}{cccc}
     0 & 1 & \frac{1}{2} & -\frac{1}{8} \\
     0 & 1 & 1 & 0 \\
     0 & 0 & -1 & 0 \\
     -8 & 8 & 4 & 0 \\
    \end{array}
    \right), \qquad
  T = \left(
    \begin{array}{cccc}
     1 & 0 & 0 & 0 \\
     0 & 1 & \frac{1}{2} & -\frac{1}{8} \\
     0 & 0 & 1 & \frac{1}{2} \\
     0 & 0 & 0 & 1 \\
    \end{array}
    \right) \ .
\end{equation}
This implies that the $SL(2, \mathbb{Z})$ and the $\Gamma^0(2)$ orbit of $\mathcal{I}_{\mathcal{N} = 4 \ SU(3)}^\text{unflavored}$ span the same linear space.

The $\mathcal{N} = 4$ theory can be viewed as an $\mathcal{N} = 2$ theory, where an $SU(2)$ subgroup of the $\mathcal{N} = 4$ $\mathcal{R}$-symmetry is a flavor symmetry from the $\mathcal{N} = 2$ point of view. We turn on the $SU(2)$ flavor symmetry $b$, and the index reads
\begin{equation}
  \mathcal{I}_{\mathcal{N} = 4 \ SU(3)}
  = - \frac{1}{8} \frac{\vartheta_4(\mathfrak{b})}{\vartheta_4(3 \mathfrak{b})}
  \biggl(
    - \frac{1}{3}
    + 4 E_1 \begin{bmatrix}
      -1 \\ b
    \end{bmatrix}^2
    - 4 E_2 \begin{bmatrix}
      +1 \\ b
    \end{bmatrix}
  \biggr) \ .
\end{equation}
The flavored index can be expanded as a $q$-series with only integer power. However, the index does contain $E_1 \big[\substack{-1 \\ b}\big]$, suggesting that the relevant modular group should really be $\Gamma^0(2)$; said differently, we expect the flavor modular differential equations that control the flavored characters of the associated chiral algebra are only quasi-modular with respect to $\Gamma^0(2)$ and not $SL(2, \mathbb{Z})$. Indeed, by direct computation we find that the $SL(2, \mathbb{Z})$ and $\Gamma^0(2)$ orbit of $\mathcal{I}_{\mathcal{N} = 4}$ span different spaces. The $\Gamma^0(2)$ orbit of $\mathcal{I}_{\mathcal{N} = 4}$ span a space with basis similar to the one we saw when unflavored,
\begin{equation}
  \mathcal{I}_{\mathcal{N} = 4 \ SU(3)}, \quad
  S_{(2)}\mathcal{I}_{\mathcal{N} = 4 \ SU(3)}, \quad
  \mathbf{T}_{(2)}S_{(2)}\mathcal{I}_{\mathcal{N} = 4 \ SU(3)}, \quad
  \mathbf{T}^2_{(2)}S_{(2)}\mathcal{I}_{\mathcal{N} = 4 \ SU(3)} \ , 
\end{equation}
All four expressions are $q$-series with integer powers (up to integer powers of $\log q$ and $\log b$). We may then compute the matrices $STS$ and $T^2$,
\begin{equation}
  STS = \begin{pmatrix}
    0 & 1 & 0 & 0 \\
    -3 & 4 & 1 & 0 \\
    4 & -4 & -1 & 0 \\
    -8 & 8 & 0 & 1 \\
  \end{pmatrix}, \qquad
  T^2 = \begin{pmatrix}
    ~1~ & 0 & 0 & 0 \\
    0 & ~1~ & ~1~ & 0 \\
    0 & 0 & ~1~ & ~1~ \\
    0 & 0 & 0 & ~1~ \\
  \end{pmatrix} \ .
\end{equation}
This is of course consistent with the product $STS$ when computed with the unflavored basis (\ref{eq:unflavored-basis-2-N4-SU3}). The $SL(2, \mathbb{Z})$-orbit, on the other hand, spans a large space, which is a 12-dimensional space with a basis
\begin{align}
  \begin{array}{llll}
    \mathcal{I}_{\mathcal{N} = 4 \ SU(3)}, &
    S_{(2)}\mathcal{I}_{\mathcal{N} = 4 \ SU(3)}, &
    \mathbf{T}_{(2)}S_{(2)} \mathcal{I}_{\mathcal{N} = 4 \ SU(3)}, &
    \mathbf{T}_{(2)}^2 S_{(2)}\mathcal{I}_{\mathcal{N} = 4 \ SU(3)},\nonumber \\
    S \mathcal{I}_{\mathcal{N} = 4 \ SU(3)}, &
    \mathbf{T}S \mathcal{I}_{\mathcal{N} = 4 \ SU(3)}, &
    \mathbf{T}^2S \mathcal{I}_{\mathcal{N} = 4 \ SU(3)}, &
    ST^2S\mathbf{T}^2S \mathcal{I}_{\mathcal{N} = 4 \ SU(3)} \\
    T\mathcal{I}_{\mathcal{N} = 4 \ SU(3)}, 
    & ST\mathcal{I}_{\mathcal{N} = 4 \ SU(3)}, &
    \mathbf{T}_{(2)}ST\mathcal{I}_{\mathcal{N} = 4 \ SU(3)}, &
    \mathbf{T}^2_{(2)}ST\mathcal{I}_{\mathcal{N} = 4 \ SU(3)}
     \ . \nonumber
  \end{array}
\end{align}
Here $S_{(2)} = STS$, $\mathbf{T}_{(2)} f \coloneqq T^2 f - f$. The three rows correspond respectively to the $\widetilde{\text{NS}}$, $\text{R}$ and $\text{NS}$ sector of the chiral algebra. So to summarize, for the $SU(3)$ case, the unflavored $SL(2, \mathbb{Z})$-orbit is identical to the flavored $\Gamma^0(2)$-orbit, and the flavored $SL(2, \mathbb{Z})$-orbit is the straightforward extension from $\widetilde{\text{NS}}$ to $\text{R}$ and $\text{NS}$ sector.

Next we turn to the $\mathcal{N} = 4$ theory with $SU(4)$ gauge group. The Schur index of this theory is given by
\begin{equation}
  \mathcal{I}_{\mathcal{N} = 4 \ SU(4)}^\text{unflavored}
  = \frac{\vartheta_4^{(2)}(0)}{48\pi \vartheta_1'(0)} + \frac{\vartheta_4^{(4)}(0)}{192\pi^3\vartheta_1'(0)} \ .
\end{equation}
The index satisfies a $\Gamma^0(2)$-modular unflavored modular differential equation of the form \cite{Beem:2017ooy}
{\small{\begin{align}
    \bigg[D_q^{(6)} & -\left(\frac{1}{4} \Theta_{0,1}\right) D_q^{(5)}-\left(\frac{565}{576} \Theta_{0,2}-\frac{413}{576} \Theta_{1,1}\right) D_q^{(4)}+\left(\frac{53}{1152} \Theta_{0,3}-\frac{23}{384} \Theta_{1,2}\right) D_q^{(3)} \nonumber \\
    & \ -\left(\frac{6329}{331776} \Theta_{0,4}-\frac{1261}{82944} \Theta_{1,3}-\frac{4823}{10292} \Theta_{2,2}\right) D_q^{(2)} \nonumber\\
    & \ -\left(\frac{5515}{3981312} \Theta_{0,5}-\frac{84145}{3981312} \Theta_{1,4}+\frac{42515}{1900656} \Theta_{2,3}\right) D_q^{(1)} \nonumber\\
    & \ +\left(\frac{405}{262144} \Theta_{0,6}-\frac{1215}{131072} \Theta_{1,5}+\frac{6075}{262144} \Theta_{2,4}-\frac{2025}{131072} \Theta_{3,3}\right) \bigg] \operatorname{ch} = 0 .
\end{align}}}
Here $\Theta_{r,s} \coloneqq \vartheta_2(0)^{4r}\vartheta_3(0)^{4s} + \vartheta_2(0)^{4s}\vartheta_3(0)^{4r}$. The linear space of the weight-$k$ $\Gamma^0(2)$-modular forms is spanned by $\Theta_{r,s}$ with $r + s = k$. Hence we may consider the $\Gamma^0(2)$ orbit of the index which is six-dimensional, and we find a basis
\begin{align}
  \mathcal{I}_{\mathcal{N} = 4 \ SU(4)}^\text{unflavored}, \quad
  S_{(2)}\mathcal{I}_{\mathcal{N} = 4 \ SU(4)}^\text{unflavored}, \quad
  \mathbf{T}_{(2)}^\ell S_{(2)}\mathcal{I}_{\mathcal{N} = 4 \ SU(4)}^\text{unflavored}\Big|_{\ell = 1, 2, 3}, \quad
  S_{(2)} \mathbf{T}_{(2)}^3 \mathcal{I}_{\mathcal{N} = 4 \ SU(4)}^\text{unflavored} \ .
\end{align}
The $STS, T^2$ matrices are
\begin{equation}
  STS = \begin{pmatrix}
  0 & 1 & 0 & 0 & 0 & 0 \\
  3 & 2 & 2 i & \frac{3}{4} & -\frac{i}{2} & \frac{i}{8} \\
  4 i & 2 i & -2 & \frac{3 i}{4} & \frac{1}{2} & -\frac{1}{8} \\
  -8 & 0 & -4 i & -2 & \frac{3 i}{2} & 0 \\
  0 & 0 & 0 & 0 & 0 & 1 \\
  -192 i & 0 & 96 & -36 i & -25 & 8 \\
  \end{pmatrix}, \quad
  T^2 = \begin{pmatrix}
  -i & 0 & 0 & 0 & 0 & 0 \\
  0 & -i & 1 & 0 & 0 & 0 \\
  0 & 0 & -i & 1 & 0 & 0 \\
  0 & 0 & 0 & -i & 1 & 0 \\
  0 & 0 & 0 & 0 & -i & 0 \\
  -96 & 0 & -48 i & 0 & 2 i & -i \\
  \end{pmatrix} \ .
\end{equation}
The basis provides the six linear independent solutions to the unflavored equation. It is straightforward to check that flavoring does not change the $\Gamma^0(2)$-orbit structure, and we may use the exact same basis to get the exact same $STS, T^2$ matrices.

We may proceed to cases with more general $SU(N)$ gauge group. Unfortunately we do not have compact flavored Schur index in terms of Eisenstein series, therefore we will focus only on the unflavored orbit. From the previous discussions, we expect that the unflavored orbit is enough to make prediction on the flavored orbit structure.

By direct computation, we tabulate the following set of basis for different cases.
\begin{center}
  \def\arraystretch{1.5}%  1 is the default, change whatever you need

  \begin{tabular}{c|c}
    gauge group & basis \\
    \hline
    $SU(3)$ & $\mathcal{I}, \mathbf{T}^\ell S\mathcal{I} \big|_{\ell = 0}^2$\\
    $SU(5)$ & $\mathcal{I}, \mathbf{T}^\ell S\mathcal{I} \big|_{\ell = 0}^2, \mathbf{T}^\ell S \mathbf{T}^2 S\mathcal{I}\big|_{\ell = 0}^4$\\
    $SU(7)$ & $\mathcal{I}, \mathbf{T}^\ell S\mathcal{I} \big|_{\ell = 0}^2, \mathbf{T}^\ell S \mathbf{T}^2 S\mathcal{I} \big|_{\ell = 0}^4, \mathbf{T}^\ell S \mathbf{T}^4 S \mathbf{T}^2 S\mathcal{I} \big|_{\ell = 0}^6$\\
    $SU(9)$& $\mathcal{I}, \mathbf{T}^\ell S\mathcal{I} \big|_{\ell = 0}^2, \mathbf{T}^\ell S \mathbf{T}^2 S\mathcal{I} \big|_{\ell = 0}^4, \mathbf{T}^\ell S \mathbf{T}^4 S \mathbf{T}^2 S\mathcal{I} \big|_{\ell = 0}^6,\mathbf{T}^\ell S\mathbf{T}^6 S \mathbf{T}^4 S \mathbf{T}^2 S\mathcal{I} \big|_{\ell = 0}^8$\\
    \hline
    $SU(4)$ & $\mathcal{I}, ~ S_{(2)}\mathcal{I}, ~\mathbf{T}_{(2)}^\ell S_{(2)}\mathcal{I} \big|_{\ell = 0}^3, ~S_{(2)}\mathbf{T}_{(2)}^3 S_{(2)}\mathcal{I} \big|_{\ell = 0}^3$\\
    $SU(6)$ & $\mathcal{I}, ~ S_{(2)}\mathcal{I}, ~\mathbf{T}_{(2)}^\ell S_{(2)}\mathcal{I} \big|_{\ell = 0}^3, ~\mathbf{T}_{(2)}^\ell S_{(2)}\mathbf{T}_{(2)}^3 S_{(2)}\mathcal{I} \big|_{\ell = 0}^5, ~S_{(2)}\mathbf{T}_{(2)}^5 S_{(2)}\mathbf{T}_{(2)}^3 S_{(2)}\mathcal{I}$\\
    \hline
  \end{tabular}
\end{center}
The dimensions inferred from these basis agree with the order of unflavored modular differential equations listed in \cite{Beem:2017ooy}, hence we have analytically constructed all the solutions to those equations. From these examples, we may also  observe a clear pattern, that the $SL(2, \mathbb{Z})$ basis of the $SU(2N + 1)$ theory is given by
\begin{equation}
  \mathcal{I}, (\mathbf{T}^\ell S)\mathcal{I}\big|_{\ell = 0}^2,
  ~(\mathbf{T}^\ell S) (\mathbf{T}^2 S )\mathcal{I}\big|_{\ell = 0}^4, ~ \dots, ~
  (\mathbf{T}^\ell S) (\mathbf{T}^{2N - 2}S) \cdots
  (\mathbf{T}^4 S) (\mathbf{T}^{2}S \mathcal{I})\big|_{\ell = 0}^{2N} \ ,
\end{equation}
hence the total dimension of the orbit is
\begin{equation}\label{eq:dimension-of-orbit-SU(2N+1)}
  \dim = 1 + 3 + 5 + \cdots + (2N + 1) = (N + 1)^2 \ .
\end{equation}
As for the $SU(2N)$ case, the basis of the $\Gamma^0(2)$-orbit is expected to be
\begin{align}
  \mathcal{I}, \quad
  (\mathbf{T}_{(2)}^\ell S_{(2)})\mathcal{I}\big|_{\ell = 0}^3,
  \quad  &(\mathbf{T}_{(2)}^\ell S_{(2)}) (\mathbf{T}_{(2)}^3 S_{(2)} )\mathcal{I}\big|_{\ell = 0}^5, \nonumber \\
  \dots, \quad
  & (\mathbf{T}_{(2)}^\ell S_{(2)}) (\mathbf{T}_{(2)}^{2N - 3}S_{(2)}) \cdots
  (\mathbf{T}_{(2)}^5 S_{(2)}) (\mathbf{T}_{(2)}^3S_{(2)} \mathcal{I})\big|_{\ell = 0}^{2N - 1} \ ,  \nonumber\\
  & S_{(2)}(\mathbf{T}_{(2)}^{2N-1} S_{(2)}) (\mathbf{T}_{(2)}^{2N - 3}S_{(2)}) \cdots
  (\mathbf{T}_{(2)}^5 S_{(2)}) (\mathbf{T}_{(2)}^3S_{(2)} \mathcal{I}) \ .
\end{align}
The dimension of the orbit is then given by
\begin{equation}\label{eq:dimension-of-orbit-SU(2N)}
  1 + \biggl(\sum_{k = 2}^{N} 2k\biggr) + 1 = N(N+1) \ .
\end{equation}
The dimension of the full $SL(2, \mathbb{Z})$-orbit is expected to be three times this number, $3N(N + 1)$, accounting for the $\widetilde{\text{NS}}$, NS and R sectors. The above two dimension formula (\ref{eq:dimension-of-orbit-SU(2N+1)}) and (\ref{eq:dimension-of-orbit-SU(2N)}) predict the minimal order of the $\mathcal{N} = 4$ unflavored modular differential equation.
%!TEX root = ../Schur index and modularity.tex

\section{High temperature behavior \label{section:high-temperature-behavior}}

\subsection{High temperature asymptotics}

We begin with the $A_1$ theories of class-$\mathcal{S}$. The closed form of Schur index $\mathcal{I}_{g,n}$ consists of familiar special functions, the Jacobi theta functions, Dedekind function and the Eisenstein series. These functions have well-known modular properties that will be handy for analyzing the high temperature behavior of $\mathcal{I}_{g, n}$. Defining $\tilde \tau \coloneqq - 1/\tau$, $\tilde{\mathfrak{b}} \coloneqq \mathfrak{b}/\tau$ (or conversely, $\tilde \tau = - \frac{1}{\tau}, \mathfrak{b} = \frac{\tilde{\mathfrak{b}}}{\tilde \tau}$), we have
\begin{equation}
	\eta(\tau) = \sqrt{-i\tilde \tau}\eta(\tilde \tau), \quad
	\vartheta _1(\mathfrak{b} |\tau ) =i\sqrt{-i\tilde{\tau}}e^{\frac{\pi i\tilde{\mathfrak{b}}^2}{\tilde{\tau}}}\vartheta _1(\tilde{\mathfrak{b}}|\tilde{\tau} ) \ ,
\end{equation}
\begin{equation}
	E_n \begin{bmatrix}
		1 \\ b
	\end{bmatrix}(\tau)
	= \sum_{k = 0}^{n} \frac{(-1)^{n - k}}{k!} \tilde{\mathfrak{b}}^k \tilde \tau^{n - k} E_{n - k} \begin{bmatrix}
		1 \\ \tilde b
	\end{bmatrix}(\tilde \tau) \ ,
\end{equation}
and
\begin{equation}
	E_n \begin{bmatrix}
		-1 \\ b
	\end{bmatrix}(\tau)
	= \sum_{k = 0}^{n} \frac{(-1)^{n - k}}{k!} \tilde{\mathfrak{b}}^k \tilde \tau^{n - k} E_{n - k} \begin{bmatrix}
		1 \\ - \tilde b
	\end{bmatrix}(\tilde \tau) \ .
\end{equation}

With these dual expressions, one can easily rewrite the Schur index $\mathcal{I}_{g, n}$ with $n = \text{even}$, 
\begin{align}
 	\mathcal{I}_{g, n} = & \ \frac{1}{2} (-i\tilde \tau)^{g - 1} \frac{\eta(\tilde \tau)^{2g - 2 + n}}{\prod_{j = 1}^{n}\vartheta_1(2 \tilde{\mathfrak{b}}_j|\tilde \tau)}
	\sum_{k = 2}^{2g - 2 + n}\lambda_k^{(2g - 2 + n)}\bigg(\\
	& \ \sum_{\ell = 0}^{k} \frac{1}{\ell!}
	\sum_{\alpha_j = \pm }\Big(\prod_{j = 1}^{n}\alpha_j\Big)\Big(\sum_{j=1}^{n}\alpha_j \tilde {\mathfrak{b}}_j\Big)^\ell (- \tilde \tau)^{k - \ell} E_{k - \ell} \begin{bmatrix}
	1 \\ \prod_j \tilde b^{\alpha_j} \nonumber
	\end{bmatrix} \bigg)\ .
\end{align}
This expression contains a sum of Eisenstein series, which are now series in $\tilde q$ starting with $1$. It also contains terms proportional to $\tilde \tau^{g - 1}, \tilde \tau^g, \dots, \tilde \tau^{3g - 3 + n}$, where the highest power coincides with the dimension of the complex moduli space of $\Sigma_{g, n}$. The factor $\frac{\eta(\tilde \tau)^{2g - 2 + n}}{\vartheta_1(2 \tilde{\mathfrak{b}}_j | \tilde \tau)}$ contributes a leading $\tilde q^{\frac{1}{12}(g - n - 1)}$, where the power gives the Cardy behavior \cite{CARDY1986186,Beem:2017ooy,ArabiArdehali:2023bpq},
\begin{equation}
	\frac{1}{12}(g - n - 1) = -2 (c_\text{4d} - a_\text{4d}) = - \frac{1}{24}c_\text{eff},
\end{equation}
where $c_\text{eff}$ is the effective central charge of the associated chiral algebra. The four dimensional central charges are related to $g, n$ by
\begin{equation}
	c_\text{4d} = \frac{1}{6}(5n + 13 g - 13), \quad
	a_\text{4d} = \frac{1}{24}(19n + 53 g - 53) \ .
\end{equation}
The associated chiral algebra has central charge $c_\text{2d} = -12 c_\text{4d}$, and therefore,
\begin{equation}
	c_\text{2d} - c_\text{eff} = - 12(2g - 2 + n) \ , \qquad
	\tilde q^{- \frac{c_\text{eff}}{24}}
	= \tilde q^{- \frac{c_\text{2d}}{24}} \tilde q^{- \frac{2g - 2 + n}{2}} \ .
\end{equation}
For theories that we will consider in this paper, $2g - 2 + n$ is positive.

As the simplest example, we first consider the 4d $\mathcal{N} = 4$ $SU(2)$ theory. The Schur index is given by
\begin{equation}
	\mathcal{I}_{\mathcal{N} = 4 \ SU(2)} = \frac{i\vartheta_4(\mathfrak{b})}{\vartheta_1(2 \mathfrak{b})} E_1 \begin{bmatrix}
		-1 \\ b
	\end{bmatrix} \ .
\end{equation}
In dual variables, the index reads
\begin{equation}
	\mathcal{I}_{\mathcal{N} = 4 \ SU(2)} = - \frac{\tilde{\mathfrak{b}}\vartheta_2(\tilde{\mathfrak{b}}|\tilde \tau)}{\vartheta_1(2 \tilde{\mathfrak{b}}|\tilde \tau)}
	+ \frac{i \tilde \tau}{2\pi} \frac{\vartheta_2'(\tilde{\mathfrak{b}}|\tilde \tau)}{\vartheta_1(2 \tilde{\mathfrak{b}}|\tilde \tau)} \ .
\end{equation}
Expanding the expression in series of $\tilde q$ and sending $\tilde b \to 1$, we find the unflavored high temperature behavior
\begin{equation}
	\mathcal{I}_{\mathcal{N} = 4 \ SU(2)} = - \frac{1}{2\pi} - \frac{2 \tilde q}{\pi} - \frac{6 \tilde q^2}{\pi} + ... - \frac{\ln \tilde q}{8\pi} - \frac{3 \tilde q \ln \tilde q}{2\pi}
	- \frac{9 \tilde q^2 \ln \tilde q}{2\pi} - \frac{27 \tilde q^3 \ln \tilde q}{2\pi} + \dots \ .
\end{equation}
which agrees with the result in \cite{ArabiArdehali:2023bpq}.

Another simple example is the Schur index of the 4d $\mathcal{N} = 2$ $SU(2)$ theory with $n_f = 4$ fundamental hypermultiplets. The index in the dual variables is given by
\begin{align}
	\label{eq:high-temperature-behavior-04}
	& \ \mathcal{I}_{0,4} = \frac{1}{2} \frac{1}{-i \tilde \tau} \frac{\eta(\tilde\tau)^2}{\prod_{j = 1}^{4}\vartheta_1(2 \tilde{\mathfrak{b}}_j)}\\
	& \ \times \sum_{\alpha_j = \pm}\bigg(\prod_j\alpha_j\bigg) \bigg[
		- \frac{(2\pi i)^2}{2}(\alpha \cdot \tilde{\mathfrak{b}})^2 - 2\pi i\tilde \tau(\alpha \cdot \tilde{\mathfrak{b}})E_1\begin{bmatrix}
		1 \\ \tilde b^\alpha
	\end{bmatrix}(\tilde \tau) + \frac{1}{2} \tilde \tau^2 E_2 \begin{bmatrix}
		1 \\ \tilde b^\alpha
	\end{bmatrix}(\tilde \tau) \bigg] \ . \nonumber
\end{align}
After unflavoring, we find the high temperature behavior by simply expanding the above analytic result in $\tilde q$ series,
\begin{align}
	= \frac{\tilde q^{-\frac{5}{12}}}{120\pi}\bigg[1 + 250 \tilde q + 4625\tilde q^2 + & \ 44250 \tilde q^3 + 305750\tilde q^4 + 1703752 \tilde{q}^5 \\
	& \ + 8150375 \tilde{q}^6 + 34673250 \tilde{q}^7 + \ldots)
	 + 60 \ln \tilde q \mathcal{I}_{0,4}(\tilde q) \bigg] \ , \nonumber
\end{align}
where
\begin{equation}
	\mathcal{I}_{0,4}(\tilde q) = q^{\frac{7}{12}} \ln \tilde q(1 + 28 \tilde q + 329 \tilde q + 2632 \tilde{q}^3 + 16380 \tilde{q}^4 + 85764 \tilde{q}^5 + 393589 \tilde{q}^6 + \dots) \ .
\end{equation}
This agrees with the result in \cite{ArabiArdehali:2023bpq}.

In the similar spirit, we turn to the defect index $\mathcal{I}^\text{defect}_{g,n}$ and its high temperature behavior. Using (\ref{eq:defect-index}), we rewrite it in the dual variables,
\begin{align}\label{eq:defect-index-dual}
  \mathcal{I}^\text{defect}_{g,n}(\kappa)
	= & \ (-1)^\kappa i^n (-i\tilde{\tau})^{g - 1}\frac{\eta(\tilde \tau)^{2g - 2 + n}}{\prod_{j = 1}^{n}\vartheta_1(2 \tilde{\mathfrak{b}}_j|\tilde \tau)}
	\sum_{k = 1}^{2g - 2 + n + 1}\tilde \lambda_k^{(2g - 2 + n + 1)}(\kappa + 1)
	\biggl( \nonumber\\
		& \ \sum_{\ell = 0}^{k}\frac{1}{\ell!}
		\sum_{\alpha_i = \pm} \biggl(\prod_{i = 1}^{n}\alpha_i\biggr)
		\bigl(\sum_{j = 1}^{n}\alpha_j \tilde{\mathfrak{b}}_j\bigr)^\ell
		(- \tilde \tau)^{k - \ell}E_{k - \ell} \begin{bmatrix}
			1 \\ (-1)^{\kappa+n}\Pi_j\tilde{b}_j^{\alpha_j}
		\end{bmatrix}
	\biggr) \ .
\end{align}
\subsection{\texorpdfstring{$S^3$ partition function}{}}

With the general result of the index in the dual variables, we can consider the high temperature limit $\tau \to +0i$ of the index. This limit implies that $\tilde \tau \to i \infty$, $\tilde q = e^{ - 2\pi i \frac{1}{\tau}} \to 0$. We again start with the example of $\mathcal{I}_{0,4}$ which has $c_\text{4d} - a_\text{4d} > 0$. From the above unflavored high temperature behavior, after multiplying $\tilde q^{- \frac{1}{12}(g - n - 1)} = \tilde q^{2(c_\text{4d} - a_\text{4d})} = \tilde q^{\frac{5}{12}}$, the index reads
\begin{equation}
	= \frac{1}{120\pi}(1 + 250 \tilde q + \ldots)
	+ 60 \tilde q \ln \tilde q (1 + 28 \tilde q + \dots) \ .
\end{equation}
In the $\tilde q \to 0$ limit, $\tilde q^k (\ln q)^\ell \to 0$ whenever $k, \ell \in \mathbb{N}_{>0}$, and therefore, the unflavored index is left with a $\tilde q$-independent constant term $\frac{1}{120\pi}$ in the high temperature limit. Let us now turn back on the flavor fugacities, keeping $\tilde b_i$ fixed in taking the limit. By direct computation,
\begin{align}
	\label{eq:high-temperature-limit-04}
	\mathcal{I}_{0,4} \xrightarrow{\tilde \tau \to i \infty}
	\left[\frac{i}{2} \frac{\eta(\tau)^2}{\prod_{j = 1}^{4}\vartheta_1(2 \tilde{\mathfrak{b}}_j)} \sum_{\alpha_j = \pm} \left(\prod_{j = 1}^{4}\alpha_j\right)(-)(\alpha \cdot \tilde{\mathfrak{b}}) E_1 \begin{bmatrix}
		1 \\ \tilde b^\alpha
	\end{bmatrix}(\tilde\tau)\right]_{\tilde q \to 0} \nonumber \\
	= - \frac{i}{2} \frac{\tilde b_1\tilde b_2\tilde b_3\tilde b_4}{\prod_{j = 1}^{4}(1 - \tilde b_j^2)}
	\sum_{\alpha_j = \pm} \left(\prod_{j = 1}^{4}\alpha_j\right)(\alpha \cdot \tilde{\mathfrak{b}_j})\left(- \frac{1}{2} + \frac{1}{1 - \prod_{j}\tilde b_j^{\alpha_j}}\right) \ .
\end{align}
Identifying $\mathfrak{m}_1 = i (\mathfrak{b}_1 + \mathfrak{b}_2)$, $\mathfrak{m}_2 = i (\mathfrak{b}_1 - \mathfrak{b}_2)$, $\mathfrak{m}_3 = i (\mathfrak{b}_3 + \mathfrak{b}_4)$, $\mathfrak{m}_4 = i (\mathfrak{b}_3 - \mathfrak{b}_4)$, it is not hard to verify that the limit agrees with the $S^3$ partition function of the $\mathcal{N} = 4$ $SU(2)$ theory with four fundamental hypermultiplets \cite{Benvenuti:2011ga},
\begin{equation}
	\mathcal{I}_{0,4} \xrightarrow{\tilde \tau \to i \infty}
	Z^{S^3}
	= \sum_{i = 1}^{4} \frac{\mathfrak{m}_i 2 \sinh(2\pi \mathfrak{m}_i)}{\prod_{j = 1, j \ne i}^{4} (2\sinh \pi \mathfrak{m}_i)^2 - (2\sinh \pi \mathfrak{m}_j)^2} \ .
\end{equation}
It is noteworthy to point out that the $\tilde \tau$ term in \eqref{eq:high-temperature-behavior-04} as a $\tilde q$ series does not start from $\tilde \tau$, but instead starts with $\tilde \tau \tilde q$. This is because
\begin{equation}
	\sum_{\alpha_j = \pm} \biggl(\prod_{j = 1}^{4}\alpha_j\biggr)E_2 \begin{bmatrix}
		1 \\ \prod_{j = 1}^{4}\tilde b_j^{\alpha_j}
	\end{bmatrix}(\tilde \tau) = {2\prod_{j = 1}^{4}(b_j - b_j^{-1})} \tilde q + O(\tilde q^2)\ ,
\end{equation}
which starts from $O(\tilde q)$.

\begin{figure}
	\centering
	\includegraphics[width=0.4\textwidth]{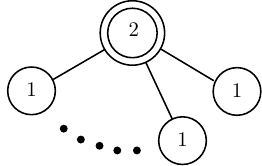}
	\caption{An $SU(2)\times U(1)^n$ star-shape quiver. A double circle denotes an $SU$ gauge group, while a single circle denotes a $U$ gauge group. Straight lines denote the bifundamental hypermultiplets. \label{fig:star-shape}}
\end{figure}

For theories with $g = 0$ and even $n = 4, 6, 8, \dots$, similar discussions follow. First we note that they all have $c_\text{4d} - a_\text{4d} > 0$. A $\tilde \tau^{K-1}$ ($0 \le K \le n - 2$) term is proportional to
\begin{align}\label{eq:tilde-tau-K-1-term}
	\sum_{k = 2}^{2g - 2 + n} \lambda_k^{(2g - 2 + n)} \frac{1}{(k - K)!}\sum_{\alpha_j = \pm} \left(\prod_{j = 1}^{n}\alpha_j\right)
	\left(\sum_{j=1}^{n}\alpha_j \tilde {\mathfrak{b}}_j\right)^{k - K} E_K \begin{bmatrix}
		1 \\  \tilde b^{\alpha}
	\end{bmatrix}
\end{align}
When $K = \text{odd}$, $E_K \big[\substack{1\\ b^\alpha}\big]$ always starts with a $\tilde q$ term except for $K = 1$, which corresponds to a $\tilde \tau^0$ term that will be dealt with later. When $K$ is even, the constant term in (\ref{eq:tilde-tau-K-1-term}) is proportional to
\begin{equation}
	\sum_{k = 2}^{- 2 + n} \frac{\lambda_k^{(2g - 2 + n)}}{(k - K)!} \sum_{\alpha_j = \pm} \left(\prod_{j = 1}^{n}\alpha_j\right)
	\left( \sum_{j = 1}^{n} \alpha_j \tilde{\mathfrak{b}}_j \right)^{k - K} \ .
\end{equation}
Since in the sum $2 \le k \le n - 2$, and therefore $k - K \le n$, the above sum over $\alpha_j$ vanishes. Therefore, after multiplying the factor $\tilde q^{- \frac{1}{12}(g - n - 1)}$, the lowest $\tilde q$-dependent terms must be proportional $\tilde q^{\ell' \ge 1}\tilde \tau^\ell \sim \tilde q^{\ell \ge 1} (\ln \tilde q)^\ell$, which vanishes in the $\tilde q \to 0$ limit. In other words, the high temperature limit of $\mathcal{I}_{g, n = \text{even}}$ keeps only the $\tilde q$-independent contribution, which reads
\begin{align}\label{eq:high-temperature-limit-even}
	& \ \mathcal{I}_{0,n} \\
	\to & \ \frac{-i}{2}\frac{\prod_{j = 1}^n \tilde b_j}{\prod_{j=1}^n ( 1- \tilde b_j^{2} )}
	\sum_{k = 2}^{2g - 2 + n}\sum_{\alpha _j=\pm}\frac{\lambda^{(2g - 2 + n)}_k}{(k-1)!}\Big(\prod_{j=1}^n{\alpha _j} \Big)(\alpha \cdot \tilde{\mathfrak{b}})^{k - 1}
	\left(\frac{1}{2} - \frac{1}{1- \prod_{j = 1}^{n}\tilde{b}_{j}^{\alpha _j}} \right)  \ .\nonumber
\end{align}
Here, we applied the leading order expansion
\begin{align}
	\tilde q^{- \frac{g - n - 1}{12}}\frac{\eta(\tilde\tau)^{2g - 2 + n}}{\prod_{j = 1}^{n}\vartheta_1(2 \tilde{\mathfrak{b}_j}|\tilde \tau)}
	= & \ - \frac{1}{\prod_{j = 1}^{n}(b_j - b_j^{-1})} + O(\tilde q)  \ ,\\
	E_1 \begin{bmatrix}
		1 \\ \prod_{j = 1}^{n}\tilde b_j^{\alpha_j}
	\end{bmatrix}(\tilde \tau)
	= & \ \frac{1}{2} - \frac{1}{1 - \prod_{j = 1}^{n}\tilde b_j^{\alpha_j}} + O(\tilde q)\ .
\end{align}
This is expected to give the a compact closed-form for the $S^3$ partition function of the 4d theories dimensionally reduced to 3d, whose 3d mirror theories are given by the $SU(2) \times U(1)^n$ quiver gauge theories. See Figure \ref{fig:star-shape}.
Concretely, the $S^3$ partition function of an $SU(2)^{n - 3}$ linear quiver theory can be written as
\begin{align}
	& \ Z^{S^3}_\text{linear quiver}
	= \\
	& \ \int \prod_{i}^{n - 3} \bigg[d\sigma_i Z_\text{VM}(\sigma_i)\bigg]
	Z_\text{HM}^{S^3}(\mathfrak{b}^\text{3d}_1, \mathfrak{b}^\text{3d}_2, \sigma_1)
	Z_\text{HM}^{S^3}(-\sigma_1, \mathfrak{b}^\text{3d}_3, \sigma_2) \cdots Z_\text{HM}^{S^3}(-\sigma_{n - 3}, \mathfrak{b}^\text{3d}_{n - 1}, \mathfrak{b}^\text{3d}_{n}) \ . \nonumber
\end{align}
Here the one-loop determinants are given by
\begin{equation}
	Z^{S^3}_\text{VM}(\sigma) \coloneqq \frac{1}{2} \biggl(2\sinh(2\pi \sigma)\biggr)^2, \quad
	Z^{S^3}_\text{HM}(a,b,c) = \frac{1}{\prod_{\beta, \gamma = \pm}2 \cosh(\pi(a + \beta b + \gamma c))} \ .\nonumber
\end{equation}
The mirror dual is an $SU(2) \times U(1)^n$ star-shape quiver with partition function
\begin{equation}
	Z^{S^3}_{SU(2) \times U(1)^n}
	= \int \frac{da}{2}(2\sinh(2\pi a))^2 \Big[\prod_{i=1}^{n} dc_i e^{2\pi i \xi_i c_i}\Big] \frac{1}{\prod_\pm\prod_{i = 1}^{n}2\cosh\pi(\pm a + c_i)} \ ,\nonumber
\end{equation}
where the parameters from both sides are related by 
\begin{equation}
	\xi_i = 2 \mathfrak{b}^\text{3d}_i \ .
\end{equation}
The $c_i$ integral can be performed, leaving a simpler integral \cite{Benvenuti:2011ga},
\begin{equation}\label{eq:Z-S3-SU2-U1n}
	Z^{S^3}_{SU(2) \times U(1)^n} = \int da \frac{1}{(2 \sinh(2\pi a))^{n - 2}} \prod_{i = 1}^{n} \frac{2 \sin(2\pi \xi_i a)}{2\sinh(\pi \xi_i)} \ .
\end{equation}

To identify the (\ref{eq:high-temperature-limit-even}) with the $S^3$ partition function, we set
\begin{equation}
	\tilde{\mathfrak{b}}_j = i \mathfrak{b}^\text{3d}_j = \frac{i}{2}\xi_j \ .
\end{equation}
We further notice that
\begin{equation}
	\sum_{k = 2}^{2g - 2 + n}\frac{\lambda_k^{(2g - 2 + n)}}{(k-1)!}x^{k - 1}
	= \frac{1}{(n-3)!} \prod_{k = - \frac{n-4}{2}}^{\frac{n-4}{2}}(x + k)\ ,
\end{equation}
hence we arrive at the closed form of the $S^3$ partition function,
\begin{align}
	& \ Z^{S^3}_{SU(2)^{n - 3}~\text{linear quiver}}(\mathfrak{b}^\text{3d} = - i \tilde{\mathfrak{b}})\\
	= & \ - \frac{i^{n + 1}}{2\prod_{j = 1}^{n}(\tilde b_j - \tilde b_j^{-1})}
	\sum_{\alpha_j = \pm}\frac{1}{(n-3)!}\left(- \frac{1}{2} + \frac{1}{1 - \prod_{j = 1}^{n}\tilde b_j^{\alpha_j}}\right)
	\prod_{k = -\frac{n - 4}{2}}^{k = \frac{n - 4}{2}}(k + \alpha \cdot \tilde {\mathfrak{b}}) \ . \nonumber
\end{align}

For $g=0$ with odd $n = 3, 5, 7, \dots$, a similar analysis can be performed, which yields the closed form
\begin{align}
	& \ Z^{S^3}_{SU(2)^{n - 3}~\text{linear quiver}}(\mathfrak{b}^\text{3d} = - i \tilde{\mathfrak{b}})\\
	= & \ + \frac{i^{n + 1}}{2\prod_{j = 1}^{n}(\tilde b_j - \tilde b_j^{-1})}
	\sum_{\alpha_j = \pm}\frac{1}{(n-3)!}\left(- \frac{1}{2} + \frac{1}{1 + \prod_{j = 1}^{n}\tilde b_j^{\alpha_j}}\right)
	\prod_{k = -\frac{n - 4}{2}}^{k = \frac{n - 4}{2}}(k + \alpha \cdot \tilde {\mathfrak{b}}) \ , \nonumber
\end{align}
where the product is over $k = - \frac{n - 4}{2}, - \frac{n-4}{2} + 1, ..., + \frac{n - 4}{2}$.

\subsection{\texorpdfstring{$S^3$ partition function with line operator}{}}

Now that we have reduced the $\mathcal{I}_{g = 0, n}$ to three dimension, it is natural to extend the calculation to $\mathcal{I}^\text{defect}_{g = 0,n}(\kappa)$ and study its $\tau \to +i0$ limit. For simplicity, we first assume $n = $ even. From the dual expression (\ref{eq:defect-index-dual}), we extract the $\tilde \tau$-independent terms,
\begin{align}
	& \ (-1)^{\kappa + 1} (-i\tau)^{g-1} \tilde q^{- \frac{g - n - 1}{12}} \frac{\eta(\tilde \tau)^{2g - 2 + n}}{\prod_{j = 1}^{n}\vartheta_1(2 \tilde{\mathfrak{b}}_j|\tilde \tau)}\\
	& \ \times \sum_{k = 1}^{2g - 2 + n + 1}
	\frac{\tilde \lambda_k^{(2g -2 + n + 1)}(\kappa + 1)}{(k-1)!}\sum_{\alpha_j = \pm} \bigg(\prod_{j = 1}^{n}\alpha_j\bigg)
	\biggl(\sum_{j = 1}^{n}\alpha_j \tilde{\mathfrak{b}}_j\biggr)^{k - 1}E_1 \begin{bmatrix}
		(-1)^\kappa \\ \tilde b^\alpha
	\end{bmatrix} \ ,\nonumber
\end{align}
and the $\tilde q$ independent term is given by
\begin{align}
	\to \frac{i\prod_{j = 1}^n \tilde b_j}{\prod_{j=1}^n ( 1- \tilde b_j^{2} )}
	\sum_{k = 1}^{2g - 2 + n + 1}
	\sum_{\alpha _j=\pm} & \ \frac{\tilde\lambda^{(2g - 2 + n + 1)}_k(\kappa + 1)}{(k-1)!}\\
	& \ \times \Big(\prod_{j=1}^n{\alpha _j} \Big)(\alpha \cdot \tilde{\mathfrak{b}})^{k - 1}
	\left(\frac{1}{2} -  \frac{1}{1- (-1)^\kappa\prod_{j = 1}^{n}\tilde{b}_{j}^{\alpha _j}} \right)  \ .\nonumber
\end{align}

We would like to understand the physical meaning of such an expression from a three dimensional perspective. A natural but slightly naive guess emerges from manipulating the $q$-YM expression of the defect index. Recall from section \ref{section:defect-index} that
\begin{equation}
  \mathcal{I}_{g,n}^\text{defect}(\kappa)
  = q^{- \frac{c_\text{2d}}{24}}
  \sum_{j \in \frac{1}{2}\mathbb{N}}
  S_{\kappa j}
  C_j(q)^{ - 2 + n} \prod_{i = 1}^{n} \psi_j(b_i) \ .
\end{equation}
We would like take the $\tau \to +i 0$ limit on the right. To do so properly, we consider the following reparametrization and the $\beta \to +0$ limit,
\begin{equation}
	q = e^{- 2\pi \beta},
	\quad
	\mathfrak{b}_i = \beta \frac{\xi_i}{2},
	\quad
	j = \frac{\sigma}{\beta}, \qquad \beta \to +0 \ .
\end{equation}
Under such limiting process,
\begin{align}
	C_j(q) \to \frac{1}{\sqrt{\beta}}e^{- \frac{\pi}{12\beta}} \frac{1}{2 \sinh(2\pi \sigma)}, \quad
	\psi_j(b_i) \to \beta^{\frac{1}{2}}e^{\frac{\pi}{4\beta}} \frac{\sin(2\pi\xi_i \sigma)}{\sinh(\pi \xi_i)} \ ,
	\quad
	S_{\kappa j} \to \frac{\sinh(2\pi(\kappa + 1)\sigma)}{\sinh(2\pi \sigma)} \ ,
\end{align}
where the reduction of $S_{\kappa j}$ is simply the $SU(2)$ character $\chi_{j = \frac{\kappa}{2}}(e^{2\pi \sigma})$.

Dropping all the $\beta$ terms which are not relevant for our purpose, we find that when $\kappa = 0$, the $q$-YM expression reduces to
\begin{equation}
	Z^{S_3}_{SU(2) \times U(1)^n} = \int d\sigma \biggl(\frac{1}{2\sinh(2\pi \sigma)}\biggr)^{n - 2} \prod_{i = 1}^{n} \frac{\sinh(2\pi \xi_i \sigma)}{\sinh (\pi \xi_i)} \ ,
\end{equation}
which is simply the $S^3$ partition function (\ref{eq:Z-S3-SU2-U1n}) of the star-shape quiver with $n$ legs. 

What about $\kappa > 0$? One might be tempted to simply conclude that
\begin{equation}
	\mathcal{I}_{g, n}^\text{defect}(\kappa) = Z^3_{SU(2)\times U(1)^n}[W_\kappa] \ , \qquad j = \frac{\kappa}{2},
\end{equation}
where the right hand side is the $S^3$ partition function of the star-shape quiver in the presence of a Wilson line charged under the central $SU(2)$ gauge node with spin $j = \frac{\kappa}{2}$,
\begin{equation}
	Z^3_{SU(2)\times U(1)^n}[W_\kappa]
	= \int d\sigma \chi_{j = \kappa/2}(e^{2\pi \sigma}) \biggl(\frac{1}{2\sinh(2\pi \sigma)}\biggr)^{n - 2} \prod_{i = 1}^{n} \frac{\sinh(2\pi \xi_i \sigma)}{\sinh (\pi \xi_i)} \ .
\end{equation}

Unfortunately, this conclusion is only correct up to a caveat. For example, it is easy to check that, as analytic functions, $\mathcal{I}^\text{defect}_{0, 4}(\kappa = 2) \propto \mathcal{I}_{0,4}$. However, the naive integral $Z^3_{SU(2)\times U(1)^n}[W_{\kappa = 2}]$ is not proportional to $Z^3_{SU(2)\times U(1)^n}$. In fact, with $\kappa = 2$ the integrand is not suppressed as $\sigma\to \pm \infty$, causing the integral to diverge. It is a general phenomenon, that for any given $g, n$, larger $\kappa$ always reduces convergence. For even $n > 0$, one may check the following facts concerning the convergence of the $S^3$-partition functions and the linear independence of the defect indices.
\begin{itemize}
	\item $\mathcal{I}_{g = 0, n}^\text{defect}(\kappa = 0)$, $\mathcal{I}_{g = 0, n}^\text{defect}(\kappa = 2) $, $\dots$, $\mathcal{I}_{g = 0, n}^\text{defect}(\kappa = n-4)$ are linear independent. In fact, they form a complete basis of all vortex defect index with even vorticity $\kappa$.
	\item $\mathcal{I}_{g = 0, n}^\text{defect}(\kappa = 1)$, $\mathcal{I}_{g = 0, n}^\text{defect}(\kappa = 3) $, $\dots$, $\mathcal{I}_{g = 0, n}^\text{defect}(\kappa = n-3)$ are linear independent; they form a complete basis of all vortex defect index with odd vorticity $\kappa$.
	\item $Z^{S^3}_{SU(2) \times U(1)^n}[W_{\kappa = 0}]$, $Z^{S^3}_{SU(2) \times U(1)^n}[W_{\kappa = 1}]$, $\dots$, $Z^{S^3}_{SU(2) \times U(1)^n}[W_{\kappa = n - 3}]$ are integrable.
	\item $\mathcal{I}_{g = 0, n}^\text{defect}(\kappa) \xrightarrow{q \to +i0} Z^{S^3}_{SU(2) \times U(1)^n}[W_{\kappa}]$, $\kappa = 0, 1, 2, ..., n - 3$.\footnote{We have performed numerical integration to verify the equalities.}
\end{itemize}
For odd integers $n>0$, similar conclusions hold:
\begin{itemize}
	\item $\mathcal{I}_{g=0,n}^{\mathrm{defect}}(\kappa=0),\mathcal{I}_{g=0,n}^{\mathrm{defect}}(\kappa=2),\ldots,\mathcal{I}_{g=0,n}^{\mathrm{defect}}(\kappa=n-3)$ form a linear independent and complete basis for the vortex defect index with even vorticity $\kappa$.
	\item 
	$\mathcal{I}_{g=0,n}^{\mathrm{defect}}(\kappa=1),\mathcal{I}_{g=0,n}^{\mathrm{defect}}(\kappa=2),\ldots,\mathcal{I}_{g=0,n}^{\mathrm{defect}}(\kappa=n-2)$ form a linear independent and complete basis for the vortex defect index with odd vorticity $\kappa$.
	\item $Z^{S^3}_{SU(2) \times U(1)^n}[W_{\kappa = 0}]$, $Z^{S^3}_{SU(2) \times U(1)^n}[W_{\kappa = 1}]$, $\dots$, $Z^{S^3}_{SU(2) \times U(1)^n}[W_{\kappa = n - 3}]$ are integrable.\footnote{The integrals involve a factor $\frac{\chi(e^{2\pi \sigma})}{(\sinh(2\pi\sigma))^{n-2}}$.  This factor tends to infinity as $\sigma \to \infty$ for $\kappa \geqslant n-1$, while it tends to a nonzero constant for $\kappa=n-2$, resulting in an oscillating integrand. Both cases lead to non-integrable integrals.}
	\item $\mathcal{I}_{g=0,n}^{\mathrm{defect}}(\kappa)\xrightarrow{q\to+i0}Z_{SU(2)\times U(1)^n}^{S^3}[W_\kappa],\kappa=0,1,2,...,n-3$.
\end{itemize}

 To summarize, for $g = 0$ and for both even and odd $ n \ge 4$,
\begin{equation}
	\mathcal{I}_{g = 0, n}^\text{defect}(\kappa) \xrightarrow{q \to +i0} Z^{S^3}_{SU(2) \times U(1)^n}[W_{\kappa}], \qquad \kappa = 0, 1, 2, ..., n - 3\ ,
\end{equation}
where the high temperature limit of the left yields the closed form expression for the Wilson line partition function,
\begin{align}
	\mathcal{I}_{g = 0, n}^\text{defect}(\kappa) \xrightarrow{q \to +i0}
	&
	\frac{i (-1)^{\frac{n - 2}{2}}}{\prod_{j = 1}^{n}(b_j - b_j^{-1})}
	\sum_{k = 1}^{2g - 2 + n + 1} \Bigg[~\frac{\tilde \lambda_k^{(2g - 2 + n + 1)}(\kappa + 1)}{(k - 1)!} \nonumber\\
	& \ \times \sum_{\alpha_j = \pm}\biggl(\prod_{j = 1}^{n}\alpha_j\biggr)(\alpha \cdot \tilde{\mathfrak{b}})^{k - 1}
	\biggl(
		\frac{1}{2} - \frac{1}{1 - (-1)^{\kappa + n}\prod_{j = 1}^{n} \tilde b_j^{\alpha_j}}
	\biggr) \Bigg] \ ,
\end{align}
where we identify $\tilde{\mathfrak{b}}_j = \frac{i}{2}\xi_j$.

\section*{Acknowledgments}
The authors would like to thank Bohan Li, Wolfger Peelaers, Kaiwen Sun, and Wenbin Yan for helpful discussions. The work of Y.P. is supported by the National Natural Science Foundation of China (NSFC) under Grant No. 11905301.
The work of P.Y. is supported by the National Natural Science Foundation of China, Grant No. 12375066, 11935009, 11975164.

%!TEX root = ../Schur index and modularity.tex
\appendix
\section{Special functions and notation \label{app:special-functions}}

\subsection{Dedekind eta function}

Throughout our paper, we use the following variable notation using the normal and fraktur font,
\begin{equation}
  a = e^{2\pi i \mathfrak{a}}, \quad
  b = e^{2\pi i \mathfrak{b}}, \quad
  \tilde b = e^{2\pi i \tilde{\mathfrak{b}}}, \quad
  \cdots, \quad
  z = e^{2\pi i \mathfrak{z}} \ ,
\end{equation}
with the exception $q = e^{2\pi i \tau}$ which is the standard practice.

The Dedekind eta function $\eta(\tau)$ is defined as
\begin{equation}
  \eta(\tau) \coloneqq q^{\frac{1}{24}} \prod_{n = 1}^{+\infty}(1 - q^n) \ .
\end{equation}
Under modular transformations,
\begin{equation}
  \eta(\tau + 1) = e^{\frac{\pi i}{12}}\eta(\tau), \qquad
  \eta( - \frac{1}{\tau}) = \sqrt{-i\tau}\eta(\tau) \ .
\end{equation}

\subsection{Jacobi theta functions}

The standard Jacobi theta functions are defined as
\begin{align}
	\vartheta_1(\mathfrak{z}|\tau) \coloneqq & \ -i \sum_{r \in \mathbb{Z} + \frac{1}{2}} (-1)^{r-\frac{1}{2}} e^{2\pi i r \mathfrak{z}} q^{\frac{r^2}{2}} ,
	& \vartheta_2(\mathfrak{z}|\tau) \coloneqq & \sum_{r \in \mathbb{Z} + \frac{1}{2}} e^{2\pi i r \mathfrak{z}} q^{\frac{r^2}{2}} \ ,\\
	\vartheta_3(\mathfrak{z}|\tau) \coloneqq & \ \sum_{n \in \mathbb{Z}} e^{2\pi i n \mathfrak{z}} q^{\frac{n^2}{2}},
	& \vartheta_4(\mathfrak{z}|\tau) \coloneqq & \sum_{n \in \mathbb{Z}} (-1)^n e^{2\pi i n \mathfrak{z}} q^{\frac{n^2}{2}} \ .
\end{align}
We will often omit the $\tau$ from the notation, and in particular,
\begin{equation}
  \vartheta_i(0) = \vartheta_i(\mathfrak{z} = 0|\tau)\ , \qquad
  \vartheta_i^{(k)}(0) = \partial_{\mathfrak{z}}^n \Big|_{\mathfrak{z} = 0} \vartheta_i(\mathfrak{z}|\tau) \ .
\end{equation}
These functions can be rewritten in infinite product form using $(z;q) \coloneqq \prod_{k = 0}^{+\infty}(1 - zq)$,
\begin{align}
  \vartheta_1(\mathfrak{z}|\tau) = & \ i q^{\frac{1}{8}} z^{-\frac{1}{2}}(q;q)(z;q)(z^{-1}q;q) \ , \\
  \vartheta_2(\mathfrak{z}|\tau) = & \ q^{\frac{1}{8}}z^{-\frac{1}{2}}(q;q)(-z;q)(- z^{-1}q;q) \ , \\
  \vartheta_3(\mathfrak{z}|\tau)= & \ (q;q)(-zq^{1/2};q)(- z^{-1}q^{1/2};q) \ , \\
  \vartheta_4(\mathfrak{z}|\tau)= & \ (q;q)(zq^{1/2};q)(z^{-1}q^{1/2};q) \ .
\end{align}
From these expressions zeros of $\vartheta_i$ can be easily read off.

The modularity of $\vartheta_i(\mathfrak{z} | \tau)$ is well-known. Under the $S$ and $T$ transformations, which act, as usual, on the nome and flavor fugacity as $(\frac{\mathfrak{z}}{\tau}, - \frac{1}{\tau})\xleftarrow{~~S~~}(\mathfrak{z}, \tau) \xrightarrow{~~T~~} (\mathfrak{z}, \tau + 1)$,
\begin{center}
  \begin{tikzpicture}

    \node(m1) at (0,0) {$\vartheta_1$};
    \node(m2) at (0,-1) {$\vartheta_2$};
    \node(m3) at (0,-2) {$\vartheta_3$};
    \node(m4) at (0,-3) {$\vartheta_4$};

    \node(l1) at (-2,0) {$-i\alpha\vartheta_1$};
    \node(l2) at (-2,-1) {$\alpha\vartheta_2$};
    \node(l3) at (-2,-2) {$\alpha\vartheta_3$};
    \node(l4) at (-2,-3) {$\alpha\vartheta_4$};

    \node(r1) at (1.5,0) {$e^{\frac{\pi i}{4}}\vartheta_1$};
    \node(r2) at (1.5,-1) {$e^{\frac{\pi i}{4}}\vartheta_2$};
    \node(r3) at (1.5,-2) {$\vartheta_3$};
    \node(r4) at (1.5,-3) {$\vartheta_4$};

    \draw[->=stealth] (m1)--node[above]{$S$}(l1);
    \draw[->=stealth] (m2)--(l4);
    \draw[->=stealth] (m3)--node[above]{$S$}(l3);
    \draw[->=stealth] (m4)--(l2);
    \draw[->=stealth] (m1)--node[above]{$T$}(r1);
    \draw[->=stealth] (m2)--node[above]{$T$}(r2);
    \draw[->=stealth] (m3)--node[above]{$T$}(r4);
    \draw[->=stealth] (m4)--(r3);
  \end{tikzpicture}
\end{center}
Here we define $\alpha \coloneqq \sqrt{-i \tau}e^{\frac{\pi i}{\tau} \mathfrak{z}^2}$. These transformation can be used to deduce the modular transformation of the Eisenstein series that we will now review.

\subsection{Eisenstein series}
The Eisenstein series\footnote{In the literature these functions are often called twisted Eisenstein series.} depend on two characteristics $\phi, \theta$, and are defined as by the infinite sum,
\begin{align}
	E_{k \ge 1}\left[\begin{matrix}
		\phi \\ \theta
	\end{matrix}\right] \coloneqq & \ - \frac{B_k(\lambda)}{k!} \\
	& \ + \frac{1}{(k-1)!}\sum_{r \ge 0}' \frac{(r + \lambda)^{k - 1}\theta^{-1} q^{r + \lambda}}{1 - \theta^{-1}q^{r + \lambda}}
	+ \frac{(-1)^k}{(k-1)!}\sum_{r \ge 1} \frac{(r - \lambda)^{k - 1}\theta q^{r - \lambda}}{1 - \theta q^{r - \lambda}} \ , \nonumber
\end{align}
where $\phi \coloneqq e^{2\pi i \lambda}$ with $0 \le \lambda < 1$, $B_k(x)$ denotes the $k$-th Bernoulli polynomial, and the prime in the sum indicates that the $r = 0$ should be omitted when $\phi = \theta = 1$. Additionally, we also define
\begin{align}
	E_0\left[\begin{matrix}
		\phi \\ \theta
	\end{matrix}\right] = -1 \ .
\end{align}
Eisenstein with even $k = 2n$ are related to the usual Eisenstein series $E_{2n}(\tau)$ by sending the $\theta, \phi \to 1$. When $k$ is odd, $\theta = \phi = 1$ is a vanishing limit, except for the case $k = 1$ where it is singular,
\begin{align}
	E_{2n}\left[\begin{matrix}
		+1 \\ +1
	\end{matrix}\right] = E_{2n} \ , \qquad E_1\left[\begin{matrix}
		+ 1 \\ z
	\end{matrix}\right] = \frac{1}{2\pi i }\frac{\vartheta'_1(\mathfrak{z})}{\vartheta_1(\mathfrak{z})}, \qquad
	E_{2n + 1 \ge 3}\left[\begin{matrix}
		+1 \\ +1
	\end{matrix}\right] = 0 \ .
\end{align}
In fact, all $E_k\big[\substack{\pm 1 \\ z}\big]$ are regular as $z \to 1$, except for $E_1\big[{\substack{1 \\ z}}\big]$ which has a simple pole.

A closely related property is the symmetry of the Eisenstein series
\begin{align}\label{Eisenstein-symmetry}
	E_k\left[\begin{matrix}
	  \pm 1 \\ z^{-1}
	\end{matrix}\right] = (-1)^k E_k\left[\begin{matrix}
	  \pm 1 \\ z
	\end{matrix}\right] \ .
\end{align}

To compute the modular property of Eisenstein series, it is useful to relate them with the Jacobi theta functions, 
\begin{align}\label{EisensteinToTheta-2}
	E_k\left[\begin{matrix}
		+ 1 \\ z
	\end{matrix}\right] = \sum_{\ell = 0}^{\lfloor k/2 \rfloor}  \frac{(-1)^{k + 1}}{(k - 2\ell)!}\left(\frac{1}{2\pi i}\right)^{k - 2\ell} \mathbb{E}_{2\ell} \frac{\vartheta_1^{(k - 2\ell)}(\mathfrak{z})}{\vartheta_1(\mathfrak{z})} \ ,
\end{align}
Here $\vartheta_i^{(k)}(\mathfrak{z})$ is the $k$-th derivative in $\mathfrak{z}$. The conversion from $E_k\left[\substack{- 1 \\ \pm z}\right]$ can be obtained by replacing $\vartheta_1$ with $\vartheta_{2,3,4}$ appropriately.

With the help from the known modular property of $\vartheta_i$, we deduce that the Eisenstein series transform according to the following formula,
\begin{align}\label{Eisenstein-S-transformation}
  E_n \begin{bmatrix}
    +1 \\ +z
  \end{bmatrix} \xrightarrow{S} &
  \left(\frac{1}{2\pi i}\right)^n \sum_{\ell = 0}^k \frac{1}{(n - \ell)!}(- \log z)^{n - \ell} (\log q)^\ell E_\ell \begin{bmatrix}
    +1 \\ z
\end{bmatrix}\ ,\\
  E_n \begin{bmatrix}
    -1 \\ +z
  \end{bmatrix} \xrightarrow{S} &
  \left(\frac{1}{2\pi i}\right)^n \sum_{\ell = 0}^k \frac{1}{(n - \ell)!}(- \log z)^{n - \ell} (\log q)^\ell E_\ell \begin{bmatrix}
    +1 \\ -z
\end{bmatrix}\ ,\\
  E_n \begin{bmatrix}
    1 \\ -z
  \end{bmatrix} \xrightarrow{S} &
  \left(\frac{1}{2\pi i}\right)^n \sum_{\ell = 0}^k \frac{1}{(n - \ell)!}(- \log z)^{n - \ell} (\log q)^\ell E_\ell \begin{bmatrix}
    -1 \\ z
\end{bmatrix}\ ,\\
  E_n \begin{bmatrix}
    -1 \\ -z
  \end{bmatrix} \xrightarrow{S} &
  \left(\frac{1}{2\pi i}\right)^n \sum_{\ell = 0}^k \frac{1}{(n - \ell)!}(- \log z)^{n - \ell} (\log q)^\ell E_\ell \begin{bmatrix}
    -1 \\ -z
\end{bmatrix}\ ,
\end{align}
while under the $T$-transformation,
\begin{align}\label{Eisenstein-T-transformation}
  E_n \begin{bmatrix}
    + 1 \\ + z
  \end{bmatrix} \xrightarrow{T}& \ E_n \begin{bmatrix}
    + 1 \\ + z
  \end{bmatrix}, & 
  E_n \begin{bmatrix}
    - 1 \\ + z
  \end{bmatrix} \xrightarrow{T}& \
  E_n \begin{bmatrix}
    - 1 \\ - z
  \end{bmatrix} \\
  E_n \begin{bmatrix}
    + 1 \\ - z
  \end{bmatrix} \xrightarrow{T}& \ E_n \begin{bmatrix}
    + 1 \\ - z
  \end{bmatrix}, & 
  E_n \begin{bmatrix}
    - 1 \\ - z
  \end{bmatrix} \xrightarrow{T}& \ 
  E_n \begin{bmatrix}
    - 1 \\ + z
  \end{bmatrix} \ .
\end{align}

Combining these transformation, we obtain
\begin{align}
  E_n \begin{bmatrix}
    -1 \\ z
  \end{bmatrix} \xrightarrow{STS}
  \left(\frac{1}{2\pi i}\right)^n\left[\bigg(\sum_{k \ge 0}\frac{1}{k!}(- \log z)^k y^k\bigg)
  \bigg(\sum_{\ell \ge 0}(\log q - 2\pi i)^\ell y^\ell E_\ell \begin{bmatrix}
    -1 \\ +z
  \end{bmatrix}\bigg)\right]_n\ . \nonumber
\end{align}

\section{Integration formula}

To compute the Schur index of Lagrangian theories analytically, we make use of a series of integration formula involving elliptic functions, Eisenstein series and polynomials.

The basic contour integral involves an elliptic function $f$, $f(\mathfrak{z}) = f(\mathfrak{z} + 1) = f(\mathfrak{z} + \tau)$,
\begin{align}\label{elliptic-integral-intro}
	\oint_{|z| = 1} \frac{dz}{2\pi i z} f(\mathfrak{z})
	= f(\mathfrak{z}_0)
	  + \sum_{\text{real } \mathfrak z_j}R_j\ E_1\left[\begin{matrix}
			-1 \\ \frac{z_j}{z_0} q^{\frac{1}{2}}
		\end{matrix}\right]
	  + \sum_{\text{imag. } \mathfrak z_j}R_j\ E_1\left[\begin{matrix}
			-1 \\ \frac{z_j}{z_0} q^{ - \frac{1}{2}}
		\end{matrix}\right] \ ,
\end{align}
The contour integral involving both an elliptic function $f$ and an Eisenstein series can also be computed, for example,
\begin{align}
	& \ \oint_{|z| = 1} \frac{dz}{2\pi i z}f(\mathfrak{z})E_k\left[
	\begin{matrix}
		-1 \\ za
	\end{matrix}\right] \nonumber\\
	= & \ - \mathcal{S}_{k} \left(f(\mathfrak{z}_0)
	  + \sum_{\text{real/imag } \mathfrak{z}_i}R_i  E_{1}\left[\begin{matrix}
			-1 \\ \frac{z_i}{z_0}q^{\pm \frac{1}{2}}
		\end{matrix}\right]
	\right)
	- \sum_{\text{real/imag } \mathfrak{z}_i} R_i  \sum_{\ell = 0}^{k - 1} \mathcal{S}_{\ell} E_{k - \ell + 1}\left[
		\begin{matrix}
			1 \\ z_i a q^{\pm \frac{1}{2}}
		\end{matrix}\right] \ , 
\end{align}
where $\mathcal{S}_{2n}$ is the $2n$-th Taylor coefficient of $\frac{y}{2}\frac{1}{\sin (y/2)}$ in the variable $y$.

The integral involving Eisenstein series and a monomial is given by the following formulas
\begin{equation}
	\begin{aligned}
		&\oint\frac{dz}{2\pi iz}z^nE_k\begin{bmatrix}1\\za\end{bmatrix}=\frac1{(k-1)!}\frac{q^n}{a^n}\frac{\operatorname{Eu}_{k-1}(q^n)}{(1-q^n)^k},\\
		& \oint\frac{dz}{2\pi iz}z^nE_k\begin{bmatrix}-1\\za\end{bmatrix}=\frac{1}{(k-1)!}\frac{q^{n/2}}{a^n}\Phi(q^n,1-k,\frac{1}{2}),
	\end{aligned}
\end{equation}
where $\operatorname{Eu}_n(t)$ represents the Eulerian polynomial defined as
\begin{equation}\sum_{n=0}^{+\infty}\operatorname{Eu}_n(t)\frac{x^n}{n!}=\frac{t-1}{t-e^{(t-1)x}},\end{equation}
and $\Phi$ denotes the Lerch transcendent function defined by
\begin{equation}\Phi(z,s,a):=\sum_{p=0}^{+\infty}\frac{z^p}{(p+a)^s}.\end{equation}

When computing Wilson line index of type-2, we need to perform the following integral with two Eisenstein series and a monomial, which satisfy the following property,
\begin{equation}\label{int1}
	\begin{aligned}
		\oint\frac{dz}{2\pi iz}z^nE_{k_1}\begin{bmatrix}\pm1\\za\end{bmatrix}E_{k_1}\begin{bmatrix}\pm1\\zb\end{bmatrix}& =b^{-n}\oint\frac{dz}{2\pi iz}z^nE_{k_1}\begin{bmatrix}\pm1\\za/b\end{bmatrix}E_{k_2}\begin{bmatrix}\pm1\\z\end{bmatrix}  \\
		&=a^{-n}\oint\frac{dz}{2\pi iz}z^nE_{k_1}\begin{bmatrix}\pm1\\z\end{bmatrix}E_{k_2}\begin{bmatrix}\pm1\\zb/a\end{bmatrix}.
\end{aligned}\end{equation}
In particular, to compute the type-2 index $\langle W_j\rangle_{g_1, n_1; g_2, n_2}$ with even $n_i$, we make use of the formula
\begin{equation}\begin{aligned}
		&\oint\frac{dz}{2\pi iz}z^nE_{k_1}\begin{bmatrix}+1\\z\end{bmatrix}E_{k_2}\begin{bmatrix}+1\\za\end{bmatrix} \\
		&=\sum_{\ell=0}^{k_{1}}\frac{1}{\ell!}\frac{q^{n}}{a^{n}}\frac{\operatorname{Eu}_{k_{2}+\ell-1}(q^{n})}{(1-q^{n})^{k_{2}+\ell}}\left[\frac{(-1)^{k_{1}-\ell}}{(k_{2}-1)!}+\frac{\ell!a^{n}}{(k_{1}-1)!(k_{2}-k_{1}+\ell)!}\right]E_{k_{1}-\ell}\begin{bmatrix}+1\\a\end{bmatrix}
\end{aligned}\end{equation}
when $n\in\mathbb{Z}_{\neq0}$ and $k_1\geq k_2$, and
\begin{equation}\begin{aligned}
		&\oint\frac{dz}{2\pi iz}z^nE_{k_1}\begin{bmatrix}+1\\z\end{bmatrix}E_{k_2}\begin{bmatrix}+1\\za\end{bmatrix}\\&=\sum_{\ell=0}^{k_2}\frac{1}{\ell!}\frac{q^n}{a^n}\frac{\operatorname{Eu}_{k_1+\ell-1}(q^n)}{(1-q^n)^{k_1+\ell}}\left[\frac{a^n}{(k_1-1)!}+\frac{(-1)^{k_2-\ell}\ell!}{(k_2-1)!(k_1-k_2+\ell)!}\right]E_{k_2-\ell}\begin{bmatrix}+1\\a\end{bmatrix}
\end{aligned}\end{equation}
when $k_1\leq k_2$. The two integration formula can be further summarized into:
\begin{equation}\oint\frac{dz}{2\pi iz}z^nE_{k_1}\begin{bmatrix}+1\\z\end{bmatrix}E_{k_2}\begin{bmatrix}+1\\za\end{bmatrix}=\sum_{\ell=1}^{\max(k_1,k_2)}\frac1{\ell!}\frac{q^n}{a^n}\mathcal{E}_{k_1,k_2;\ell}(a^n,q^n)E_{\max(k_1,k_2)-\ell}\begin{bmatrix}+1\\a\end{bmatrix},\end{equation}
with rational functions $\mathcal{E}_{k_1,k_2;\ell}(a^n,q^n)$ that can be read off from the two separate situations. For other cases involving odd $n_i$, the contour integrals contain $E_k\big[\substack{-1 \\ z}\big]$. Such integrals can be obtained by solving the following set of equations,
{\begin{equation}
	q^{-n/2}\oint \frac{dz}{2\pi iz}z^n E_{k_1}\begin{bmatrix}1\\z\end{bmatrix}E_{k_2}\begin{bmatrix}1\\za\end{bmatrix}=\sum_{\ell_1=0}^{k_1}\sum_{\ell_2=0}^{k_2}\frac{1}{2^{\ell_1+\ell_2}}\frac{1}{\ell_1!\ell_2!}\oint\frac{dz}{2\pi iz}z^n E_{k_1}\begin{bmatrix}-1\\z\end{bmatrix}E_{k_2}\begin{bmatrix}-1\\z a\end{bmatrix}.
\end{equation}}
{\begin{equation}
	\begin{aligned}
		q^{-n/2}\oint \frac{dz}{2\pi iz}z^n E_{k_1}\begin{bmatrix}1\\z\end{bmatrix}E_{k_2}\begin{bmatrix}1\\za q^{1/2}\end{bmatrix}=&\sum_{\ell_1=0}^{k_1}\sum_{\ell_2=0}^{k_2}\sum_{\ell_3=0}^{k_2-\ell_2}\frac{1}{2^{\ell_1+\ell_2+\ell_3}}\frac{1}{\ell_1!\ell_2!\ell_3!}\oint\frac{dz}{2\pi iz}z^n\\
		&E_{k_1}\begin{bmatrix}1\\z\end{bmatrix}E_{k_2-\ell_2-\ell_3}\begin{bmatrix}-1\\z a\end{bmatrix}.
	\end{aligned}
\end{equation}}

Without writing down the result explicitly, we note that the integrals involve some linear combinations of
\begin{equation}\label{twom}
	\oint \frac{dz}{2\pi iz}z^n E_{k}\begin{bmatrix}1\\z\end{bmatrix}E_{i_2}\begin{bmatrix}1\\za\end{bmatrix}, ~ \oint \frac{dz}{2\pi iz}z^n E_{k}\begin{bmatrix}-1\\z\end{bmatrix}, ~ \oint \frac{dz}{2\pi iz}z^n E_{k}\begin{bmatrix}-1\\z a\end{bmatrix} 
\end{equation}
when both $n_i$ are odd, and
\begin{equation}\label{onem}
	\oint \frac{dz}{2\pi iz}z^n E_{i_1}\begin{bmatrix}1\\z\end{bmatrix}E_{i_2}\begin{bmatrix}1\\za q^{1/2}\end{bmatrix},
  ~\oint \frac{dz}{2\pi iz}z^n E_{i_1}\begin{bmatrix}1\\z\end{bmatrix},
  ~ \oint \frac{dz}{2\pi iz}z^n E_{i_1}\begin{bmatrix}-1\\z a\end{bmatrix} \ 
\end{equation}
and when only one $n_i$ is odd,

\bibliographystyle{utphys2}

\bibliography{ref}

\providecommand{\href}[2]{#2}\begingroup\raggedright\begin{thebibliography}{10}\setlength{\parskip}{1pt}\setlength{\itemsep}{0pt
  plus 0.3ex}

\bibitem{Verlinde:1988sn}
E.~P. Verlinde, ``{Fusion Rules and Modular Transformations in 2D Conformal
  Field Theory},'' \href{http://dx.doi.org/10.1016/0550-3213(88)90603-7}{{\em
  Nucl. Phys. B} {\bfseries 300} (1988) 360--376}.

\bibitem{CARDY1986186}
J.~L. Cardy, ``Operator content of two-dimensional conformally invariant
  theories,''
  \href{http://dx.doi.org/https://doi.org/10.1016/0550-3213(86)90552-3}{{\em
  Nuclear Physics B} {\bfseries 270} (1986) 186--204}.
  \url{https://www.sciencedirect.com/science/article/pii/0550321386905523}.

\bibitem{Mathur:1988rx}
S.~D. Mathur, S.~Mukhi, and A.~Sen, ``{Differential Equations for Correlators
  and Characters in Arbitrary Rational Conformal Field Theories},''
  \href{http://dx.doi.org/10.1016/0550-3213(89)90022-9}{{\em Nucl. Phys. B}
  {\bfseries 312} (1989) 15--57}.

\bibitem{Eguchi:1987qd}
T.~Eguchi and H.~Ooguri, ``{Differential Equations for Conformal Characters in
  Moduli Space},'' \href{http://dx.doi.org/10.1016/0370-2693(88)91567-5}{{\em
  Phys. Lett. B} {\bfseries 203} (1988) 44}.

\bibitem{Naculich:1989NuPhB.323..423N}
S.~G. {Naculich}, ``{Differential equations for rational conformal
  characters},'' \href{http://dx.doi.org/10.1016/0550-3213(89)90150-8}{{\em
  Nuclear Physics B} {\bfseries 323} no.~2, (Sept., 1989) 423--440}.

\bibitem{Chandra:2018pjq}
A.~R. Chandra and S.~Mukhi, ``{Towards a Classification of Two-Character
  Rational Conformal Field Theories},''
  \href{http://dx.doi.org/10.1007/JHEP04(2019)153}{{\em JHEP} {\bfseries 04}
  (2019) 153}, \href{http://arxiv.org/abs/1810.09472}{{\ttfamily
  arXiv:1810.09472 [hep-th]}}.

\bibitem{Bae:2020xzl}
J.-B. Bae, Z.~Duan, K.~Lee, S.~Lee, and M.~Sarkis, ``{Fermionic rational
  conformal field theories and modular linear differential equations},''
  \href{http://dx.doi.org/10.1093/ptep/ptab033}{{\em PTEP} {\bfseries 2021}
  no.~8, (2021) 08B104}, \href{http://arxiv.org/abs/2010.12392}{{\ttfamily
  arXiv:2010.12392 [hep-th]}}.

\bibitem{Das:2020wsi}
A.~Das, C.~N. Gowdigere, and J.~Santara, ``{Wronskian Indices and Rational
  Conformal Field Theories},''
  \href{http://dx.doi.org/10.1007/JHEP04(2021)294}{{\em JHEP} {\bfseries 04}
  (2021) 294}, \href{http://arxiv.org/abs/2012.14939}{{\ttfamily
  arXiv:2012.14939 [hep-th]}}.

\bibitem{Mukhi:2020gnj}
S.~Mukhi, R.~Poddar, and P.~Singh, ``{Rational CFT with three characters: the
  quasi-character approach},''
  \href{http://dx.doi.org/10.1007/JHEP05(2020)003}{{\em JHEP} {\bfseries 05}
  (2020) 003}, \href{http://arxiv.org/abs/2002.01949}{{\ttfamily
  arXiv:2002.01949 [hep-th]}}.

\bibitem{Kaidi:2021ent}
J.~Kaidi, Y.-H. Lin, and J.~Parra-Martinez, ``{Holomorphic modular bootstrap
  revisited},'' \href{http://dx.doi.org/10.1007/JHEP12(2021)151}{{\em JHEP}
  {\bfseries 12} (2021) 151}, \href{http://arxiv.org/abs/2107.13557}{{\ttfamily
  arXiv:2107.13557 [hep-th]}}.

\bibitem{Bae:2021mej}
J.-B. Bae, Z.~Duan, K.~Lee, S.~Lee, and M.~Sarkis, ``{Bootstrapping fermionic
  rational CFTs with three characters},''
  \href{http://dx.doi.org/10.1007/JHEP01(2022)089}{{\em JHEP} {\bfseries 01}
  (2022) 089}, \href{http://arxiv.org/abs/2108.01647}{{\ttfamily
  arXiv:2108.01647 [hep-th]}}.

\bibitem{Das:2021uvd}
A.~Das, C.~N. Gowdigere, and J.~Santara, ``{Classifying three-character RCFTs
  with Wronskian index equalling 0 or 2},''
  \href{http://dx.doi.org/10.1007/JHEP11(2021)195}{{\em JHEP} {\bfseries 11}
  (2021) 195}, \href{http://arxiv.org/abs/2108.01060}{{\ttfamily
  arXiv:2108.01060 [hep-th]}}.

\bibitem{Mahanta:2022fvl}
R.~Mahanta and T.~Sengupta, ``Modular linear differential equations for
  four-point sphere conformal blocks,''
  \href{http://dx.doi.org/10.1007/JHEP02(2023)158}{{\em JHEP} {\bfseries 02}
  (2023) 158}, \href{http://arxiv.org/abs/2211.05158}{{\ttfamily
  arXiv:2211.05158 [hep-th]}}.

\bibitem{Eleftheriou:2022kkv}
G.~Eleftheriou, ``{Root of unity asymptotics for Schur indices of 4d Lagrangian
  theories},'' \href{http://dx.doi.org/10.1007/JHEP01(2023)081}{{\em JHEP}
  {\bfseries 01} (2023) 081}, \href{http://arxiv.org/abs/2207.14271}{{\ttfamily
  arXiv:2207.14271 [hep-th]}}.

\bibitem{ArabiArdehali:2023bpq}
A.~Arabi~Ardehali, M.~Martone, and M.~Rossell\'o, ``{High-temperature expansion
  of the Schur index and modularity},''
  \href{http://arxiv.org/abs/2308.09738}{{\ttfamily arXiv:2308.09738
  [hep-th]}}.

\bibitem{Jiang:2024baj}
H.~Jiang, ``{Modularity in Argyres-Douglas Theories with $a=c$},''
  \href{http://arxiv.org/abs/2403.05323}{{\ttfamily arXiv:2403.05323
  [hep-th]}}.

\bibitem{Beem:2013sza}
C.~Beem, M.~Lemos, P.~Liendo, W.~Peelaers, L.~Rastelli, and B.~C. van Rees,
  ``{Infinite Chiral Symmetry in Four Dimensions},''
  \href{http://dx.doi.org/10.1007/s00220-014-2272-x}{{\em Commun. Math. Phys.}
  {\bfseries 336} no.~3, (2015) 1359--1433},
\href{http://arxiv.org/abs/1312.5344}{{\ttfamily arXiv:1312.5344 [hep-th]}}.
%%CITATION = ARXIV:1312.5344;%%.

\bibitem{Beem:2014rza}
C.~Beem, W.~Peelaers, L.~Rastelli, and B.~C. van Rees, ``{Chiral algebras of
  class S},'' \href{http://dx.doi.org/10.1007/JHEP05(2015)020}{{\em JHEP}
  {\bfseries 05} (2015) 020}, \href{http://arxiv.org/abs/1408.6522}{{\ttfamily
  arXiv:1408.6522 [hep-th]}}.

\bibitem{Lemos:2014lua}
M.~Lemos and W.~Peelaers, ``{Chiral Algebras for Trinion Theories},''
  \href{http://dx.doi.org/10.1007/JHEP02(2015)113}{{\em JHEP} {\bfseries 02}
  (2015) 113}, \href{http://arxiv.org/abs/1411.3252}{{\ttfamily arXiv:1411.3252
  [hep-th]}}.

\bibitem{Xie:2016evu}
D.~Xie, W.~Yan, and S.-T. Yau, ``{Chiral algebra of the Argyres-Douglas theory
  from M5 branes},'' \href{http://dx.doi.org/10.1103/PhysRevD.103.065003}{{\em
  Phys. Rev. D} {\bfseries 103} no.~6, (2021) 065003},
  \href{http://arxiv.org/abs/1604.02155}{{\ttfamily arXiv:1604.02155
  [hep-th]}}.

\bibitem{Kiyoshige:2020uqz}
K.~Kiyoshige and T.~Nishinaka, ``{The Chiral Algebra of Genus Two Class
  $\mathcal{S}$ Theory},'' \href{http://arxiv.org/abs/2009.11629}{{\ttfamily
  arXiv:2009.11629 [hep-th]}}.

\bibitem{Cordova:2015nma}
C.~Cordova and S.-H. Shao, ``{Schur Indices, BPS Particles, and Argyres-Douglas
  Theories},'' \href{http://dx.doi.org/10.1007/JHEP01(2016)040}{{\em JHEP}
  {\bfseries 01} (2016) 040}, \href{http://arxiv.org/abs/1506.00265}{{\ttfamily
  arXiv:1506.00265 [hep-th]}}.

\bibitem{Cordova:2016uwk}
C.~Cordova, D.~Gaiotto, and S.-H. Shao, ``{Infrared Computations of Defect
  Schur Indices},'' \href{http://dx.doi.org/10.1007/JHEP11(2016)106}{{\em JHEP}
  {\bfseries 11} (2016) 106}, \href{http://arxiv.org/abs/1606.08429}{{\ttfamily
  arXiv:1606.08429 [hep-th]}}.

\bibitem{Cordova:2017mhb}
C.~Cordova, D.~Gaiotto, and S.-H. Shao, ``{Surface Defects and Chiral
  Algebras},'' \href{http://dx.doi.org/10.1007/JHEP05(2017)140}{{\em JHEP}
  {\bfseries 05} (2017) 140}, \href{http://arxiv.org/abs/1704.01955}{{\ttfamily
  arXiv:1704.01955 [hep-th]}}.

\bibitem{Nishinaka:2018zwq}
T.~Nishinaka, S.~Sasa, and R.-D. Zhu, ``{On the Correspondence between Surface
  Operators in Argyres-Douglas Theories and Modules of Chiral Algebra},''
  \href{http://dx.doi.org/10.1007/JHEP03(2019)091}{{\em JHEP} {\bfseries 03}
  (2019) 091}, \href{http://arxiv.org/abs/1811.11772}{{\ttfamily
  arXiv:1811.11772 [hep-th]}}.

\bibitem{Pan:2021mrw}
Y.~Pan and W.~Peelaers, ``{The exact Schur index in closed form},''
  \href{http://arxiv.org/abs/2112.09705}{{\ttfamily arXiv:2112.09705
  [hep-th]}}.

\bibitem{Guo:2023mkn}
Z.~Guo, Y.~Li, Y.~Pan, and Y.~Wang, ``{$\mathcal{N}=2$ Schur index and line
  operators},'' \href{http://dx.doi.org/10.1103/PhysRevD.108.106002}{{\em Phys.
  Rev. D} {\bfseries 108} no.~10, (2023) 106002},
  \href{http://arxiv.org/abs/2307.15650}{{\ttfamily arXiv:2307.15650
  [hep-th]}}.

\bibitem{Kac:1988qc}
V.~G. Kac and M.~Wakimoto, ``{Modular invariant representations of infinite
  dimensional Lie algebras and superalgebras},''
  \href{http://dx.doi.org/10.1073/pnas.85.14.4956}{{\em Proc. Nat. Acad. Sci.}
  {\bfseries 85} (1988) 4956--5960}.

\bibitem{Beem:2017ooy}
C.~Beem and L.~Rastelli, ``{Vertex operator algebras, Higgs branches, and
  modular differential equations},''
  \href{http://dx.doi.org/10.1007/JHEP08(2018)114}{{\em JHEP} {\bfseries 08}
  (2018) 114}, \href{http://arxiv.org/abs/1707.07679}{{\ttfamily
  arXiv:1707.07679 [hep-th]}}.

\bibitem{Kaidi:2022sng}
J.~Kaidi, M.~Martone, L.~Rastelli, and M.~Weaver, ``{Needles in a haystack. An
  algorithmic approach to the classification of 4d $ \mathcal{N} $ = 2
  SCFTs},'' \href{http://dx.doi.org/10.1007/JHEP03(2022)210}{{\em JHEP}
  {\bfseries 03} (2022) 210}, \href{http://arxiv.org/abs/2202.06959}{{\ttfamily
  arXiv:2202.06959 [hep-th]}}.

\bibitem{Beemetal}
C.~Beem, S.~S. Razamat, and P.~Singh, ``{Schur Indices of Class $\mathcal{S}$
  and Quasimodular Forms},'' \href{http://arxiv.org/abs/2112.10715}{{\ttfamily
  arXiv:2112.10715 [hep-th]}}.

\bibitem{2016arXiv161207423K}
V.~G. {Kac} and M.~{Wakimoto}, ``{A remark on boundary level admissible
  representations},'' \href{http://dx.doi.org/10.48550/arXiv.1612.07423}{{\em
  arXiv e-prints} (Dec., 2016) arXiv:1612.07423},
  \href{http://arxiv.org/abs/1612.07423}{{\ttfamily arXiv:1612.07423
  [math.RT]}}.

\bibitem{Arakawa:2016hkg}
T.~Arakawa and K.~Kawasetsu, ``{Quasi-lisse vertex algebras and modular linear
  differential equations},''
\href{http://arxiv.org/abs/1610.05865}{{\ttfamily arXiv:1610.05865 [math.QA]}}.
%%CITATION = ARXIV:1610.05865;%%.

\bibitem{2023arXiv230409681L}
B.~{Li}, H.~{Li}, and W.~{Yan}, ``{Spectral flow, twisted modules and MLDE of
  quasi-lisse vertex algebras},''
  \href{http://dx.doi.org/10.48550/arXiv.2304.09681}{{\em arXiv e-prints}
  (Apr., 2023) arXiv:2304.09681},
  \href{http://arxiv.org/abs/2304.09681}{{\ttfamily arXiv:2304.09681
  [math.QA]}}.

\bibitem{Hatsuda:2023iwi}
Y.~Hatsuda and T.~Okazaki, ``{Exact $ \mathcal{N} $ = 2$^{*}$ Schur line defect
  correlators},'' \href{http://dx.doi.org/10.1007/JHEP06(2023)169}{{\em JHEP}
  {\bfseries 06} (2023) 169}, \href{http://arxiv.org/abs/2303.14887}{{\ttfamily
  arXiv:2303.14887 [hep-th]}}.

\bibitem{Zheng:2022zkm}
H.~Zheng, Y.~Pan, and Y.~Wang, ``{Surface defects, flavored modular
  differential equations, and modularity},''
  \href{http://dx.doi.org/10.1103/PhysRevD.106.105020}{{\em Phys. Rev. D}
  {\bfseries 106} no.~10, (2022) 105020},
  \href{http://arxiv.org/abs/2207.10463}{{\ttfamily arXiv:2207.10463
  [hep-th]}}.

\bibitem{Pan:2023jjw}
Y.~Pan and Y.~Wang, ``{Flavored modular differential equations},''
  \href{http://dx.doi.org/10.1103/PhysRevD.108.085027}{{\em Phys. Rev. D}
  {\bfseries 108} no.~8, (2023) 085027},
  \href{http://arxiv.org/abs/2306.10569}{{\ttfamily arXiv:2306.10569
  [hep-th]}}.

\bibitem{Bourdier:2015wda}
J.~Bourdier, N.~Drukker, and J.~Felix, ``{The exact Schur index of
  $\mathcal{N}=4$ SYM},'' \href{http://dx.doi.org/10.1007/JHEP11(2015)210}{{\em
  JHEP} {\bfseries 11} (2015) 210},
  \href{http://arxiv.org/abs/1507.08659}{{\ttfamily arXiv:1507.08659
  [hep-th]}}.

\bibitem{Huang:2022bry}
M.-x. Huang, ``{Modular Anomaly Equation for Schur Index of $\mathcal{N}=4$
  Super-Yang-Mills},'' \href{http://arxiv.org/abs/2205.00818}{{\ttfamily
  arXiv:2205.00818 [hep-th]}}.

\bibitem{Hatsuda:2022xdv}
Y.~Hatsuda and T.~Okazaki, ``{$ \mathcal{N} $ = 2$^{*}$ Schur indices},''
  \href{http://dx.doi.org/10.1007/JHEP01(2023)029}{{\em JHEP} {\bfseries 01}
  (2023) 029}, \href{http://arxiv.org/abs/2208.01426}{{\ttfamily
  arXiv:2208.01426 [hep-th]}}.

\bibitem{Du:2023kfu}
B.-n. Du, M.-x. Huang, and X.~Wang, ``{Schur indices for $ \mathcal{N} $ = 4
  super-Yang-Mills with more general gauge groups},''
  \href{http://dx.doi.org/10.1007/JHEP03(2024)009}{{\em JHEP} {\bfseries 03}
  (2024) 009}, \href{http://arxiv.org/abs/2311.08714}{{\ttfamily
  arXiv:2311.08714 [hep-th]}}.

\bibitem{Beccaria:2024szi}
M.~Beccaria and A.~Cabo-Bizet, ``{Giant graviton expansion of Schur index and
  quasimodular forms},'' \href{http://arxiv.org/abs/2403.06509}{{\ttfamily
  arXiv:2403.06509 [hep-th]}}.

\bibitem{Gaiotto:2021xce}
D.~Gaiotto and J.~H. Lee, ``{The Giant Graviton Expansion},''
  \href{http://arxiv.org/abs/2109.02545}{{\ttfamily arXiv:2109.02545
  [hep-th]}}.

\bibitem{Gaiotto:2012xa}
D.~Gaiotto, L.~Rastelli, and S.~S. Razamat, ``{Bootstrapping the superconformal
  index with surface defects},''
  \href{http://dx.doi.org/10.1007/JHEP01(2013)022}{{\em JHEP} {\bfseries 01}
  (2013) 022}, \href{http://arxiv.org/abs/1207.3577}{{\ttfamily arXiv:1207.3577
  [hep-th]}}.

\bibitem{Gaiotto:2009we}
D.~Gaiotto, ``{N=2 dualities},''
  \href{http://dx.doi.org/10.1007/JHEP08(2012)034}{{\em JHEP} {\bfseries 08}
  (2012) 034}, \href{http://arxiv.org/abs/0904.2715}{{\ttfamily arXiv:0904.2715
  [hep-th]}}.

\bibitem{Pan:2019bor}
Y.~Pan and W.~Peelaers, ``{Schur correlation functions on $S^3\times S^1$},''
  \href{http://dx.doi.org/10.1007/JHEP07(2019)013}{{\em JHEP} {\bfseries 07}
  (2019) 013},
\href{http://arxiv.org/abs/1903.03623}{{\ttfamily arXiv:1903.03623 [hep-th]}}.
%%CITATION = ARXIV:1903.03623;%%.

\bibitem{Dedushenko:2019yiw}
M.~Dedushenko and M.~Fluder, ``{Chiral Algebra, Localization, Modularity,
  Surface defects, And All That},''
\href{http://arxiv.org/abs/1904.02704}{{\ttfamily arXiv:1904.02704 [hep-th]}}.
%%CITATION = ARXIV:1904.02704;%%.

\bibitem{Ardehali:2015bla}
A.~Arabi~Ardehali, ``{High-temperature asymptotics of supersymmetric partition
  functions},'' \href{http://dx.doi.org/10.1007/JHEP07(2016)025}{{\em JHEP}
  {\bfseries 07} (2016) 025},
\href{http://arxiv.org/abs/1512.03376}{{\ttfamily arXiv:1512.03376 [hep-th]}}.
%%CITATION = ARXIV:1512.03376;%%.

\bibitem{Gadde:2011ia}
A.~Gadde and W.~Yan, ``{Reducing the 4d Index to the $S^3$ Partition
  Function},'' \href{http://dx.doi.org/10.1007/JHEP12(2012)003}{{\em JHEP}
  {\bfseries 12} (2012) 003}, \href{http://arxiv.org/abs/1104.2592}{{\ttfamily
  arXiv:1104.2592 [hep-th]}}.

\bibitem{Nishioka:2011dq}
T.~Nishioka, Y.~Tachikawa, and M.~Yamazaki, ``{3d Partition Function as Overlap
  of Wavefunctions},'' \href{http://dx.doi.org/10.1007/JHEP08(2011)003}{{\em
  JHEP} {\bfseries 08} (2011) 003},
  \href{http://arxiv.org/abs/1105.4390}{{\ttfamily arXiv:1105.4390 [hep-th]}}.

\bibitem{Buican:2015hsa}
M.~Buican and T.~Nishinaka, ``{Argyres--Douglas theories, S$^1$ reductions, and
  topological symmetries},''
  \href{http://dx.doi.org/10.1088/1751-8113/49/4/045401}{{\em J. Phys.}
  {\bfseries A49} no.~4, (2016) 045401},
\href{http://arxiv.org/abs/1505.06205}{{\ttfamily arXiv:1505.06205 [hep-th]}}.
%%CITATION = ARXIV:1505.06205;%%.

\bibitem{Dedushenko:2016jxl}
M.~Dedushenko, S.~S. Pufu, and R.~Yacoby, ``{A one-dimensional theory for Higgs
  branch operators},'' \href{http://dx.doi.org/10.1007/JHEP03(2018)138}{{\em
  JHEP} {\bfseries 03} (2018) 138},
\href{http://arxiv.org/abs/1610.00740}{{\ttfamily arXiv:1610.00740 [hep-th]}}.
%%CITATION = ARXIV:1610.00740;%%.

\bibitem{Dedushenko:2019mzv}
M.~Dedushenko, ``{From VOAs to short star products in SCFT},''
  \href{http://dx.doi.org/10.1007/s00220-021-04066-2}{{\em Commun. Math. Phys.}
  {\bfseries 384} no.~1, (2021) 245--277},
  \href{http://arxiv.org/abs/1911.05741}{{\ttfamily arXiv:1911.05741
  [hep-th]}}.

\bibitem{Pan:2020cgc}
Y.~Pan and W.~Peelaers, ``{Deformation quantizations from vertex operator
  algebras},'' \href{http://dx.doi.org/10.1007/JHEP06(2020)127}{{\em JHEP}
  {\bfseries 06} (2020) 127}, \href{http://arxiv.org/abs/1911.09631}{{\ttfamily
  arXiv:1911.09631 [hep-th]}}.

\bibitem{Dedushenko:2019mnd}
M.~Dedushenko and Y.~Wang, ``{4d/2d $\rightarrow $ 3d/1d: A song of protected
  operator algebras},'' \href{http://arxiv.org/abs/1912.01006}{{\ttfamily
  arXiv:1912.01006 [hep-th]}}.

\bibitem{Benini:2010uu}
F.~Benini, Y.~Tachikawa, and D.~Xie, ``{Mirrors of 3d Sicilian theories},''
  \href{http://dx.doi.org/10.1007/JHEP09(2010)063}{{\em JHEP} {\bfseries 09}
  (2010) 063}, \href{http://arxiv.org/abs/1007.0992}{{\ttfamily arXiv:1007.0992
  [hep-th]}}.

\bibitem{Gadde:2011ik}
A.~Gadde, L.~Rastelli, S.~S. Razamat, and W.~Yan, ``{The 4d Superconformal
  Index from q-deformed 2d Yang-Mills},''
  \href{http://dx.doi.org/10.1103/PhysRevLett.106.241602}{{\em Phys. Rev.
  Lett.} {\bfseries 106} (2011) 241602},
  \href{http://arxiv.org/abs/1104.3850}{{\ttfamily arXiv:1104.3850 [hep-th]}}.

\bibitem{Alday:2013kda}
L.~F. Alday, M.~Bullimore, M.~Fluder, and L.~Hollands, ``{Surface defects, the
  superconformal index and q-deformed Yang-Mills},''
  \href{http://dx.doi.org/10.1007/JHEP10(2013)018}{{\em JHEP} {\bfseries 10}
  (2013) 018}, \href{http://arxiv.org/abs/1303.4460}{{\ttfamily arXiv:1303.4460
  [hep-th]}}.

\bibitem{MUKHI1990263}
S.~Mukhi and S.~Panda, ``Fractional-level current algebras and the
  classification of characters,''
  \href{http://dx.doi.org/https://doi.org/10.1016/0550-3213(90)90632-N}{{\em
  Nuclear Physics B} {\bfseries 338} no.~1, (1990) 263--282}.
  \url{https://www.sciencedirect.com/science/article/pii/055032139090632N}.

\bibitem{Buican:2017rya}
M.~Buican and Z.~Laczko, ``{Nonunitary Lagrangians and unitary non-Lagrangian
  conformal field theories},''
  \href{http://dx.doi.org/10.1103/PhysRevLett.120.081601}{{\em Phys. Rev.
  Lett.} {\bfseries 120} no.~8, (2018) 081601},
  \href{http://arxiv.org/abs/1711.09949}{{\ttfamily arXiv:1711.09949
  [hep-th]}}.

\bibitem{MdAbhishek:2023vgp}
M.~Abhishek, S.~Grover, D.~P. Jatkar, and K.~Singh, ``{Finding $G_2$ Higgs
  branch of 4D rank 1 SCFTs},''
  \href{http://arxiv.org/abs/2312.00275}{{\ttfamily arXiv:2312.00275
  [hep-th]}}.

\bibitem{Gadde:2011uv}
A.~Gadde, L.~Rastelli, S.~S. Razamat, and W.~Yan, ``{Gauge Theories and
  Macdonald Polynomials},''
  \href{http://dx.doi.org/10.1007/s00220-012-1607-8}{{\em Commun. Math. Phys.}
  {\bfseries 319} (2013) 147--193},
\href{http://arxiv.org/abs/1110.3740}{{\ttfamily arXiv:1110.3740 [hep-th]}}.
%%CITATION = ARXIV:1110.3740;%%.

\bibitem{Kinney:2005ej}
J.~Kinney, J.~M. Maldacena, S.~Minwalla, and S.~Raju, ``{An Index for 4
  dimensional super conformal theories},''
  \href{http://dx.doi.org/10.1007/s00220-007-0258-7}{{\em Commun. Math. Phys.}
  {\bfseries 275} (2007) 209--254},
  \href{http://arxiv.org/abs/hep-th/0510251}{{\ttfamily arXiv:hep-th/0510251}}.

\bibitem{Rastelli:2014jja}
L.~Rastelli and S.~S. Razamat, {\em {The Superconformal Index of Theories of
  Class $\mathcal {S}$}},
  \href{http://dx.doi.org/10.1007/978-3-319-18769-3_9}{pp.~261--305}.
\newblock Springer, 2016.
\newblock \href{http://arxiv.org/abs/1412.7131}{{\ttfamily arXiv:1412.7131
  [hep-th]}}.

\bibitem{Bourdier:2015sga}
J.~Bourdier, N.~Drukker, and J.~Felix, ``{The $\mathcal{N}=2$ Schur index from
  free fermions},'' \href{http://dx.doi.org/10.1007/JHEP01(2016)167}{{\em JHEP}
  {\bfseries 01} (2016) 167}, \href{http://arxiv.org/abs/1510.07041}{{\ttfamily
  arXiv:1510.07041 [hep-th]}}.

\bibitem{Mathur:1988na}
S.~D. Mathur, S.~Mukhi, and A.~Sen, ``{On the Classification of Rational
  Conformal Field Theories},''
  \href{http://dx.doi.org/10.1016/0370-2693(88)91765-0}{{\em Phys. Lett. B}
  {\bfseries 213} (1988) 303--308}.

\bibitem{Duan:2022kxr}
Z.~Duan, K.~Lee, S.~Lee, and L.~Li, ``{On classification of fermionic rational
  conformal field theories},''
  \href{http://dx.doi.org/10.1007/JHEP02(2023)079}{{\em JHEP} {\bfseries 02}
  (2023) 079}, \href{http://arxiv.org/abs/2210.06805}{{\ttfamily
  arXiv:2210.06805 [hep-th]}}.

\bibitem{Arakawa:2018egx}
T.~Arakawa, ``{Chiral algebras of class $\mathcal{S}$ and Moore-Tachikawa
  symplectic varieties},'' \href{http://arxiv.org/abs/1811.01577}{{\ttfamily
  arXiv:1811.01577 [math.RT]}}.

\bibitem{Arakawa:2023cki}
T.~Arakawa, T.~Kuwabara, and S.~M\"oller, ``{Hilbert Schemes of Points in the
  Plane and Quasi-Lisse Vertex Algebras with $\mathcal{N}=4$ Symmetry},''
  \href{http://arxiv.org/abs/2309.17308}{{\ttfamily arXiv:2309.17308
  [math.RT]}}.

\bibitem{Arakawa:2024ejd}
T.~Arakawa, X.~Dai, J.~Fasquel, B.~Li, and A.~Moreau, ``{On a series of simple
  affine VOAs at non-admissible level arising from rank One 4D SCFTs},''
  \href{http://arxiv.org/abs/2403.04472}{{\ttfamily arXiv:2403.04472
  [math.RT]}}.

\bibitem{Peelaers}
W.~Peelaers , unpublished.

\bibitem{Pan:2021ulr}
Y.~Pan, Y.~Wang, and H.~Zheng, ``{Defects, modular differential equations, and
  free field realization of N=4 vertex operator algebras},''
  \href{http://dx.doi.org/10.1103/PhysRevD.105.085005}{{\em Phys. Rev. D}
  {\bfseries 105} no.~8, (2022) 085005},
  \href{http://arxiv.org/abs/2104.12180}{{\ttfamily arXiv:2104.12180
  [hep-th]}}.

\bibitem{zhu1996modular}
Y.~Zhu, ``Modular invariance of characters of vertex operator algebras,'' {\em
  Journal of the American Mathematical Society} {\bfseries 9} no.~1, (1996)
  237--302.

\bibitem{Gaberdiel:2008pr}
M.~R. Gaberdiel and C.~A. Keller, ``{Modular differential equations and null
  vectors},'' \href{http://dx.doi.org/10.1088/1126-6708/2008/09/079}{{\em JHEP}
  {\bfseries 09} (2008) 079},
\href{http://arxiv.org/abs/0804.0489}{{\ttfamily arXiv:0804.0489 [hep-th]}}.
%%CITATION = ARXIV:0804.0489;%%.

\bibitem{Tuite:2014fha}
M.~P. Tuite and H.~D. Van, ``{On Exceptional Vertex Operator (Super)
  Algebras},'' \href{http://arxiv.org/abs/1401.5229}{{\ttfamily arXiv:1401.5229
  [math.QA]}}.

\bibitem{Bianchi:2019sxz}
L.~Bianchi and M.~Lemos, ``{Superconformal surfaces in four dimensions},''
  \href{http://dx.doi.org/10.1007/JHEP06(2020)056}{{\em JHEP} {\bfseries 06}
  (2020) 056}, \href{http://arxiv.org/abs/1911.05082}{{\ttfamily
  arXiv:1911.05082 [hep-th]}}.

\bibitem{Gang:2012yr}
D.~Gang, E.~Koh, and K.~Lee, ``{Line Operator Index on $S^{1}\times S^{3}$},''
  \href{http://dx.doi.org/10.1007/JHEP05(2012)007}{{\em JHEP} {\bfseries 05}
  (2012) 007}, \href{http://arxiv.org/abs/1201.5539}{{\ttfamily arXiv:1201.5539
  [hep-th]}}.

\bibitem{Bonetti:2018fqz}
F.~Bonetti, C.~Meneghelli, and L.~Rastelli, ``{VOAs labelled by complex
  reflection groups and 4d SCFTs},''
  \href{http://dx.doi.org/10.1007/JHEP05(2019)155}{{\em JHEP} {\bfseries 05}
  (2019) 155},
\href{http://arxiv.org/abs/1810.03612}{{\ttfamily arXiv:1810.03612 [hep-th]}}.
%%CITATION = ARXIV:1810.03612;%%.

\bibitem{Benvenuti:2011ga}
S.~Benvenuti and S.~Pasquetti, ``{3D-partition functions on the sphere: exact
  evaluation and mirror symmetry},''
  \href{http://dx.doi.org/10.1007/JHEP05(2012)099}{{\em JHEP} {\bfseries 05}
  (2012) 099},
\href{http://arxiv.org/abs/1105.2551}{{\ttfamily arXiv:1105.2551 [hep-th]}}.
%%CITATION = ARXIV:1105.2551;%%.

\end{thebibliography}\endgroup

\end{document}